\documentclass[11pt,twoside]{report}

\usepackage[german,british]{babel}
\usepackage[T1]{fontenc}
\usepackage{mathpazo}
\usepackage{helvet}
\usepackage{courier}
\usepackage{amssymb}
\usepackage{amsmath}
\usepackage{amsthm}
\usepackage{amscd}
\usepackage{bbm}
\usepackage{dsfont}
\usepackage{exscale}
\usepackage{fancyhdr}
\usepackage[all]{xy}
\usepackage{graphicx}
\usepackage[flushmargin]{footmisc}

\makeatletter
\renewcommand{\@endtheorem}{\endtrivlist}
\makeatother

\makeatletter
\newdimen\bibindent
\def\@biblabel#1{#1.}
\def\@lbibitem[#1]#2{\item[{[#1]}\hfill]\if@filesw
      {\let\protect\noexpand
       \immediate
       \write\@auxout{\string\bibcite{#2}{#1}}}\fi\ignorespaces}
\renewenvironment{thebibliography}[1]
     {\section*{\refname
        \@mkboth{\refname}{\refname}}\small
      \list{\@biblabel{\@arabic\c@enumiv}}%
           {\settowidth\labelwidth{\@biblabel{#1}}%
            \leftmargin\labelwidth
            \advance\leftmargin\labelsep
            \@openbib@code
            \usecounter{enumiv}%
            \let\p@enumiv\@empty
            \renewcommand\theenumiv{\@arabic\c@enumiv}}%
      \sloppy\clubpenalty4000\widowpenalty4000%
      \sfcode`\.\@m}
     {\def\@noitemerr
       {\@latex@warning{Empty `thebibliography' environment}}%
      \endlist}
\makeatother

\theoremstyle{definition}
\newtheorem{Def}{Definition}[chapter]
\newtheorem*{Def*}{Definition}
\newtheorem{Ex}[Def]{Example}

\theoremstyle{plain}
\newtheorem{The}[Def]{Theorem}
\newtheorem{Pro}[Def]{Proposition}
\newtheorem{Cor}[Def]{Corollary}
\newtheorem{Lem}[Def]{Lemma}

\theoremstyle{remark}
\newtheorem{Rem}[Def]{Remark}

\DeclareMathAlphabet{\mathib}{T1}{ptm}{b}{it}
\DeclareMathAlphabet{\mathscr}{U}{rsfs}{m}{n}
\DeclareMathAlphabet{\mathcat}{T1}{phv}{b}{n}
\DeclareMathAlphabet{\mathpb}{OT1}{ppl}{b}{it}

\DeclareMathOperator{\adv}{{adv}}
\DeclareMathOperator{\App}{App}
\DeclareMathOperator{\Aut}{Aut}
\DeclareMathOperator{\id}{id}
\DeclareMathOperator{\Hom}{hom}
\DeclareMathOperator{\obj}{Obj}
\DeclareMathOperator{\ret}{{ret}}
\DeclareMathOperator{\singsupp}{sing\thinspace supp}
\DeclareMathOperator{\supp}{supp}
\DeclareMathOperator{\trace}{Tr\,}
\DeclareMathOperator{\WF}{WF}

\newcommand{\set}[1]{\{ #1 \}}
\newcommand{\bset}[1]{\bigl\{ #1 \bigr\}}

\newcommand{\setf}[2]{\{ #1 \medspace \vert \medspace #2 \}}
\newcommand{\bsetf}[2]{\bigl\{ #1 \medspace \big\vert \medspace #2 \bigr\}}
\newcommand{\Bsetf}[2]{\Bigl\{ #1 \medspace \Big\vert \medspace #2 \Bigr\}}

\newcommand{\comm}[2]{[ #1 , #2 ]}
\newcommand{\bcomm}[2]{\bigl[ #1 , #2 \bigr]}

\newcommand{\abs}[1]{\lvert #1 \rvert}
\newcommand{\babs}[1]{\bigl\lvert #1 \bigr\rvert}

\newcommand{\norm}[1]{\lVert #1 \rVert}

\newcommand{\bra}[1]{\langle #1 \vert}

\newcommand{\ket}[1]{\vert #1 \rangle}

\newcommand{\scp}[2]{\langle #1 \vert #2 \rangle}

\newcommand{\dpair}[2]{\langle #1 \thinspace , \thinspace #2 \rangle}
\newcommand{\bdpair}[2]{\bigl\langle #1 \thinspace , \thinspace #2 \bigr\rangle}

\newcommand{\eg}{\textit{e.\,g.}}
\newcommand{\etc}{\textit{etc.}}
\newcommand{\ie}{\textit{i.\,e.}}
\newcommand{\viz}{\textit{viz.}}
\newcommand{\vv}{\textit{vice versa}}

\newcommand{\Cbb}{\mathds{C}}
\newcommand{\Mbb}{\mathds{M}}
\newcommand{\Nbb}{\mathds{N}}
\newcommand{\Rbb}{\mathds{R}}

\newcommand{\algunit}{\mathbbm{1}}

\newcommand{\Afrak}{\mathfrak{A}}
\newcommand{\Bfrak}{\mathfrak{B}}
\newcommand{\Cfrak}{\mathfrak{C}}
\newcommand{\Lfrak}{\mathfrak{L}}

\newcommand{\pib}{\mathib{p}}

\newcommand{\Ascr}{\mathscr{A}}
\newcommand{\Bscr}{\mathscr{B}}
\newcommand{\Cscr}{\mathscr{C}}
\newcommand{\Dscr}{\mathscr{D}}
\newcommand{\Escr}{\mathscr{E}}
\newcommand{\Fscr}{\mathscr{F}}
\newcommand{\Gscr}{\mathscr{G}}
\newcommand{\Hscr}{\mathscr{H}}
\newcommand{\Kscr}{\mathscr{K}}
\newcommand{\Mscr}{\mathscr{M}}
\newcommand{\Nscr}{\mathscr{N}}
\newcommand{\Oscr}{\mathscr{O}}
\newcommand{\Pscr}{\mathscr{P}}
\newcommand{\Rscr}{\mathscr{R}}
\newcommand{\Sscr}{\mathscr{S}}
\newcommand{\Uscr}{\mathscr{U}}
\newcommand{\Vscr}{\mathscr{V}}
\newcommand{\Xscr}{\mathscr{X}}
\newcommand{\Zscr}{\mathscr{Z}}

\newcommand{\unit}{\mathbf{1}}

\newcommand{\Rn}{\mathds{R}^n}

\newcommand{\fwcone}{V_{\negthinspace +}}
\newcommand{\clfwcone}{\overline{V}_{\negthinspace +}}

\newcommand{\BH}{\mathscr{B} ( \mathscr{H} )}
\newcommand{\BHzero}{\mathscr{B} ( \mathscr{H}_0 )}

\newcommand{\Poin}{\mathsf{P}_{\negthinspace +}^\uparrow}

\newcommand{\aLax}{\alpha_{( \Lambda , x )}}

\newcommand{\AO}{\mathfrak{A} ( \mathscr{O} )}
\newcommand{\AOone}{\mathfrak{A} ( \mathscr{O}_1 )}
\newcommand{\AOtwo}{\mathfrak{A} ( \mathscr{O}_2 )}
\newcommand{\AOperp}{\mathfrak{A} ( \mathscr{O}^\perp )}

\newcommand{\BM}{\mathfrak{B} ( \mathscr{M} )}
\newcommand{\BMone}{\mathfrak{B} ( \mathscr{M}_1 )}
\newcommand{\BMtwo}{\mathfrak{B} ( \mathscr{M}_2 )}

\newcommand{\Loc}{\mathcat{Loc}}

\newcommand{\Obs}{\mathcat{Obs}}
\newcommand{\Sts}{\mathcat{Sts}}
\newcommand{\Sym}{\mathcat{Sym}}
\newcommand{\TAlg}{\mathcat{TAlg}}
\newcommand{\Test}{\mathcat{Test}}
\newcommand{\Top}{\mathcat{Top}}

\newcommand{\Ss}{\mathcal{S}}

\newcommand{\Cinf}{C^\infty}
\newcommand{\Czeroinf}{C_0^\infty}

\newcommand{\mnarg}{\medspace . \medspace}
\newcommand{\nnarg}{\thinspace . \thinspace}

\newcommand{\Zuk}{\mathscr{Z}^1 ( \mathfrak{A}_\mathscr{K} )}
\newcommand{\Zup}{\mathscr{Z}^1 ( \mathfrak{A}_\mathscr{P} )}

\newcommand{\Zutp}{\mathscr{Z}^1_t ( \mathfrak{A}_\mathscr{P} )}

\newcommand{\bfF}{\boldsymbol{F}}
\newcommand{\bfS}{\boldsymbol{S}}
\newcommand{\f}{\mathrm{f}}
\newcommand{\F}{\mathrm{F}}
\newcommand{\RE}{\mathrm{R}}

\newcounter{defitem}

\newenvironment{deflist}{\begin{list}{(\alph{defitem})}%
  {\usecounter{defitem} \setlength{\topsep}{0ex}%
   \setlength{\parsep}{0.2ex} \setlength{\itemsep}{0.4ex}%
   \setlength{\leftmargin}{0em} \setlength{\itemindent}{0.5em}%
   }}{\end{list}}

\newcounter{proofitem}

\newenvironment{prooflist}{\begin{list}{(\roman{proofitem})}%
  {\usecounter{proofitem} \setlength{\topsep}{0ex}%
   \setlength{\parsep}{0.2ex} \setlength{\itemsep}{0.4ex}%
   \setlength{\leftmargin}{0em} \setlength{\itemindent}{0.5em}%
   \setlength{\listparindent}{1em}}}{\qed \end{list}}

\newcounter{propitem}

\newenvironment{proplist}{\begin{list}{(\roman{propitem})}%
  {\usecounter{propitem} \setlength{\topsep}{0ex}%
   \setlength{\parsep}{0.2ex} \setlength{\itemsep}{0.4ex}%
   \setlength{\leftmargin}{0em} \setlength{\itemindent}{0.5em}%
   }}{\end{list}}

\newcounter{remalphaitem}

\newenvironment{remalphalist}{\begin{list}{(\Alph{remalphaitem})}%
  {\usecounter{remalphaitem} \setlength{\topsep}{0ex}%
   \setlength{\parsep}{0.2ex} \setlength{\itemsep}{0.4ex}%
   \setlength{\leftmargin}{0em} \setlength{\itemindent}{0.5em}%
   }}{\end{list}}

\newcounter{exitem}

\newenvironment{exlist}{\begin{list}{(\arabic{exitem})}%
  {\usecounter{exitem} \setlength{\topsep}{0ex}%
   \setlength{\parsep}{0.2ex} \setlength{\itemsep}{0.4ex}%
   \setlength{\leftmargin}{0em} \setlength{\itemindent}{0.5em}%
   }}{\end{list}}

\newcounter{romanitem}

\newenvironment{romanlist}{\begin{list}{(\roman{romanitem})}%
  {\usecounter{romanitem} \setlength{\topsep}{0ex}%
   \setlength{\parsep}{0.2ex} \setlength{\itemsep}{0.4ex}%
   \setlength{\leftmargin}{2em}}}{\end{list}}

\newcounter{abcitem}

\newenvironment{abclist}{\begin{list}{(\Alph{abcitem})}%
  {\usecounter{abcitem} \setlength{\topsep}{0ex}%
   \setlength{\parsep}{0.2ex} \setlength{\itemsep}{0.4ex}%
   \setlength{\leftmargin}{2em}}}{\end{list}}

\newenvironment{indentbulletlist}{\begin{list}%
  {$\mspace{-50mu} \bullet$}%
  {\setlength{\topsep}{0ex} \setlength{\parsep}{0.2ex}%
   \setlength{\itemsep}{0.4ex} \setlength{\leftmargin}{1.5em}%
   \setlength{\rightmargin}{1.5em}\setlength{\itemindent}{1em}}}%
   {\end{list}}

\newenvironment{lindentbulletlist}{\begin{list}%
  {$\mspace{-50mu} \bullet$}%
  {\setlength{\topsep}{0ex} \setlength{\parsep}{0.2ex}%
   \setlength{\itemsep}{0.4ex} \setlength{\leftmargin}{1.5em}%
   \setlength{\itemindent}{1em}}}{\end{list}}

\begin{document}

\thispagestyle{plain}

~

\vspace{3cm}

\begin{center}
  {\huge General Covariance in\\[3mm]
    Algebraic Quantum Field Theory}\\[2cm]
  {\Large Romeo Brunetti,$^a$ Martin Porrmann$^a$ and Giuseppe
    Ruzzi$^b$}\\[5mm]
  \texttt{romeo.brunetti@desy.de}, \texttt{martin.porrmann@desy.de},
  \texttt{ruzzi@mat.uniroma2.it}

\vfill

$^a$
\begin{minipage}[t]{10.6cm}
  II. Institut f\"ur Theoretische Physik, Universit\"at Hamburg\\
  Luruper Chaussee 149, 22761 Hamburg, Germany
\end{minipage}

\vspace{3mm}

$^b$ 
\begin{minipage}[t]{10.6cm}
  Dipartimento di Matematica, Universit\`a di Roma "Tor Vergata''\\
  Via della Ricerca Scientifica, 00133 Roma, Italy
\end{minipage}

\end{center}

\newpage

\thispagestyle{empty}

\tableofcontents

\chapter{Introduction}
\label{chap:intro}

\section{The Problem of General Covariance}
\label{sec:gen-covariance}

The question of \emph{general covariance} of physical theories in
space and time can be traced back to the famous debate between
Gottfried Wilhelm Leibniz and Samuel Clarke (the latter assisted by
Sir Isaac Newton) on the ontological status of space in the years
1715--1716 \cite{alexander:1998}, the central question being if space
exists as a substance or as an absolute being and absolute motion is
present (Clarke) or if it is constituted only in relation to
co-existent things allowing for relativism in motions only (Leibniz).
This kind of problems also played an important role when the general
theory of relativity was being developed in the years around 1910.
While Albert Einstein first characterized generally covariant field
equations as inadmissible since they did not determine the metric
field uniquely as shown in the hole argument (\emph{Lochbetrachtung})
in the appendix of \cite{einstein/grossmann:1914a}, he later accepted
\cite{einstein:1916a} that all physical laws had to be expressed by
equations that are valid in all coordinate systems, \ie, which are
covariant (generally covariant) under arbitrary
substitutions.\footnote{\emph{``Die allgemeinen Naturgesetze sind
    durch Gleichungen auszudr\"ucken, die f\"ur alle
    Koordinatensysteme gelten, d. h.  die beliebigen Substitutionen
    gegen\"uber kovariant (allgemein kovariant) sind.''}
  \cite[p.\,776]{einstein:1916a}} The hole argument was dismissed by
the reasoning that it is not the spacetime metric that has to be fixed
uniquely by the field equations, but \label{page:gen-cov}\emph{only
  the physical phenomena that occur in spacetime need to be given a
  unique expression with reference to any description of spacetime}.
All physical statements are given in terms of spacetime coincidences;
measurements result in statements on meetings of material points of
the measuring rods with other material points or in coincidences
between watch hands and points on the clockface. The introduction of a
reference system merely serves the easy description of the totality of
all these coincidences (point-coincidence argument)
\cite[p.\,776f]{einstein:1916a}.\footnote{\emph{``Da sich alle unsere
    physikalischen Erfahrungen letzten Endes auf solche Koinzidenzen
    zur\"uckf\"uhren lassen, ist zun\"achst kein Grund vorhanden,
    gewisse Koordinatensysteme vor anderen zu bevorzugen, d.~h. wir
    gelangen zu der Forderung der allgemeinen Kovarianz.''}
  \cite[p.\,777]{einstein:1916a}}

In this paper we give an account of new results on how this concept of
general covariance is approached within the framework of algebraic
quantum field theory, a task already put forward in the Gibbs lecture
on \emph{Missed Opportunities} held by Freeman Dyson in 1972
\cite{dyson:1972} (including a warning on the necessity of a
\emph{``delicate weaving together of concepts from differential
  geometry, functional analysis, and abstract algebra''}). A couple of
articles on this problem have been published since then
\cite{fredenhagen/haag:1987,bannier:1988,bannier:1994}. But the
introduction of notions from category theory into these attempts sheds
new light on the problem at hand, opening up a new and promising
perspective \cite{brunetti/fredenhagen/verch:2003}. We will turn to a
presentation of these ideas after a short account of the ideas of
algebraic quantum field theory in this Introduction.

\section{Algebraic Quantum Field Theory}
\label{sec:aqft}

The use of algebraic concepts has been one of the
great achievements in the development of a theory for the
description of microscopical phenomena in the middle of the 1920's,
when a consistent description as quantum mechanics was presented by
Max Born, Pascual Jordan, Werner Heisenberg and Paul Adrien Maurice
Dirac \cite{heisenberg:1925,born/jordan:1925,dirac:1925,%
  born/heisenberg/jordan:1926}. For quantum mechanics an algebraic
generalization of the formalism has been presented in
\cite{jordan/neumann/wigner:1934}. With respect to physical systems of
infinitely many degrees of freedom (quantum field theory), a
mathematically sound foundation for an algebraic approach has been
given by Irving Segal in the 1940's \cite{segal:1947}, which later,
following the introduction of the concept of \emph{weak equivalence}
of representations of the algebras by James Michael Gardner Fell
\cite{fell:1960}, was supplemented by the principle of locality for
physical reasons in the work of Rudolf Haag \cite{haag:1955} and
Rudolf Haag, Hans-J\"urgen Borchers and Bert Schroer
\cite{borchers/haag/schroer:1963} and axiomatized in the innovative
publication \emph{An Algebraic Approach to Quantum Field Theory} by
Rudolf Haag and Daniel Kastler \cite{haag/kastler:1964} in 1964.

In the following years the mathematical framework thus acquired proved
to be a suitable basis for the formulation and investigation of
structural questions of quantum field theory. Among these are counted
the understanding of the multiparticle structure of field theory in
the Haag-Ruelle scattering theory \cite{haag:1958,ruelle:1962}, the
structure of inner symmetries in the Doplicher-Haag-Roberts (DHR)
theory \cite{doplicher/haag/roberts:1969a,%
doplicher/haag/roberts:1969b,doplicher/haag/roberts:1971,%
doplicher/haag/roberts:1974} and in the Doplicher-Roberts (DR) duality
theory \cite{doplicher/roberts:1989,doplicher/roberts:1990}. Since the
general assumptions of the algebraic approach were not restrictive
enough to exclude quantum field theoretic models exhibiting unphysical
behaviour, attempts were made to supplement the structure by specific
information about the phase-space structure of quantum field theory.
These fall into two different groups: First, there are qualitative
requirements in terms of \emph{compactness} conditions
\cite{haag/swieca:1965,fredenhagen/hertel:1979}; second, quantitative
requirements using the mathematical notion of \emph{nuclearity} were
introduced \cite{buchholz/wichmann:1986,buchholz/porrmann:1990}. The
former ones proved to be successful in eliminating quantum field
theoretic models without a particle interpretation, while the second
group was devised with thermodynamical applications in mind (cf. also
\cite{buchholz/junglas:1986,buchholz/jacobi:1987,%
  buchholz/junglas:1989}). They are an essential ingredient also in
the analysis of scattering states in theories with long-range
interactions via the concept of particle weights
\cite{buchholz/porrmann/stein:1991} in the attempt to generalize an
ansatz by Araki and Haag \cite{araki/haag:1967} for the description of
scattering states in terms of local observables
(cf.~\cite{porrmann:2004a,porrmann:2004b} for a thorough exposition).

The investigation of thermodynamic equilibrium states within the
framework of algebraic quantum field theory is based on the algebraic
reformulation of the Gibbs ansatz for statistical mechanics
\cite{haag/hugenholtz/winnink:1967} based on the so-called
Kubo-Martin-Schwinger (KMS) condition
\cite{kubo:1957,martin/schwinger:1959}. Whereas the Gibbs ansatz
requires the spectrum of the Hamiltonian to be discrete (system in a
box), the KMS condition is mathematically well-defined without this
restriction, giving direct access to the thermodynamic limit.
Considerable efforts have been undertaken since then to further
generalize this condition to relativistic systems
\cite{bros/buchholz:1994} and also to develop a mathematically sound
description of states not in equilibrium
\cite{buchholz/ojima/roos:2002}. Closely connected with these
questions is the Tomita-Takesaki modular theory, a cornerstone in the
development of the theory of operator algebras and important also in
applications to physical problems
(cf.~\cite{borchers:1995,borchers:2000} for an overview).

A comprehensive exposition of the principles and achievements of the
algebraic approach can be found in the monograph by Rudolf Haag
\cite{haag:1996}. To the open problems of algebraic quantum field
theory belongs the task to exhibit physically relevant interacting
models that meet the axioms (this is taken up in the approach of
constructive quantum field theory). Furthermore, the incorporation of
gravity into this framework is a major desideratum with all the
pertaining questions: description of equilibrium and nonequilibrium
states in curved spacetimes, their implications on cosmological
questions, and a host of other exciting problems.

\section{Observables, States and Experiments}
\label{sec:obsstat}

Information on physical systems is gathered by performing experiments,
\ie, one prepares the physical system in a clear-cut way to have it in
a certain \emph{state} $\alpha$ (or, more generally, in a
\emph{mixture of states} as a result of an incomplete preparation)
which is the embodiment of all information that can possibly be
gained, and then one has it interact with a measuring apparatus, the
\emph{observable} $Q$, which upon interaction reveals a certain change
of its (macroscopic) properties. The measurement results are
conveniently represented by real numbers so that formally we have to
deal with a mapping
\begin{equation}
  \label{eq:measurement}
  ( \alpha , Q ) \mapsto r \in \Rbb
\end{equation}
as the basis of any theoretical attempts to the description of nature.
A thorough analysis of this scheme has been given by Huzihiro Araki in
his lectures at the \emph{ETH} Zurich \cite{araki:1961,araki:1962} and
constitutes the basis for the corresponding discussion in
\cite{araki:1999}. Due to the fact that different procedures might
result in the preparation of the same state of the physical system and
that different measuring devices might return the same information
about this state at hand, these notions refer to equivalence classes
of preparation and measurement procedures rather than to individual
ones.

Moreover, the science of physics is concerned with phenomena that can
be reproduced (the theory of a \emph{single} event being a futile
venture) so that measurements are always concerned with ensembles
representing a physical system in the same state. Apart from the
limited accuracy of any actual measurement, this fact forces one to
look for a probabilistic description of experiments; not the
individual measurement result is decisive, but rather the probability
distribution as the outcome of many (in principle, indefinitely many)
experiments is the fact to be explained by a consistent theory. While
in classical physics this indeterminacy of the results is ascribed
solely to the limited accuracy in the preparation and measurement
procedures, the developed theory giving definite predictions on the
outcome of future experiments, this situation is changed drastically
in quantum physics, where \emph{in principle} only statistical
assertions are possible.

States of a physical system either being the result of a complete
preparation as \emph{pure} ones or else the upshot of an incomplete
preparation as a \emph{mixture}, \ie, a convex combination of the
former pure states, the mathematical structure of the state space
$\Sigma$ as a convex set is evident. The general formulation
\eqref{eq:measurement} takes the shape of a dual pairing,
\begin{equation}
  \label{eq:experiment}
  \Sigma \times \Ascr \ni ( \alpha , Q ) \mapsto \alpha ( Q ) \in \Rbb
  \text{,}
\end{equation}
where the structure of the set $\Ascr$ of (equivalence classes of)
observables is not fixed like that of $\Sigma$. One proposal for its
structure is that of a Jordan algebra
\cite{jordan/neumann/wigner:1934} with the product defined by
\begin{equation}
  \label{eq:jordan}
  A \circ B \doteq \frac{1}{4} \big( ( A + B )^2 - ( A - B ) \big) =
  \frac{1}{2} ( A B + B A ) \text{,} \quad A , B \in \Ascr \text{.}
\end{equation}
Note that the first expression of $A \circ B$ only involves scalar
multiples, sums and powers of elements of $\Ascr$, not the product of
$A$ and $B$ like the second one. A comprehensive presentation of
Jordan algebras and their state spaces can be found in
\cite[Part~I]{alfsen/shultz:2003} (see also \cite{landsman:1998}). A
theory for propositions about quantum systems has been developed under
the term \emph{quantum logic}. Garrett Birkhoff and John von Neumann
revealed their structure of an orthocomplemented lattice
\cite{birkhoff/neumann:1936}. If this is supplemented by the
assumptions of semi- or orthomodularity, one can establish that the
propositions are isomorphic to the set of orthogonal projection
operators on a Hilbert space. A comprehensive presentation of
orthocomplemented modular lattices is to be found in
\cite{varadarajan:1985}. In view of these investigations it is not a
far-fetched assumption to consider the structure of
\eqref{eq:experiment} as that of a dual pair of a $^*$-algebra $\Ascr$
with a state space $\Sigma$ as the normalized part of its (algebraic
or topological) dual. Normalization is required here in order for a
probabilistic interpretation to be possible.

\section{$C^*$-Algebras, States and Representations}
\label{sec:alg}

Quantum mechanics can conveniently be formulated with reference to a
certain Hilbert space, the unit rays of which represent the states of
the physical system under consideration while the self-adjoint
operators stand for the measurements. In this way the set of
measurements acquires the structure of the self-adjoint part of a
concrete $^*$-algebra of operators on a Hilbert space. Segal
\cite{segal:1947} had pointed to the fact that questions of physical
interest could be answered without having to select a certain Hilbert
space, if the measurements were part of a $C^*$-algebra $\Afrak$, \ie,
a Banach$^*$-algebra with the additional property
\begin{equation}
  \label{eq:cstar}
  \norm{A^* A} = \norm{A}^2 \text{,} \quad A \in \Afrak \text{.}
\end{equation}
This idea was absorbed by Haag and Kastler in 1964
\cite{haag/kastler:1964} who proposed not to consider just one single
$C^*$-algebra but rather a \emph{net of $C^*$-algebras} associated
with bounded regions $\Oscr$ of spacetime,
\begin{equation}
  \label{eq:local-net}
  \Oscr \mapsto \AO \text{,}
\end{equation}
claiming that the physical content of a theory is encoded in this
mapping. Thereby they stressed the local nature of measurements in
much the same spirit as Einstein did in 1916 \cite{einstein:1916a} by
stating \cite[p.\,851]{haag/kastler:1964} that \emph{``ultimately all
  physical processes are analyzed in terms of geometric relations of
  (unresolved) phenomena.''}

The net \eqref{eq:local-net} of local $C^*$-algebras is supposed to be
subject to the following conditions (the existence of the quasi-local
algebra is in fact not an assumption, but a consequence of the assumed
isotony):
\begin{subequations}
  \begin{lindentbulletlist}
  \item Isotony: For any two bounded regions $\Oscr_1$ and $\Oscr_2$
    \begin{equation}
      \label{eq:isotony}
      \Oscr_1 \subseteq \Oscr_2 \Rightarrow \AOone \subseteq \AOtwo
      \text{.}
    \end{equation}
  \item Locality: If the bounded regions $\Oscr_1$ and $\Oscr_2$ are
    spacelike separated, \ie, $\Oscr_1$ belongs to the spacelike
    complement of $\Oscr_2$, formally $\Oscr_1 \subseteq \Oscr_2'$,
    then
    \begin{equation}
      \label{eq:locality}
      \bcomm{\AOone}{\AOtwo} = \set{0} \text{.}
    \end{equation}
    Note that at this point the \emph{net structure} of $\Oscr \mapsto
    \AO$ enters, \ie~the directedness of the index set, implying that
    any two local algebras are contained in a larger one pertaining to
    a region $\Oscr$ that comprises both $\Oscr_1$ and $\Oscr_2$.
  \item Quasi-Local Algebra: The set-theoretic union of all local
    algebras $\AO$ generates a normed $^*$-algebra which upon
    completion is again a $C^*$-algebra $\Afrak$ called the
    \emph{quasi-local algebra}.\footnote{This minimal $C^*$-algebra is
      well-defined on account of isotony and called the inductive
      limit of the family $\bsetf{\AO}{\Oscr \text{~a bounded
          region}}$.}
  \item Special Relativistic Covariance: If there is a representation
    of the proper orthochronous Poincar\'e group $\Poin$ on the
    quasi-local algebra $\Afrak$ by automorphisms, $\Poin \ni (
    \Lambda , x ) \mapsto \aLax \in \Aut ( \Afrak )$, then these
    automorphisms should act covariantly, \ie
    \begin{equation}
      \label{eq:spec-rel-cov}
      \Afrak \big( (\Lambda , x ) \Oscr \big) = \aLax \AO \text{,}
    \end{equation}
    where $( \Lambda , x ) \Oscr$ denotes the image of $\Oscr$ under
    the Poincar\'e transformation.
  \end{lindentbulletlist}
\end{subequations}
The actual measurements are represented by the self-adjoint elements
of the corresponding $C^*$-algebras, while the states are the
positive, normalized, linear functionals $\omega$ on the quasi-local
algebra $\Afrak$.
\begin{subequations}
  \begin{indentbulletlist}
  \item Positivity: For any element $A \in \Afrak$
    \begin{equation}
      \label{eq:state-positivity}
      \omega ( A^* A ) \geqslant 0 \text{.}
    \end{equation}
  \item Normalization: The norm of $\omega$ as an element of the
    topological dual $\Afrak^*$ of $\Afrak$ is $1$,
    \begin{equation}
      \label{eq:state-normalization}
      \norm{\omega} \doteq \sup_{A \in \Afrak_1} \abs{\omega ( A )} =
      1 \text{.}
    \end{equation}
  \end{indentbulletlist}
\end{subequations}
In view of the probabilistic interpretation, the real number $\omega (
A )$ that the state $\omega$ returns when applied to a
\emph{self-adjoint} element $A$ of $\Afrak$ is considered as the
expectation value (mean value) resulting from measuring the
corresponding observable in the ensemble represented by $\omega$.

The algebra $\Afrak$ is not considered as a concrete algebra, \ie, an
algebra of bounded linear operators on a given Hilbert space. Instead
it can be represented as such in various ways on different Hilbert
spaces, where the term \emph{representation} refers to a
$^*$-homomorphism $\pi : \Afrak \rightarrow \Bscr ( \Hscr_\pi )$
(denoted $( \Hscr_\pi , \pi )$) that associates bounded operators $\pi
( A )$ on a certain representation Hilbert space $\Hscr_\pi$ with
elements $A \in \Afrak$, respecting the algebraic structure of
$\Afrak$ as well as its $^*$-operation. In general there will exist an
abundance of different representations of $\Afrak$ that are not
unitarily equivalent, \ie~connected by a unitary operator $U : \Hscr_1
\rightarrow \Hscr_2$, while some of them might be \emph{physically
  equivalent} in the weak sense of Fell \cite{fell:1960}
(cf.~\cite[p.\,851]{haag/kastler:1964}). In fact, via the GNS
(Gel'fand, Naimark, Segal) construction, every state $\omega$ on
$\Afrak$ is associated with its own representation $( \Hscr_\omega,
\pi_\omega )$. In the years to come this general framework,
supplemented by physically motivated further requirements depending on
the investigations to be performed, has proven to constitute a
flexible foundation to tackle structural questions of quantum field
theory. Among the most important contributions rank investigations on
nuclearity and the split property
\cite{buchholz/dantoni/longo:1990,doplicher/longo:1984}, equilibrium
and nonequilibrium physics
\cite{bros/buchholz:1994,buchholz/ojima/roos:2002}, modular theory
\cite{borchers:2000}, conformal field theory models
\cite{kawahigashi/longo:2004}, energy inequalities
\cite{fewster/verch:2003} and the approach to noncommutative spacetime
\cite{doplicher/fredenhagen/roberts:1995}. Apart from that, a sound
formulation of perturbation theory has been given within the algebraic
approach to quantum field theory
\cite{brunetti/fredenhagen:1997,brunetti/fredenhagen:2000}. This
review will concentrate on the investigation of general covariance as
one of the many examples referred to above for research in this
domain.

\chapter{Locally Covariant Quantum Field Theory}
\label{chap:lcqft}

\section{Preliminaries}
\label{sec:pre}

The following exposition will be concerned with four-dimensional,
globally hyperbolic spacetimes. Therefore, also to fix our notation,
we will summarize some of their basic properties; for a thorough
discussion the reader is referred to
\cite{hawking/ellis:1975,wald:1984}. The condition of global
hyperbolicity does not seem to be very restrictive on physical
grounds, its main purpose being to rule out certain causal
pathologies.

A spacetime is denoted by the pair $( \Mscr, g )$, where $\Mscr$ is a
smooth (\ie~$C^\infty$, Hausdorff, paracompact and connected)
four-dimensional manifold and $g$ is a Lorentzian metric on it with
signature $( +1 , -1 , -1 , -1 )$. Not to overburden the formalism, we
will usually only write $\Mscr$ for the manifold, implicitly including
its metric $g$. This also applies to other manifolds of the said type,
where the metric shares the diacritical signs (primes, subscripts,
\dots) differentiating the manifolds. The spacetimes are moreover
assumed to be oriented and time-oriented. Time orientability requires
the existence of a $C^\infty$ vector field $u$ on $\Mscr$ which is
timelike everywhere, \ie, $g ( u , u ) > 0$ at all points of $\Mscr$.
Nonspacelike tangent vectors $t$ are called \emph{future directed} or
\emph{past directed} if, in relation to the vector field $u$, one has
$g ( u , t ) > 0$ or $g ( u , t ) < 0$, respectively. This defines a
time direction on $\Mscr$ which is globally consistent. A smooth curve
$\gamma : I \to \Mscr$, $I$ a connected subset of $\Rbb$, with tangent
vector $\dot{\gamma}$ is called \emph{future directed timelike} if $g
( \dot{\gamma} , \dot{\gamma} ) > 0$ and $g (u , \dot{\gamma} ) > 0$
for all points of $\gamma$; it is called \emph{future directed causal}
if $g ( \dot{\gamma} , \dot{\gamma} ) \geqslant 0$ and $g (u ,
\dot{\gamma} ) > 0$ on all of $\gamma$. $\gamma$ is said to be
\emph{past directed timelike} or \emph{past directed causal} if in the
above definitions the Lorentz product of $u$ and $\dot{\gamma}$ is
negative. If the nonspacelike curve $\gamma$ is future directed and
the point $lim_{t \rightarrow \sup I} \, \gamma ( t )$ exists in
$\Mscr$ it is called the \emph{future endpoint}. The definition of a
\emph{past endpoint} requires $\gamma$ to be past directed and $\sup$
replaced by $\inf$. A curve $\gamma$ is called \emph{future
  inextendible} (\emph{past inextendible}) if the corresponding
endpoints do not exist; it is called \emph{inextendible} if it is
future and past inextendible \cite{oneill:1983}. Given a point $x \in
\Mscr$, the set $I^+ ( x )$ consists of all points in $\Mscr$ that can
be connected to $x$ by a future directed timelike curve, while $J^+ (
x )$ consists of all points in $\Mscr$ that are connectable to $x$ by
some future directed causal curve $\gamma : I \to \Mscr$ with $x =
\gamma ( \inf I )$.  Replacing the requirement of future-directedness
by past-directedness, one gets the sets $I^- ( x )$ and $J^- ( x )$,
respectively.

Two subsets $\Oscr_1$ and $\Oscr_2$ in $\Mscr$ are called causally
separated if it is impossible to connect them by a causal curve, \ie,
if for all $x \in \overline{\Oscr_1}$, $J^+ ( x ) \cup J^- ( x )$ has
empty intersection with $\overline{\Oscr_2}$ (the bar denotes closure
with respect to the induced topology of the manifold $\Mscr$). The
largest open set in $\Mscr$ that is causally separated from $\Oscr$ is
called the causal complement of $\Oscr$ denoted $\Oscr^\perp$. A
subset $\Sscr$ of $\Mscr$ is called \emph{achronal} if it does
\emph{not} contain any pair of points $x$, $y$ with $x \in I^+ ( y )$.
For a closed achronal set $\Sscr$ the \emph{future domain of
  dependence} $D^+ ( \Sscr )$ is the set of all points $x \in \Mscr$
with the property that \emph{any} past inextendible causal curve
through $x$ intersects $\Sscr$; the \emph{past domain of dependence}
$D^- ( \Sscr )$ is defined using future inextendible curves through
$x$ intersecting $\Sscr$. The \emph{full domain of dependence} $D (
\Sscr )$ is the union of $D^+ ( \Sscr )$ and $D^- ( \Sscr )$.

An oriented and time-oriented spacetime $\Mscr$ is called globally
hyperbolic if for each pair of points $x$, $y \in \Mscr$ the
intersection of $J^- ( y )$ and $J^+ ( x )$ is compact. This property
has been shown to be equivalent to the existence of a smooth foliation
of $\Mscr$ in terms of Cauchy surfaces by Bernal and S\'anchez
\cite{bernal/sanchez:2003}. A Cauchy surface is a smooth hypersurface
of $\Mscr$ that any inextendible causal curve intersects exactly once.
In globally hyperbolic spacetimes the global Cauchy problem (inital
value problem) for linear hyperbolic wave equations is well-posed, and
these wave equations possess unique retarded and advanced fundamental
solutions. While the property of global hyperbolicity does not refer
to any spacetime isometries, the concept of isometric embeddings will
be important later. Let $( \Mscr_1 , g_1 )$ and $( \Mscr_2 , g_2 )$ be
two globally hyperbolic spacetimes, then we call a mapping $\psi :
\Mscr_1 \rightarrow \Mscr_2$ an isometric embedding of $\Mscr_1$ into
$\Mscr_2$ if $\psi$ is a diffeomorphism onto its range $\psi ( \Mscr_1
)$, \ie, the map $\bar{\psi} : \Mscr_1 \rightarrow \psi ( \Mscr_1 )
\subseteq \Mscr_2$ is a diffeomorphism, and if $\psi$ is an isometry,
\ie, $\psi_* g_1 = g_2 \restriction \psi ( \Mscr_1 )$.

Given a globally hyperbolic spacetime $\Mscr$, we want to introduce
families of open subsets which will be used as an index set for nets
of local algebras according to \eqref{eq:local-net} and prove their
stability under isometric embeddings.
\paragraph*{Globally hyperbolic regions} Let $\Kscr^h ( \Mscr )$ be
the collection of subsets $\Oscr \subseteq \Mscr$ satisfying the
following properties:
\begin{romanlist}
\item $\Oscr$ is an open, arcwise-connected, relatively compact set,
  and $\Oscr^\perp \ne \emptyset$.
\item If $x_1$, $x_2 \in \Oscr$, then $J^+ ( x_1 ) \cap J^- ( x_2 )$
  is either empty or contained in $\Oscr$.
\end{romanlist}
It turns out by this definition that $\Kscr^h ( \Mscr )$ is a basis
for the topology of $\Mscr$ and that any element $\Oscr$ of $\Kscr^h (
\Mscr )$ with metric $g \restriction \Oscr$ and induced orientation
and time orientation is again a globally hyperbolic spacetime. There a
some straightforward geometrical results for $\Kscr^h ( \Mscr )$.
\begin{Lem}
  \label{lem:index-base}
  Let $\Mscr$ be a globally hyperbolic spacetime, then the following
  assertions hold:
  \begin{proplist}
  \item\label{list:base1} Let $\Oscr$, $\Oscr_1 \in \Kscr^h ( \Mscr )$
    be such that $\overline{\Oscr_1} \perp \Oscr$. Then there exists
    $\Oscr_2 \in \Kscr^h ( \Mscr )$ such that $\overline{\Oscr} \cup
    \overline{\Oscr_1} \subset \Oscr_2$.
  \item\label{list:base2} For any $\Oscr \in \Kscr^h ( \Mscr )$ there
    are $\Oscr_1$, $\Oscr_2 \in \Kscr^h ( \Mscr )$ such that
    $\overline{\Oscr_1} \perp \Oscr$ and $\overline{\Oscr} \cup
    \overline{\Oscr_1} \subseteq \Oscr_2$.
  \item\label{list:base3} Let $\Oscr \in \Kscr^h ( \Mscr )$ and let
    $\Sscr \subset \Mscr$ be an open set with $\overline{\Oscr}
    \subset \Sscr$, then $\Oscr^\perp \cap \Sscr \ne \emptyset$.
  \end{proplist}
\end{Lem}
This lemma can be used to show that the only obstruction to the
directedness of the family $\Kscr^h ( \Mscr )$, \viz~the fact that for
any pair $\Oscr_1$, $\Oscr_2 \in \Kscr^h ( \Mscr )$ there is $\Oscr
\in \Kscr^h ( \Mscr )$ such that $\Oscr_1 \cup \Oscr_2 \subseteq
\Oscr$, is the compactness of Cauchy surfaces.
\begin{Lem}
  \label{lem:noncomp-cauchy}
  Let the globally hyperbolic spacetime $\Mscr$ possess noncompact
  Cauchy surfaces, then $\Kscr^h ( \Mscr )$ is directed.
\end{Lem}
\begin{Lem}
  \label{lem:emb-index}
  Consider globally hyperbolic spacetimes $\Mscr_1$ and $\Mscr$ with
  an isometric embedding $\psi : \Mscr_1 \rightarrow \Mscr$ and define
  \begin{align*}
    \psi \big( \Kscr^h ( \Mscr_1 ) \big) & \doteq \bsetf{\psi ( \Oscr
      ) \subseteq \Mscr}{\Oscr \in \Kscr^h ( \Mscr_1 )} \text{,} \\
    \Kscr^h ( \Mscr ) \restriction \psi ( \Mscr_1 ) & \doteq
    \bsetf{\Oscr \subseteq \Mscr}{\overline{\Oscr} \subseteq \psi (
      \Mscr_1 )} \text{.}
  \end{align*}
  Then
  \begin{equation*}
    \Kscr^h ( \Mscr ) \restriction \psi ( \Mscr_1 ) = \psi \big(
    \Kscr^h ( \Mscr_1 ) \big) \text{.}
  \end{equation*}
\end{Lem}
\begin{proof}
  It is clear that if $\Oscr_1 \in \Kscr^h ( \Mscr_1 )$, then $\psi (
  \Oscr_1 ) \in \Kscr^h ( \Mscr ) \restriction \psi ( \Mscr_1 )$.
  Conversely, let $\Oscr \in \Kscr^h ( \Mscr ) \restriction \psi (
  \Mscr_1)$. By Lemma \ref{lem:index-base}\ref{list:base3},
  $\Oscr^\perp \cap \psi ( \Mscr_1 ) \ne \emptyset$. Then the causal
  complement of $\psi^{- 1} ( \Oscr )$ in $\Mscr_1$ is nonempty, and
  one arrives at $\psi^{- 1} ( \Oscr ) \in \Kscr^h ( \Mscr_1 )$.
\end{proof}
We stress that the collection $\Kscr^h ( \Mscr )$ possesses elements
which are not simply connected subsets of $\Mscr$ and elements whose
causal complement is not arcwise-connected. Problems associated with
this topological feature can be avoided by passing to certain
subcollections like diamond regions.
\paragraph*{Diamond regions} The set $\Kscr^d ( \Mscr )$ of diamonds
of $\Mscr$ \cite{ruzzi:2005b} consists of open subsets $\Oscr$ of
$\Mscr$, called \emph{diamonds}, for which there exist a spacelike
Cauchy surface $\Cscr$, a chart $( \Uscr , \phi )$ of $\Cscr$ and an
open ball $B$ of $\Rbb^3$ such that
\begin{equation*}
  \Oscr = D \big( \phi^{- 1} ( B ) \big) \text{,} \quad \overline{B}
  \subset \phi ( \Uscr ) \subset \Rbb^3 \text{.}
\end{equation*}
$\Oscr$ is said to be \emph{based} on $\Cscr$, \ie, $\phi^{- 1} ( B )$
is the \emph{base} of $\Oscr$. Any diamond $\Oscr$ is an open,
relatively compact, arcwise- and simply connected subset of $\Mscr$,
and its causal complement $\Oscr^\perp$ is likewise arcwise-connected.
Furthermore, given a diamond $\Oscr$, there exists a pair of diamonds
$\Oscr_1$, $\Oscr_2$ with
\begin{equation}
  \label{eq:diamond-relations}
  \overline{\Oscr} \text{,~} \overline{\Oscr_1} \subset \Oscr_2
  \text{,} \quad \Oscr \perp \Oscr_1 \text{.}
\end{equation}
The set of diamonds $\Kscr^d ( \Mscr )$ is a basis for the topology of
$\Mscr$.
\begin{Lem}
  \label{lem:emb-diamond}
  Consider globally hyperbolic spacetimes $\Mscr_1$ and $\Mscr$ with
  an isometric embedding $\psi : \Mscr_1 \rightarrow \Mscr$ and define
  \begin{align*}
    \psi \big( \Kscr^d ( \Mscr_1 ) \big) & \doteq \bsetf{\psi ( \Oscr
      ) \subseteq  \Mscr}{\Oscr \in \Kscr^d ( \Mscr_1 )} \text{,} \\
    \Kscr^d ( \Mscr ) \restriction \psi ( \Mscr_1 ) & \doteq
    \bsetf{\Oscr \subseteq \Mscr}{\overline{\Oscr} \subseteq \psi (
      \Mscr_1 )} \text{,}
  \end{align*}
 then
 \begin{equation*}
   \Kscr^d ( \Mscr ) \restriction \psi ( \Mscr_1 ) = \psi \big(
   \Kscr^d ( \Mscr_1 ) \big) \text{.}
 \end{equation*}
\end{Lem}
\begin{proof}
  We prove the inclusion $\supseteq$. Let $\Oscr$ be a diamond of
  $\Mscr_1$, \ie, there exist a spacelike Cauchy surface $\Cscr_1$ of
  $\Mscr_1$, a chart $( \Uscr , \phi_1 )$ such that $\Oscr_1 = D \big(
  \phi_1^{- 1} ( B ) \big)$, where $B$ is a ball of $\Rbb^3$ such that
  $\overline{\phi_1^{- 1} ( B )} \subseteq \Uscr$. Let $B_1$ be a ball
  of $\Rbb^3$ with $\overline{B} \subseteq \overline{B_1}$ and
  $\overline{\phi_1^{- 1} ( B_1 )} \subseteq \Uscr$. Note that $\psi
  \phi_1^{- 1} ( B_1 )$ is a compact, spacelike acausal open set of
  $\Mscr$ with boundaries and with a nonempty complement. By
  \cite{bernal/sanchez:2005}, there exists in $\Mscr$ a spacelike
  Cauchy surface $\Cscr$ such that $\overline{\psi \phi_1^{- 1} ( B_1
    )} \subseteq \Cscr$. Define $\Vscr \doteq \psi \phi_1^{- 1} ( B_1
  )$ and $\phi \doteq \phi_1 \psi^{- 1}$. The pair $( \Vscr , \phi )$
  is a chart of $\Cscr$ and $\overline{\phi^{- 1} ( B )} \subset
  \Vscr$. Finally, observe that, due to the properties of $\psi$, we
  have $\psi \big( D \big( \phi_1^{- 1} ( B ) \big) \big) = D \big(
  \psi \phi_1^{- 1} ( B ) \big) = D \big( \phi^{- 1} ( B ) \big)$,
  \viz~$\Kscr^d ( \Mscr ) \restriction \psi ( \Mscr_1 ) \supseteq
  \psi \big( \Kscr^d ( \Mscr_1 ) \big)$. The proof of the reverse
  inclusion is very similar.
\end{proof}

There are other interesting collections of subsets of $\Mscr$, \eg~the
family of \emph{regular diamonds} used in
\cite{guido/longo/roberts/verch:2001} that contains the family of
diamonds introduced above as a subset and has the same stability
property.

\section{Quantum Field Theories as Covariant Functors}
\label{sec:qftfunc}

The investigations to be presented draw upon the theory of categories
and functors. The reader not familiar with these concepts can find an
introduction to the field in \cite{maclane:1998}. The two categories
to be used in this chapter are
\begin{description}
\item[$\Loc$] This category has as objects the collection $\obj ( \Loc
  )$ consisting of all four-dimensional, globally hyperbolic
  spacetimes $( \Mscr , g )$ which are oriented and time-oriented. The
  morphisms between two such objects $( \Mscr_1 , g_1 )$ and $(
  \Mscr_2 , g_2 ) \in \obj ( \Loc )$ constitute the collection
  $\Hom_\Loc ( \Mscr_1 , \Mscr_2 )$ and are those isometric embeddings
  $\psi : \Mscr_1 \rightarrow \Mscr_2$ which satisfy the following
  additional constraints:
  \begin{proplist}
  \item\label{list:feat1} if $\gamma : [ a , b ] \rightarrow \Mscr_2$
    is any causal curve and $\gamma ( a )$, $\gamma ( b ) \in \psi
    (\Mscr_1)$ then the whole curve must be contained in the image
    $\psi ( \Mscr_1 )$, \ie, $\gamma ( t ) \in \psi ( \Mscr_1 )$ for
    all $t \in ] a , b [$;
  \item\label{list:feat2} the isometric embedding preserves
    orientation and time-orientation of the embedded spacetime.
  \end{proplist} 
  The composition for any morphisms $\psi$ and $\psi'$ in $\Hom_\Loc (
  \Mscr_1 , \Mscr_2 )$ and $\Hom_\Loc ( \Mscr_2 , \Mscr_3 )$,
  respectively, is the set-theoretic composition of maps $\psi' \circ
  \psi$. $\psi' \circ \psi : \Mscr_1 \rightarrow \Mscr_3$ is thus a
  well-defined map which obviously is a diffeomorphism onto its range
  $\psi' \big( \psi ( \Mscr_1 ) \big)$ and is evidently isometric.
  Likewise, the properties \ref{list:feat1} and \ref{list:feat2} are
  obviously fulfilled so that indeed $\psi' \circ \psi \in \Hom_\Loc (
  \Mscr_1 , \Mscr_3 )$. The associativity of the composition rule is
  an immediate consequence of that of the set-theoretic composition of
  maps. The requirement for categories that each $\Hom_\Loc ( \Mscr ,
  \Mscr )$ has to possess a unit element is fulfilled by the identity
  map $\id_\Mscr : x \mapsto x$, $x \in \Mscr$.
\item[$\Obs$] This category has as objects the class $\obj ( \Obs )$
  formed by all unital $C^*$-algebras, while the collection of
  morphisms between the objects (algebras) $\Afrak_1$ and $\Afrak_2$
  are the faithful (injective) unit-preserving $^*$-homomor\-phisms.
  Given such morphisms $\alpha$ and $\alpha'$ belonging to $\Hom_\Obs
  ( \Afrak_1 , \Afrak_2 )$ and $\Hom_\Obs ( \Afrak_2 , \Afrak_3 )$,
  respectively, their composition $\alpha' \circ \alpha$ is again the
  composition of maps and easily seen to be an element of $\Hom_\Obs (
  \Afrak_1 , \Afrak_3 )$. The unit element of $\Hom_\Obs ( \Afrak ,
  \Afrak )$ for any algebra $\Afrak \in \obj ( \Obs )$ is the identity
  $\id_\Afrak : A \mapsto A$, $A \in \Afrak$.
\end{description}
\begin{Rem}
  \begin{remalphalist}
  \item The first requirement \ref{list:feat1} that the morphisms of
    $\Loc$ have to satisfy enforces the induced and intrinsic causal
    structures to coincide for the embedded spacetime $\psi ( \Mscr_1)
    \subseteq \Mscr_2$ (cf.~\cite{kay:1997}). The second condition
    \ref{list:feat2} could be relaxed so that the possible reversal of
    orientation in space and time allows for a a discussion of PCT
    theorems.
  \item The framework presented above is open to variations depending
    on the problems to be investigated. The algebras that constitute
    the objects of $\Obs$ could be replaced by general $^*$-algebras,
    Borchers algebras or von Neumann algebras. With respect to $\Loc$,
    the spacetimes could have less specific properties or even be
    endowed with additional features like spin structures as in
    \cite{dimock:1982,verch:2001}.
  \end{remalphalist}
\end{Rem}
Now we introduce the concept of a locally covariant quantum field
theory which encodes the generally covariant principle of locality.
\begin{Def}
  \label{def:covfunctor}
  \begin{deflist}
  \item\label{list:lcqft} A \textbf{locally covariant quantum field
      theory} is a covariant functor $\Ascr$ between the two
    categories $\Loc$ and $\Obs$, \ie, writing $\alpha_\psi$ for the
    image $\Ascr ( \psi )$ of the morphism $\psi$ under the functor
    $\Ascr$, the following mappings
    \begin{equation*}
      \begin{CD}
        ( \Mscr , g ) @> \psi >> ( \Mscr' , g' )\\
        @V{\Ascr}VV     @VV{\Ascr}V\\
        \Ascr( \Mscr , g ) @> {\alpha_\psi} >> \Ascr( \Mscr' , g' )
      \end{CD}
   \end{equation*}
   have to satisfy the covariance properties
   \begin{subequations}
     \begin{align}
       \alpha_{\psi'} \circ \alpha_\psi & = \alpha_{\psi' \circ \psi}
       \text{,} \\
       \alpha_{\id_\Mscr} & = \id_{\Ascr ( \Mscr )}
     \end{align}
   \end{subequations}
   for all morphisms $\psi \in \Hom_\Loc ( \Mscr_1 , \Mscr_2 )$,
   $\psi' \in \Hom_\Loc ( \Mscr_2 , \Mscr_3 )$ and all spacetimes
   $\Mscr \in \obj ( \Loc )$.
 \item\label{list:caus-lcqft} A locally covariant quantum field theory
   given by the covariant functor $\Ascr$ is called \textbf{causal} if
   for morphisms $\psi_j \in \Hom_\Loc ( \Mscr_j , \Mscr )$, $j = 1$,
   $2$, embedding $\Mscr_1$ and $\Mscr_2$ in the common spacetime
   $\Mscr$ such that the sets $\psi_1 ( \Mscr_1)$ and $\psi_2 (
   \Mscr_2 )$ are causally separated in $\Mscr$, the corresponding
   algebras in $\Ascr ( \Mscr )$ commute, \ie,
   \begin{equation}
     \label{eq:causal-funct}
     \bcomm{\alpha_{\psi_1} \big( \Ascr ( \Mscr_1 )
       \big)}{\alpha_{\psi_2} \big( \Ascr ( \Mscr_2 ) \big)} = \set{0}
     \text{,}
   \end{equation}
   where $\comm{\Afrak}{\Bfrak} = \set{ A B - B A : A \in \Afrak, B
     \in \Bfrak}$ denotes the commutator of any pair $\Afrak$ and
   $\Bfrak$ of $C^*$-subalgebras contained in a common larger one.
 \item\label{list:slice} A locally covariant quantum field theory
   given by the covariant functor $\Ascr$ obeys the \textbf{time slice
     axiom} if for any $\psi \in \Hom_\Loc ( \Mscr , \Mscr' )$ with
   the property that $\psi ( \Mscr )$ contains a Cauchy surface for
   $\Mscr'$ we have
   \begin{equation}
     \label{eq:time-slice}
     \alpha_\psi \big( \Ascr ( \Mscr ) \big) = \Ascr ( \Mscr' )
     \text{.} 
   \end{equation}
 \end{deflist}
\end{Def}

A locally covariant quantum field theory given by the functor $\Ascr$
assigns to any globally hyperbolic spacetime a corresponding
$C^*$-algebra in such a way that the algebras can be identified when
the spacetimes are isometric. This is the mathematically precise way
to express the requirement for a unique expression for physical
phenomena in spacetime regardless of the selected coordinate system as
explained on p.\,\pageref{page:gen-cov}. The term \emph{local} is used
here in the sense of \emph{geometrically local} which is not to be
confused with locality in the sense of Einstein causality. Causal
properties are specified only in \ref{list:caus-lcqft} and
\ref{list:slice} of Definition~\ref{def:covfunctor}. Here, causality
means that elements of the algebras $\alpha_{\psi_1} \big( \Ascr (
\Mscr_1 ) \big)$ and $\alpha_{\psi_2} \big( \Ascr ( \Mscr_2 ) \big)$,
respectively, commute in the larger algebra $\Ascr ( \Mscr )$ when the
subregions $\psi_1 ( \Mscr_1 )$ and $\psi_2 ( \Mscr_2 )$ of $\Mscr$
are causally separated with respect to the metric $g$ on $\Mscr$. This
property is expected to hold generally for observable quantities which
can be localized in certain subregions of spacetimes. The time slice
axiom \ref{list:slice} (also called strong Einstein causality, or
existence of a causal dynamical law, cf.~\cite{verch:2001}) states
that the algebra of observables on a globally hyperbolic spacetime is
already determined by the algebra of observables localized in a
neighbourhood of any Cauchy surface.

\section{Recovering Algebraic Quantum Field Theory}
\label{sec:raqft}

In this section we will demonstrate that and how the setting of
algebraic quantum field theory on a fixed globally hyperbolic
spacetime as described in Section~\ref{sec:alg} can be regained from a
locally covariant quantum field theory given by a covariant functor
$\Ascr$ with the properties listed in the preceding
Section~\ref{sec:qftfunc}. The general assumptions on the net of local
algebras \eqref{eq:local-net}, \ie~isotony \eqref{eq:isotony} and
locality \eqref{eq:locality}, are supplemented here by the requirement
that all the local algebras $\AO$ as well as the quasi-local algebra
$\Afrak$ contain a common unit $\algunit$.

Let $( \Mscr , g )$ be an object in $\obj ( \Loc )$. Given $\Oscr \in
\Kscr^h ( \Mscr )$, $g_\Oscr$ denotes the Lorentzian metric restricted
to $\Oscr$ so that $( \Oscr , g_\Oscr)$ (with the induced orientation
and time orientation) is a member of $\obj ( \Loc )$. The
corresponding injection $\iota_{\Mscr , \Oscr} : ( \Oscr , g_\Oscr )
\rightarrow ( \Mscr , g )$, \ie~the identity map restricted to
$\Oscr$, is an element in $\Hom_\Loc ( \Oscr , \Mscr )$. Using this
notation we can formulate the following assertion.
\begin{Pro} 
  \label{pro:localnet}
  Let $\Ascr$ be a covariant functor with the properties stated in
  Definition~\ref{def:covfunctor}. Define a map $\Kscr^h ( \Mscr ) \ni
  \Oscr \mapsto \AO \subseteq \Ascr ( \Mscr )$ via
  \begin{equation*}
    \AO \doteq \alpha_{\Mscr , \Oscr} \big( \Ascr ( \Oscr ) \big)
    \text{,}  
  \end{equation*}
  where $\alpha_{\Mscr , \Oscr}$ is an abbreviation for
  $\alpha_{\iota_{\Mscr , \Oscr}}$. Then the following statements
  hold:
  \begin{proplist}
  \item The map fulfills isotony, \ie, for all $\Oscr_1$, $\Oscr_2 \in
    \Kscr^h ( \Mscr )$
    \begin{equation*}
      \Oscr_1 \subseteq \Oscr_2 \Rightarrow \AOone \subseteq \AOtwo
      \text{.}
    \end{equation*}
  \item Assume that there exists a group $G$ of isometric
    diffeomorphisms $\kappa : \Mscr \rightarrow \Mscr$ (\ie~$\kappa_*
    g = g$) preserving orientation and time orientation. Then there
    also exists a representation $G \ni \kappa \mapsto
    \tilde{\alpha}_\kappa$ of the group $G$ by $C^*$-algebra
    automorphisms $\tilde{\alpha}_\kappa : \Afrak \rightarrow \Afrak$
    of the $C^*$-algebra $\Afrak$ generated by the net
    $\bsetf{\AO}{\Oscr \in \Kscr^h ( \Mscr )}$ such that
    \begin{equation}
      \label{eq:cov1}
      \tilde{\alpha}_\kappa \big( \AO \big) = \Afrak \big( \kappa (
      \Oscr ) \big) \text{,} \quad \Oscr \in \Kscr ^h( \Mscr )
      \text{.} 
    \end{equation}
  \item If in addition the functor $\Ascr$ is causal, then for all
    subsets $\Oscr_1$, $\Oscr_2 \in \Kscr ^h ( \Mscr )$ that are
    causally separated from each other the corresponding algebras
    commute,
    \begin{equation*}
      \bcomm{\AOone}{\AOtwo} = \set{0} \text{.}
    \end{equation*}
  \end{proplist}
\end{Pro}
\begin{proof}
  \begin{prooflist}
  \item The proof of the first statement uses the covariance
    properties of the functor $\Ascr$. To demonstrate isotony, let
    $\Oscr_1$ and $\Oscr_2$ be elements of $\Kscr^h ( \Mscr )$ with
    $\Oscr_1 \subseteq \Oscr_2$. Let $\iota_{\Oscr_2 , \Oscr_1} :
    \Oscr_1 \rightarrow \Oscr_2$ be the canonical embedding obtained
    by restricting the identity map on $\Oscr_2$ to $\Oscr_1$, then
    $\iota_{\Oscr_2 , \Oscr_1} \in \Hom_\Loc ( \Oscr_1 , \Oscr_2 )$.
    With $\alpha_{\Mscr , \Oscr_1} \doteq \alpha_{\iota_{\Mscr ,
        \Oscr_1}}$, \etc, covariance of the functor $\Ascr$ implies
    $\alpha_{\Mscr , \Oscr_1} = \alpha_{\Mscr , \Oscr_2} \circ
    \alpha_{\Oscr_2 , \Oscr_1}$, and therefore
    \begin{align*}
      \AOone = \alpha_{\Mscr , \Oscr_1} \big( \Ascr ( \Oscr_1 ) \big)
      & = \alpha_{\Mscr , \Oscr_2} \big( \alpha_{\Oscr_2 , \Oscr_1}
      \big( \Ascr ( \Oscr_1 ) \big) \big) \\
      & \subseteq \alpha_{\Mscr , \Oscr_2} \big( \Ascr ( \Oscr_2 )
      \big) = \AOtwo \text{,}
    \end{align*}
    since $\alpha_{\Oscr_2 , \Oscr_1} \big( \Ascr ( \Oscr_1 ) \big)
    \subseteq \Ascr ( \Oscr_2 )$ by the very properties of the functor
    $\Ascr$.
  \item To prove the second statement, let $\kappa : \Mscr \rightarrow
    \Mscr $ be a diffeomorphism preserving the metric as well as time
    orientation and orientation. The functor $\Ascr$ assigns an
    automorphism $\alpha_\kappa : \Ascr ( \Mscr ) \rightarrow \Ascr(
    \Mscr )$ to this special diffeomorphism. To the map
    $\tilde{\kappa} : \Oscr \rightarrow \kappa ( \Oscr )$ with $x
    \mapsto \kappa ( x )$ the functor $\Ascr$ associates a morphism
    $\alpha_{\tilde{\kappa}} : \Ascr ( \Oscr ) \rightarrow \Ascr \big(
    \kappa ( \Oscr ) \big)$. Hence
    \begin{align*}
      \alpha_\kappa \big( \AO \big) & = \alpha_\kappa \circ
      \alpha_{\Mscr , \Oscr} \big( \Ascr ( \Oscr ) \big) =
      \alpha_{\kappa \circ \iota_{\Mscr , \Oscr}} \big( \Ascr ( \Oscr
      ) \big) = \alpha_{\iota_{\Mscr , \kappa ( \Oscr )} \circ
        \tilde{\kappa}} \big( \Ascr ( \Oscr ) \big) \\ 
      & = \alpha_{\Mscr , \kappa ( \Oscr )} \circ
      \alpha_{\tilde{\kappa}} \big( \Ascr ( \Oscr ) \big) = \alpha_{
        \Mscr , \kappa ( \Oscr )} \big( \Ascr \big( \kappa ( \Oscr )
      \big) \big) = \Afrak \big( \kappa ( \Oscr ) \big) \text{.}
    \end{align*}
    Since $\Afrak \subseteq \Ascr( \Mscr )$, the definition of
    $\tilde{\alpha}_{\kappa}$ as the restriction of $\alpha_{\kappa}$
    to $\Afrak$ yields an automorphism with the required properties.
    The feature of a group representation is an immediate consequence
    of the covariance properties of the functor which imply
    $\alpha_{\kappa_1} \circ \alpha_{\kappa_2} = \alpha_{\kappa_1
      \circ \kappa_2}$ for any pair of elements $\kappa_1$, $\kappa_2
    \in G$ together with \eqref{eq:cov1}. Therefore, one concludes
    that indeed $\tilde{\alpha}_{\kappa_1} \circ
    \tilde{\alpha}_{\kappa_2} = \tilde{\alpha}_{\kappa_1 \circ
      \kappa_2}$.
  \item If $\Oscr_1$ and $\Oscr_2$ are causally separated elements of
    $\Kscr^h ( \Mscr )$, one can find a Cauchy surface $\Sigma$ in
    $\Mscr$ and a pair of disjoint subsets $\Sscr_1$ and $\Sscr_2$ of
    $\Sigma$, both connected and relatively compact, which satisfy
    $\Oscr_j \subseteq \Sscr_j^{\perp \perp}$, $j = 1$, $2$. The sets
    $\Sscr_j^{\perp \perp}$ are causally separated members of $\Kscr^h
    ( \Mscr )$, and, when equipped with the appropriate restrictions
    of the metric $g$, they are globally hyperbolic spacetimes in
    their own right, naturally embedded into $\Mscr$. According to the
    causality assumption on $\Ascr$, the algebras $\Afrak \big(
    \Sscr_j^{\perp \perp} \big) = \alpha_{\Mscr , \Sscr_j^{\perp
        \perp}} \big( \Ascr \big( \Sscr_j^{\perp \perp} \big) \big)$
    are pairwise commuting subalgebras of $\Ascr ( \Mscr )$ and, due
    to isotony, $\Afrak( \Oscr_j ) \subseteq \Afrak \big(
    \Sscr_j^{\perp \perp} \big)$ which implies that also
    $\bcomm{\Afrak ( \Oscr_1 )}{\Afrak (\Oscr_2 )} = \set{0}$. This
    completes the proof.
  \end{prooflist}
  \renewcommand{\qed}{}
\end{proof}
Proposition~\ref{pro:localnet} shows that the Haag-Kastler framework
of quantum field theory can be recovered within the functorial
approach presented in Section~\ref{sec:qftfunc}.

\section{Quantum Fields as Natural Transformations}
\label{sec:qfnattrans}

In the previous section a quantum field \emph{theory} has been defined
in terms of a covariant functor: an algebra is mapped into another
algebra via the endomorphism $\alpha_\psi = \Ascr ( \psi )$, but
\emph{a priori} nothing is said about how specific elements of the
algebras are mapped onto each other by that transformation. In this
section the possibility to define locally covariant \emph{fields} is
considered, their importance lying in the potential construction of
fields that only locally depend on the geometry. The physical meaning
underlying this construction is the idea that such fields might serve
as carriers of information from one point of a spacetime to another in
absence of global isometries like translations, or even from one
spacetime to another one. The definition of locally covariant quantum
fields to be given below generalizes the approach of G{\aa}rding and
Wightman, who characterized quantum fields as operator-valued
distributions. Here the distributions are not specified explicitly in
this way, but can take values in a topological $^*$-algebra instead.

We consider a family $\bset{\Afrak ( \Mscr )}$ of topological
$^*$-algebras that are indexed by all spacetimes $\Mscr$ in $\obj (
\Loc )$.  For each spacetime a quantum field is defined as a
\emph{generalized algebra-valued distribution}, \ie~a map $\Phi_\Mscr
: C_0^\infty ( \Mscr ) \rightarrow \Afrak ( \Mscr )$ which is supposed
to be continuous, but not necessarily linear. The collection of all
these mappings constitutes the family $\Phi \doteq \bset{\Phi_\Mscr}$.
In addition, we demand that for any $\psi \in \Hom_\Loc ( \Mscr_1 ,
\Mscr_2 )$ there exists a continuous endomorphism $\alpha_\psi :
\Afrak ( \Mscr_1 ) \rightarrow \Afrak( \Mscr_2 )$ so that
\begin{equation*}
  \alpha_\psi \big( \Phi_{\Mscr_1} ( f ) \big) = \Phi_{\Mscr_2} \big(
  \psi_* ( f ) \big) \text{,}
\end{equation*}
where $f \in C_0^\infty ( \Mscr_1 )$ is any test function and $\psi_*
( f ) = f \circ \psi^{- 1}$. The family $\bset{\Phi_\Mscr}$ with these
covariance conditions is called a \emph{locally covariant quantum
  field}. This simple description has a functorial translation to be
outlined next.

Again we consider the category $\Loc$ and furthermore introduce the
category $\TAlg$ consisting of topological $^*$-algebras with unit
elements as objects and of continuous $^*$-endomorphisms as morphisms.
This means that $\alpha : \Afrak_1 \rightarrow \Afrak_2$ is an element
of $\Hom_\TAlg ( \Afrak_1 , \Afrak_2 )$ if it is a continuous,
unit-preserving, injective $^*$-morphism. Let $\Test$ denote the
category containing all possible test function spaces over $\Loc$ as
objects, \ie, the objects are all spaces $C_0^\infty ( \Mscr )$ of
smooth, compactly supported test functions on $\Mscr$ for all $\Mscr
\in \obj ( \Loc )$, and the morphisms are all possible push-forwards
$\psi_*$ of isometric embeddings $\psi : \Mscr_1 \rightarrow \Mscr_2$.
The action of any push-forward $\psi_*$ on an element of a test
function space has been defined in the preceding paragraph, and it
clearly satisfies the requirements for morphisms between test function
spaces.

Now, let the locally covariant quantum field \emph{theory} $\Ascr$ be
defined as a functor according to Definition~\ref{def:covfunctor} with
the modification that the category $\TAlg$ replaces the category
$\Obs$; again $\alpha_\psi$ stands for $\Ascr ( \psi )$ whenever
$\psi$ is any morphism in $\Loc$. Moreover, let $\Dscr$ be the
covariant functor between $\Loc$ and $\Test$ assigning to each $\Mscr
\in \obj ( \Loc )$ the test function space $\Dscr ( \Mscr ) =
C_0^\infty ( \Mscr )$ and to each morphism $\psi$ of $\Loc$ its
push-forward: $\Dscr ( \psi ) = \psi_*$. $\Test$ as well as $\TAlg$
are regarded as subcategories of the category of all topological
spaces, $\Top$, leading to the following definition.
\begin{Def}
  \label{def:nattrans}
  A \textbf{locally covariant quantum field} $\Phi$ is a natural
  transformation between the functors $\Dscr$ and $\Ascr$, \ie, for
  any object $\Mscr \in \Loc$ there exists a morphism $\Phi_\Mscr :
  \Dscr ( \Mscr ) \rightarrow \Ascr( \Mscr )$ in $\Top$ such that for
  each morphism $\psi \in \Hom_\Loc ( \Mscr_1 , \Mscr_2 )$ the
  following diagram commutes:
  \begin{equation*}
    \begin{CD}
      \Dscr ( \Mscr_1 ) @> \Phi_{\Mscr_1} >> \Ascr ( \Mscr_1 ) \\
      @V{\psi_*}VV     @VV{\alpha_\psi}V \\
      \Dscr ( \Mscr_2 ) @>> \Phi_{\Mscr_2} > \Ascr ( \Mscr_2 ) 
    \end{CD}
  \end{equation*}
  Explicitly,
  \begin{equation*}
    \alpha_\psi \circ \Phi_{\Mscr_1} = \Phi_{\Mscr_2} \circ \psi_*
    \text{,} 
  \end{equation*}
  which is the requirement of covariance for fields.
\end{Def}
\begin{Rem}
  \begin{remalphalist}
  \item This definition is open to extensions. The test function
    spaces $C_0^\infty ( \Mscr )$ could be replaced by smooth,
    compactly supported sections of vector bundles, and as
    endomorphisms of these test section spaces one takes suitable
    pull-backs of vector-bundle endomorphisms. One can also include
    conditions on the wave front set of the field operators
    (cf.~Definition~\ref{def:wavefrontset} below).
  \item The notion of causality is introduced in an obvious way: A
    locally covariant quantum field is \emph{causal} if $\Phi_\Mscr (
    f )$ and $\Phi_\Mscr ( h )$ commute for all $f$, $h \in \Dscr (
    \Mscr )$ such that $\supp f \perp \supp h$.
  \item Admitting nonlinear fields in Definition~\ref{def:nattrans}
    opens up the possibility to apply it to more general objects,
    \eg~the definition of a locally covariant $S$-matrix patterned
    according to the definition of a \emph{local} $S$-matrix of
    Epstein and Glaser \cite{brunetti/fredenhagen:2000} (see
    also\cite{brunetti/fredenhagen:2004}).
  \item We give some examples in Chapter~\ref{chap:examples}.
  \end{remalphalist}
\end{Rem}

\section{On the Notion of State Space}
\label{sec:statespace}
 
We have seen previously that part of the physical description of a
theory proceeds through the selection (preparation) of a suitable
class of states. Here we attempt at a definition of a state space
suitable for the description of a generally covariant quantum field
theory.

Indeed, suppose that our theory is given in terms of a covariant
functor $\Ascr$. The question arises what the concept of a state might
be in this case. The first, quite natural idea is to say that a state
is a family $\bsetf{\omega_\Mscr}{\Mscr \in \obj ( \Loc )}$ indexed by
the members in the object class $\Loc$, where each $\omega_\Mscr$ is a
state on the $C^*$-algebra $\Ascr ( \Mscr )$. One might wonder if
there are families of states $\bsetf{\omega_\Mscr}{\Mscr \in \obj (
  \Loc )}$ that are distinguished by a property which in our framework
would correspond to \emph{local diffeomorphism invariance}, \viz,
\begin{equation*}
  \omega_\Mscr' \circ \alpha_\psi = \omega_\Mscr \quad \text{on $\Ascr
    ( \Mscr )$}
\end{equation*}
for all $\psi \in \Hom_\Loc ( \Mscr , \Mscr' )$.  However, there is a
simple argument that shows that the above property will, in general,
not be physically realistic. We refer to
\cite{brunetti/fredenhagen/verch:2003} for a thorough explanation.
 
A crucial question is whether there exists a more general concept of
\emph{invariance} that can be attributed to families of states
$\bsetf{\omega_\Mscr}{\Mscr \in \obj ( \Loc )}$ for a locally
covariant quantum field theory given by a functor $\Ascr$. We will
argue that there is a positive answer to this question. To arrive at
an explanation, let us fix some concepts that will turn out to be
useful also in the subsequent chapters.
\paragraph*{Folium of a representation} Let $\Afrak$ be a
$C^*$-algebra and $\pi : \Afrak \rightarrow \BH$ be a
$^*$-representation of $\Afrak$ by bounded linear operators on a
Hilbert space $\Hscr$.  The \emph{folium of $\pi$}, denoted $\bfF (
\pi )$, is the set of all states $\omega'$ on $\Afrak$ which can be
written as
\begin{equation*}
  \omega' ( A ) = \trace \big( \rho \cdot \pi ( A ) \big) \text{,}
  \quad A \in \Afrak \text{,}
\end{equation*}
where $\trace$ denotes the trace in $\Hscr$. In other words, the
folium of a representation consists of all density matrix states in
that representation.
\paragraph*{Local quasi-equivalence and local normality}  Let $\Ascr$ be a
locally covariant quantum field theory and let, for fixed $\Mscr$,
$\omega$ and $\tilde{\omega}$ be two states on $\Ascr ( \Mscr )$. We
call these states (or their GNS representations, denoted $\pi$ and
$\tilde{\pi}$, respectively) \emph{locally quasi-equivalent} if for
all $\Oscr \in \Kscr^h ( \Mscr )$ the relation
\begin{equation}
  \label{eq:F1}
  \bfF ( \pi \circ \alpha_{\Mscr , \Oscr} ) = \bfF ( \tilde{\pi} \circ
  \alpha_{\Mscr,\Oscr} )
\end{equation}
is valid, where $\alpha_{\Mscr , \Oscr} = \alpha_{\iota_{\Mscr ,
    \Oscr}}$ and $\iota_{\Mscr , \Oscr} : \Oscr \rightarrow \Mscr$ is
the natural embedding.

Moreover, we say that \emph{$\omega$ is locally normal to
  $\tilde{\omega}$} (or to the corresponding GNS representation
$\tilde{\pi}$) if
\begin{equation}
  \label{eq:F2}
  \omega \circ \alpha_{\Mscr , \Oscr} \in \bfF ( \tilde{\pi} \circ
  \alpha_{\Mscr , \Oscr} )
\end{equation}
for all $\Oscr \in \Kscr^h ( \Mscr )$.

It is known that quasifree states of the free scalar field on globally
hyperbolic spacetimes which fulfill the microlocal spectrum condition
(cf.~Section~\ref{sec:microlocal}) are locally quasi-equivalent. We
also note that the property of a state to fulfill the microlocal
spectrum condition is a locally covariant property (owing to the
covariant behaviour of wavefront sets of distributions under
diffeomorphisms \cite{hoermander:2003}). Thus, for a locally covariant
quantum field theory it is natural to assume that, if $\omega_\Mscr'$
fulfills (any suitable variant of) the microlocal spectrum condition,
then so does $\omega_\Mscr' \circ \alpha_\psi$ for any $\psi \in
\Hom_\Loc ( \Mscr , \Mscr' )$.  In the case where also the folia of
states (\ie~the folia of their GNS representations) satisfying the
microlocal spectrum condition coincide locally, one thus obtains the
invariance of local folia under local diffeomorphisms, more precisely,
at the level of the GNS representations of $\omega_\Mscr$ and
$\omega_\Mscr'$, one has
\begin{equation*}
  \bfF ( \pi_\Mscr' \circ \alpha_\psi \circ \alpha_{\Mscr , \Oscr} ) =
  \bfF ( \pi_\Mscr \circ \alpha_{\Mscr , \Oscr} )
\end{equation*}
for all $\psi \in \Hom_\Loc ( \Mscr , \Mscr')$ and all $\Oscr \in
\Kscr^h ( \Mscr )$. All these properties are known to hold for
quasifree states of the free scalar field fulfilling the microlocal
spectrum condition on globally hyperbolic spacetimes.

Thus one sees that local diffeomorphism invariance really occurs at
the level of local folia of states for $\Ascr$. In this light it
appears natural to give a functorial description of the space of
states taking this form of local diffeomorphism invariance into
account. To this end, it is convenient to first introduce a new
category, the category of the set of states.
\begin{description}
\item[$\Sts$] An object $\bfS \in \obj ( \Sts )$ is a set of states on
  a $C^*$-algebra $\Afrak$. Morphisms between members $\bfS'$ and
  $\bfS$ of $\obj ( \Sts )$ are positive maps $\gamma^* : \bfS'
  \rightarrow \bfS$. In the present work, $\gamma^*$ always arises as
  the dual map of a faithful $C^*$-algebra endomorphism $\gamma :
  \Afrak \rightarrow \Afrak$ via
  \begin{equation*}
    \gamma^* \omega' ( A ) = \omega' ( \gamma ( A ) ) \text{,} \quad
    \omega' \in \bfS' \text{,~} A \in \Afrak \text{.}
  \end{equation*}
  The category $\Sts$ is therefore \emph{dual} to the category $\Obs$.
  The composition rules for morphisms are thus obvious.
\end{description}
Now we can define a state space for a locally covariant quantum field
theory in a functorial manner.
\begin{Def}
  \label{def:state-space}
  Let $\Ascr$ be a locally covariant quantum
  field theory.
  \begin{deflist}
  \item\label{list:statespace1} A \textbf{state space} for $\Ascr$ is
    a contravariant functor $\Sscr$ between $\Loc $ and $\Sts$:
    \begin{equation*}
      \begin{CD}
        (\Mscr,g) @>\psi>> (\Mscr',g')\\
        @V{\Sscr}VV     @VV{\Sscr}V\\
        \Sscr(\Mscr,g)@<{\alpha_\psi^*}<< \Sscr(\Mscr',g')
      \end{CD}
    \end{equation*}
    where $\Sscr ( \Mscr )$ is a set of states on $\Ascr ( \Mscr )$
    and $\alpha_\psi^*$ is the dual map of $\alpha_\psi$; the
    contravariance property is
    \begin{equation*}
      \alpha_{\tilde{\psi} \circ \psi}^* = \alpha_\psi^* \circ
      \alpha_{\tilde{\psi}}^*
    \end{equation*}
    together with the requirement that unit morphisms are mapped to
    unit morphisms.
  \item\label{list:statespace2} We say that a state space $\Sscr$ is
    \textbf{locally quasi-equivalent} if \eqref{eq:F1} holds for any
    pair of states $\omega$, $\tilde{\omega} \in \Sscr ( \Mscr )$
    (with GNS representations $\pi$, $\tilde{\pi}$, respectively)
    whenever $\Mscr \in \Loc$ and $\Oscr \in \Kscr^h ( \Mscr )$.
  \item\label{list:statespace3} A state space $\Sscr$ is called
    \textbf{locally normal} if there exists a locally quasi-equivalent
    state space $\tilde{\Sscr}$ so that for each $\omega \in \Sscr (
    \Mscr )$ there is some $\tilde{\omega} \in \tilde{\Sscr} ( \Mscr
    )$ (with GNS representation $\tilde{\pi}$) so that \eqref{eq:F2}
    holds for all $\Mscr \in \Loc$ and $\Oscr \in \Kscr^h ( \Mscr )$.
  \end{deflist}
\end{Def}
Some straightforward consequences of these definitions can be found in
\cite{brunetti/fredenhagen/verch:2003}. Further properties are
outlined in the last chapter.

\chapter{Examples}
\label{chap:examples}

\section{Microlocal Analysis}
\label{sec:microlocal}

This section serves the purpose of giving a self-contained
introduction to definitions and results of H\"ormander's microlocal
analysis that will be needed in the sequel.  After having defined the
wave front set of a distribution, we recall H\"ormander's results on
the multiplication of distributions which extends to the composition
of distribution-valued operators. Further details of this mathematical
theory can be found in H\"ormander's monograph \cite{hoermander:2003}
and in the original sources.

The theory of wave front sets was developed in the 1970's by
H\"ormander and Duistermaat
\cite{hoermander:1971,duistermaat/hoermander:1972}, following the work
of Sato~\cite{sato:1970,sato:1971}. Mathematicians use wave front sets
($\WF$) mainly as a tool in partial differential equations. These sets
are refinements of the notion of the singular support of a
distribution. One advantage of the use of wave front sets over
singular supports is their providing a simple characterization for the
existence of products of distributions, eliminating the difference
between local and global results. Duistermaat and H\"ormander
\cite{duistermaat/hoermander:1972} recognized a link between
microlocal analysis and quantum field theory, but these results
were rarely used in the physics literature.

In microlocal analysis the study of singularities is shifted from the
base space to the cotangent bundle by localizing the distribution
around the singularity followed by an analysis of the result in
\emph{Fourier space}. Let $u \in \Dscr' ( \Rn )$ be a distribution and
let $\phi \in \Czeroinf ( V )$ be a smooth function with support in $V
\subseteq \Rn$. By a well-known argument from the theory of
distributions, the Fourier transform of $\phi u$ yields a smooth
function in frequency space with the following relation,
\begin{equation*}
  \widehat{\phi u} ( \xi )= \bdpair{u}{e^{-i \dpair{\nnarg}{\xi}} \phi}
  \text{,} 
\end{equation*}
where $\dpair{\nnarg}{\nnarg}$ denotes dual pairing. This result implies
\begin{Lem}
  \label{lem:growestimate}
  Let $u \in \Dscr' ( V )$ and let $W$ be an open subset of $V$. Then
  $u \restriction W \in \Cinf ( W )$ if and only if for each $\phi \in
  \Czeroinf ( W )$ and each integer $N \geqslant 0$ there is a constant
  $C_{\phi , N}$ such that
  \begin{equation*}
    \babs{\dpair{u}{e^{-i \dpair{\nnarg}{\xi}}\phi}} \leqslant  C_{\phi , N}
    ( 1 + \abs{\xi} )^{-N} \text{,} \quad \xi \in \Rn \text{}.
  \end{equation*}
\end{Lem}
The \emph{singular support}, $\singsupp u$, of $u \in \Dscr' ( V )$ is
the complement of the largest open subset of $V$ where $u$ is smooth.
Motivated by the previous lemma, the notion of wave front sets is a
refinement of that of the singular support, taking into account the
direction in which the Fourier transform does not strongly decay.
\begin{Def}
  \label{def:wavefrontset}
  The \emph{\ wave front set}, $\WF ( u )$, of $u \in \Dscr' ( V )$ is
  the complement in $V \times \Rn \setminus \set{0}$ of the set of
  points $( x_0 , \xi_0 ) $ in $V \times \Rn \setminus \set{0}$ such
  that for some neighbourhood $U$ of $x_0$ and some conic
  neighbourhood $\Sigma$ of $\xi_0$ we have for each $\phi \in \Czeroinf
  ( U )$ and each integer $N \geqslant 0$ a constant $C_{\phi , N}$
  such that
  \begin{equation*}
    \babs{\bdpair{u}{e^{- i \dpair{\nnarg}{\xi}}\phi}} \leqslant
    C_{\phi , N} ( 1 + \abs{\xi} )^{-N} \text{,} \quad \xi \in \Sigma
    \text{.} 
  \end{equation*}
\end{Def}

Note that a conic set $\Sigma$ has the property that with $( x , \xi
)$ also the pair $( x , t \xi )$ belongs to $\Sigma$ for all $t > 0$.
The following remarks can easily be proved using
Lemma~\ref{lem:growestimate}.
\begin{Rem}
  \label{page:remark}
  \begin{remalphalist}
  \item Let $V$ be an open subset of $\Rn$, then for $v \in \Dscr' ( V
    )$ with wave front set $\WF ( v )$ the projection of this wave
    front set to the base point gives the singular support of $v$.
  \item $\WF ( v )$ is a closed subset of $V \times \Rn \setminus
    \set{0}$ since, by definition, each point $( x , k ) \notin \WF (
    v )$ has an open neighbourhood in $V \times \Rn \setminus \set{0}$
    consisting of such points, too.
  \item\label{rem:wfinklusion} For all smooth test functions $\phi$
    with compact support one has $\WF ( \phi v ) \subseteq \WF ( v )$.
  \item For any distribution $v$ with wave front set $\WF ( v )$ the
    wave front sets of its partial derivatives are contained in $\WF (
    v )$.
  \end{remalphalist}
\end{Rem}
\begin{Ex}
  \begin{exlist}
  \item Let $f \in \Cinf ( V ) \subseteq \Dscr' ( V )$ be a smooth
    function, then its wave front set is empty: $\WF ( f ) =
    \emptyset$.
  \item Consider the Dirac $\delta$-distribution on $\Rbb^2$. Its wave
    front set is $\WF \big( \delta ( x , y ) \big) = \bsetf{( x , k ;
      y , k' ) \in \Rbb^2 \times \Rbb^{2n} \setminus \set{0}}{x = y \;
      ; \; k = - k'}$.
  \end{exlist}
\end{Ex} 
It is worth recalling that the set of normal coordinates $( x_1 ,
\dots , x_n , \xi_1 , \dots , \xi_n )$ of the cotangent bundle $T^* V$
over the base coordinates $( x_1 , \dots ,x_n )$ in $V$ allows for the
identification of $V \times \Rn$ with $T^* V$ and to consider $\WF ( u
)$ as a subset of the cotangent bundle. Since the definition of wave
front sets is local, it can be lifted to manifolds. An intrinsic
definition of wave front sets will be given below.

In this chapter the manifolds considered are assumed to be
$n$-dimensional carrying a smooth Riemannian or semi-Riemannian
metric.
\begin{Def}
  \label{def:reg_dir_points}
  Let $\Mscr$ be an $n$-dimensional smooth manifold with cotangent
  bundle $T^* \Mscr$ and let $u \in \Dscr' ( \Mscr )$, the
  distributions on $\Mscr$. The point $( x_0 , k_0 ) \in T^* \Mscr
  \setminus \set{0}$ is called a \emph{regular directed point} if and
  only if for all $s \geqslant 1$, for all $\lambda_0 \in \Rbb^s$ and
  for any function $\phi \in \Cinf ( \Mscr \times \Rbb^s , \Rbb )$
  with $d_x \phi ( x_0 , \lambda_0 ) = k_0$ there exists a
  neighbourhood $\Vscr$ of $x_0$ in $\Mscr$ and a neighbourhood
  $\Lambda$ of $\lambda_0$ in $\Rbb^s$ such that, for all $\rho \in
  \Czeroinf ( \Vscr )$ and all $N \geqslant 0$, one has, uniformly in
  $\lambda \in \Lambda$,
  \begin{equation*}
     \babs{\bdpair{u}{\rho e^{- i \tau \phi ( \nnarg , \lambda )}}} =
     O ( \tau^{- N} ) \quad \text{if} \quad \tau \rightarrow \infty
     \text{.}
  \end{equation*}
\end{Def}
Now, the wave front set, $\WF ( u )$, of $u \in \Dscr' ( \Mscr )$ is
the complement in $T^* \Mscr \setminus \set{0}$ of the set of all
regular directed points of $u$.

A useful application of wave front sets is the definition of products
of distributions. Wave front sets provide a simple characterization
for the existence of such products, which turns out to be sequentially
continuous provided the wave front sets of the corresponding
distributions are contained in a suitable cone in $T^* \Mscr \setminus
\set{0}$.
\begin{Def}
  \label{def:seqcontdef}
  Let $\Gamma$ be a closed cone in $T^* \Mscr \setminus \set{0}$ and
  let $\Dscr_\Gamma' ( \Mscr )$ denote the subspace of distributions
  with wave front set contained in $\Gamma$. A sequence $\set{u_j}$ of
  distributions in $\Dscr_\Gamma' ( \Mscr )$ converges to a
  distribution $u \in \Dscr_\Gamma' ( \Mscr )$ if and only if the
  following conditions hold:
  \begin{deflist}
  \item $\set{u_j}$ converges to $u$ in $\Dscr' ( \Mscr )$ (weakly);
  \item for all $( x_0 , k_0 ) \in ( T^* \Mscr \setminus \set{0} )
    \setminus \Gamma$, there exist a test function $\rho\in \Czeroinf
    ( \Mscr )$ with $\rho ( x_0 ) \neq 0$, a conic neighbourhood $W$
    of $k_0$ in $T^* \Mscr \setminus \set{0}$ and a function $\phi$ as
    in Definition~\ref{def:reg_dir_points} such that for all $N$
    \begin{equation*}
      \sup_{\tau \in \Rbb_+} \sup_{k \in W} \left( \tau^N
        \babs{\bdpair{u - u_j}{\rho e^{- i \tau \phi ( \nnarg , k
              )}}}\right) \rightarrow 0 \quad \text{if}\quad j
      \rightarrow \infty \text{.}
     \end{equation*}
  \end{deflist}
\end{Def}

Note that every subspace $\Dscr'_\Gamma ( \Mscr )$ contains all smooth
test functions with compact support. Moreover, for $u \in
\Dscr_\Gamma' ( \Mscr )$ there exists a sequence $\set{u_j}$ of
compactly supported smooth functions such that $u_j \rightarrow u$ in
$\Dscr_\Gamma' ( \Mscr )$. Thus it is possible to choose their
supports in an arbitrary neighbourhood of the support of $u$.
\begin{The}
  \label{the:prodDist}
  Let $\Mscr$ be a smooth manifold and let $\Gamma$, $\Sigma \subseteq
  T^* \Mscr \setminus \set{0}$ be two closed cones such that $\Gamma
  \oplus \Sigma \doteq \setf{( x , k + l )}{( x , k ) \in \Gamma
    \text{;~} ( x , l ) \in \Sigma} \subseteq T^* \Mscr \setminus
  \set{0}$. Then the multiplication operator
  \begin{equation*}
    \Czeroinf ( \Mscr ) \times \Czeroinf ( \Mscr ) \ni ( u , v )
    \mapsto u \cdot v \in \Czeroinf ( \Mscr )
  \end{equation*}
  extends to a unique sequentially continuous operator from the set
  $\Dscr'_\Gamma ( \Mscr ) \times \Dscr'_\Sigma ( \Mscr )$ to
  $\Dscr'_\Lambda ( \Mscr )$, where $\Lambda \doteq ( \Gamma \oplus
  \Sigma ) \cup \Gamma \cup \Sigma$. In particular, the wave front
  sets are related by
  \begin{equation}
    \WF ( u \cdot v ) \subseteq \WF ( u ) \oplus \WF ( v ) \cup \WF (
    u ) \cup \WF ( v ) \text{.}
 \end{equation}
\end{The}
\begin{Rem}
  Note that for the product of two distributions $u$ and $v$ to exist
  it is sufficient that $\WF ( u ) \oplus \WF ( v )$ does not contain
  terms of the form $( x , 0 )$.
\end{Rem}

\section{The Scalar Klein-Gordon Field on a Fixed Spacetime
  Background} 
\label{sec:kgfield-fixed}

We describe quantum fields propagating on a four-dimensional
Lorentzian manifold $( \Mscr , g )$ in terms of the general theory of
quantized fields (cf.~\cite{fredenhagen:1992,haag:1996,wald:1994}).
The manifold is assumed to be globally hyperbolic, \ie, it admits
spacelike Cauchy hypersurfaces. We deal with the G{\aa}rding-Wightman
approach to quantum fields \cite{wightman/garding:1964} and its
algebraic formulation by Borchers and Uhlmann
\cite{borchers:1962,uhlmann:1962}. To each differentiable manifold
$\Mscr$ is assigned a topological $^*$-algebra $\BM$ constructed as
follows: Elements in $\BM$ are sequences $( f_n )$, $n \in \Nbb_0$,
where $f_0 \in \Cbb$ and $f_n \in C_0^\infty ( \Mscr^n )$ for $n > 0$.
Addition and scalar multiplication are defined in the usual way for
sequences with values in vector spaces, and the product $( f_n ) ( h_n
)$ in $\BM$ is determined as the sequence $( j_n )$ with
\begin{equation*}
  j_n ( x_1, \dots ,x_n ) \doteq \sum_{i + j = n} f_i ( x_1 , \dots ,
  x_i ) h_j ( x_{i + 1} , \dots , x_n ) \text{,} \quad ( x_1 , \dots ,
  x_n ) \in \Mscr^n \text{.}
\end{equation*}
The $^*$-operation is $( f_n )^* \doteq ( \bar{\bar{f_n}} )$, where
the elements of the sequence on the right-hand side are
$\bar{\bar{f_n}} ( x_1 , \dots , x_n ) \doteq \overline{f_n ( x_n ,
  \dots , x_1 )}$ with the bar denoting complex conjugation. The unit
element is $\unit \doteq ( 1 , 0 , 0 , \dots )$. This algebra can be
equipped with a fairly natural locally convex topology with respect to
which it is complete. See \cite{borchers:1962,uhlmann:1962} for a
further discussion of the Borchers-Uhlmann algebra (and also
\cite{fredenhagen/haag:1987,sahlmann/verch:2000} for the context of
curved spacetime manifolds). A state $\omega$, defined as a positive
linear functional over $\BM$, consists of a hierarchy of $m$-point
distributions, $\omega = \set{\omega_m}_{m \in \Nbb}$. Via the
following GNS Reconstruction Theorem, every state satisfying local
commutativity determines a Hilbert space, a \emph{vacuum vector} in it
and a representation of the algebra $\BM$, thus linking the algebraic
approach to the Hilbert space setting of G{\aa}rding and Wightman.
\begin{The}[GNS Reconstruction]
  \label{the:gns-reconstruction}
  For every state $\omega = \set{\omega_m}_{m \in \Nbb}$ on the
  Bor\-chers-Uhlmann algebra there is a GNS quadruple $( \Hscr_\omega
  ,\Dscr_\omega , \phi_\omega , \Omega_\omega)$, unique up to unitary
  equivalence, such that for each $m \geqslant 1$ and any test
  functions $f_1 , \dots , f_m \in \Czeroinf ( \Mscr , \Omega_1 )$ one
  has
  \begin{equation*}
    \omega_m ( f_1 , \dots , f_m ) = \scp{\Omega_\omega}{\phi_\omega (
      f_1 ) \cdots \phi_\omega ( f_m ) \Omega_\omega} \text{.}
  \end{equation*}
\end{The}

Recall that a GNS quadruple $( \Hscr_\omega , \Dscr_\omega ,
\phi_\omega , \Omega_\omega )$ satisfies the following properties,
usually referred to as the (curved spacetime) G{\aa}rding-Wightman
axioms:
\begin{abclist}
\item $\Hscr_\omega$ is a separable Hilbert space, $\Dscr_\omega$ is a
  dense subspace of $\Hscr_\omega$, and the GNS vacuum $\Omega_\omega$
  is a distinguished vector in $\Hscr_\omega$.
\item The fields $\phi_\omega$ are operator-valued distributions, \ie,
  for all $\Phi$, $\Psi \in \Dscr_\omega$ the linear mapping
  \begin{equation*}
    \scp{\Psi}{\phi_\omega ( \nnarg ) \Phi} : \Czeroinf ( \Mscr ) \ni
    f \mapsto \scp{\Psi}{\phi_\omega ( f ) \Phi}
  \end{equation*}
  belongs to $\Dscr' ( \Mscr )$.
\item The vacuum $\Omega_\omega$ belongs to the subspace
  $\Dscr_\omega$, and this space is an invariant domain for the
  fields, \ie, for each $f \in \Czeroinf ( \Mscr )$ the domain of
  $\phi_\omega ( f )$ contains $\Dscr_\omega$ and $\phi_\omega ( f )
  \Dscr_\omega \subseteq \Dscr_\omega$.
\item The fields are Hermitian, \ie, for each $f \in \Czeroinf ( \Mscr
  )$ the domain of the adjoint of $\phi_\omega ( f )$, denoted
  $\phi^*_\omega ( f )$, contains $\Dscr_\omega$ and $\phi^*_\omega (
  f ) \supset \phi_\omega ( \overline{f} )$.
\item The subspace $\Dscr_\omega$ is generated by applying finitely
  many smeared field operators to $\Omega_\omega$.
\end{abclist}
In flat Minkowski spacetime these axioms are supplemented by requiring
covariance with respect to the Poincar\'e (inhomogeneous Lorentz)
group and a condition on the spectrum stating that the Fourier
transform $\tilde{\omega} ( p_1 , \dots , p_m )$ of the $m$-point
function is concentrated at $p_k + \dots + p_m \in - \clfwcone$ for $k
= 2$, $3$, \dots, $m$.

The simplest and best studied example of a quantum field theory in
curved spacetime is the scalar Klein-Gordon field. As shown by Dimock
\cite{dimock:1980}, its local $C^*$-algebras can easily be constructed
on each globally hyperbolic spacetime. As mentioned before, global
hyperbolicity of the spacetime $( \Mscr , g )$ entails well-posedness
of the global Cauchy problem for the scalar Klein-Gordon equation,
\begin{equation}
  \label{eq:klein-gordon}  
  ( \nabla^a \nabla_a + m^2 + \xi R ) \varphi = 0
\end{equation}
(for smooth, real-valued $\varphi$), where $\nabla$ is the covariant
derivative of $g$, $m \ge 0$ and $\xi \ge 0$ are constants, and $R$ is
the scalar curvature of $g$. Moreover, there exist uniquely determined
advanced and retarded fundamental solutions of the Klein-Gordon
equation, $E^{\adv/\ret}: \Czeroinf ( \Mscr ) \rightarrow \Cinf (
\Mscr )$. Their difference $E \doteq E^{\adv} - E^{\ret}$ is called
the causal propagator of the Klein-Gordon equation. Let its range $E \big(
\Czeroinf ( \Mscr ) \big)$ be denoted  $\Rscr$. It can be shown
\cite{dimock:1980} that the definition
\begin{equation*}
  \sigma ( Ef , Eh ) \doteq \int_\Mscr d\mu_g \;  f ( Eh ) \text{,}
  \quad f , h \in \Czeroinf ( \Mscr ) \text{,}
\end{equation*}
with $d\mu_g$ the metric-induced volume form on $\Mscr$, endowes
$\Rscr$ with a symplectic form, so that $( \Rscr , \sigma )$ is a
symplectic space.

If the field $\phi_\omega$ arising from
Theorem~\ref{the:gns-reconstruction} satisfies the Klein-Gordon
equation and its commutator is given by
\begin{equation*}
  \bcomm{\phi_\omega ( f )}{\phi_\omega ( g )} = E ( f \otimes g )
  \text{,} \quad f , g \in \Czeroinf ( \Mscr ) \text{,}
\end{equation*}
we call $\omega$ a state of the \emph{Klein-Gordon field} over
$\Mscr$. Not all states $\omega$ are believed to be physically
meaningful. A condition expected to be satisfied by physically
reasonable states is the Hadamard condition \cite{dewitt/brehme:1960}
which has been studied by various authors (cf.~the references in
Fulling's book~\cite{fulling:1989}). A mathematically precise
definition of the Hadamard condition in terms of boundary values of
certain complex-valued functions has been given by Kay and Wald in
\cite{kay/wald:1991}. Radzikowski discovered \cite{radzikowski:1992}
that an equivalent characterization of Hadamard states is possible in
terms of their wave front sets. Using his results, Junker
\cite{junker:1995,junker:1996} (note also the erratum to
\cite{junker:1996} in \cite{junker:2002}) constructed Hadamard states
for free scalar fields on arbitrary globally hyperbolic spacetimes.
Radzikowski's result can be recast in the following definition.
\begin{Def}
  \label{def:wavefrontscalar}
  Let $\omega$ be a quasifree state of the Klein-Gordon field over a
  globally hyperbolic manifold $( \Mscr , g )$. Then $\omega$ is a
  Hadamard state if and only if its $2$-point distribution $\omega_2$
  has the wave front set
  \begin{equation}
    \label{eq:wfhadamscalar}
    \WF ( \omega_2 ) = \bsetf{( x_1 , k_1 ; x_2 , - k_2 ) \in T^*
      \Mscr^2 \setminus \set{0}}{( x_1 , k_1 ) \sim ( x_2 , k_2 )
      \text{~and~} k_1^0 \geqslant 0} \text{,}
  \end{equation}
  where $( x_1 , k_1 ) \sim ( x_2 , k_2 )$ means that there exists a
  lightlike geodesic $\gamma$ connecting $x_1$ and $x_2$ with
  cotangent vectors $k_1$ at $x_1$ and $k_2$ at $x_2$.
\end{Def}
A state $\omega$ is called \emph{quasifree} if and only if all its odd
$m$-point distributions vanish and the even ones satisfy
\begin{equation}
  \label{eq:qfprop}
  \omega_m (x_1 , \dots ,x_m ) = \sum_P \prod_r \omega_2 \big( x_{( r
    , 1 )} , x_{( r , 2 )} \big) \text{,}
\end{equation}
where $P$ denotes a partition of the set $\set{1 , \dots , m}$ into
subsets which are pairings of points, labelled by $r$. The ordering of
points in $\omega_2$ is preserved, \eg, $( r , 1 ) < ( r , 2 )$, and
two arguments never coincide so that the product $\prod_r$ exists
whenever $\omega_2 ( x_i , x_j )$ are distributions.

Using Theorem~\ref{the:prodDist} in connection with \eqref{eq:qfprop},
the wave front set of $\omega_m$ is seen to satisfy
\begin{equation}
  \label{eq:wf_omega_n}
  \WF ( \omega_m ) \subseteq \left( \bigcup_Q \bigoplus_{r \in Q} \WF
    ( \omega_2^r ) \right) \text{,}
\end{equation}
where $Q$ denotes a nonempty set of disjoint pairs and $\omega_2^r$ is
the $2$-point distribution in the variables $x_{( r , 1 )}$, $x_{( r ,
  2 )}$ that, considered as a distribution on $\Mscr^n$, has the wave
front set
\begin{multline}
  \label{eq:wf_omega_2_r}
  \WF ( \omega_2^r ) = \bsetf{( x_1 , 0 ; \dots ; x_{( r , 1 )} ,
    k_{( r , 1 )} ; \dots ; x_{( r , 2 )} , k_{( r , 2 )} ; \dots ;
    x_n , 0 )}{\\
    ( x_{( r , 1 )} , k_{( r , 1 )} ; x_{( r , 2 )} , k_{( r , 2
      )}) \in \WF (\omega_2 )} \text{.}
\end{multline}

A generalization of the usual spectrum condition in Minkowski space to
curved spacetimes can be given in terms of the wave front sets of a
quantum field on the globally hyperbolic spacetime $\Mscr$. Its
formulation requires some definitions from graph theory: Let $\Gscr_n$
denote the set of all finite graphs with vertices $\set{1 , \dots ,
  n}$, such that for every element $G \in \Gscr_n$ all edges occur in
both admissible directions. An \emph{immersion} of a graph $G \in
\Gscr_n$ into $\Mscr$ is an assignment of the vertices of $G$ to
points in $\Mscr$, $\nu \mapsto x ( \nu )$, and of the edges of $G$ to
piecewise smooth curves in $\Mscr$, $e \mapsto \gamma ( e )$ with
source $s ( \gamma ( e ) ) = x ( s ( e ) )$ and range $r ( \gamma ( e
) ) = x ( r ( e ) )$, respectively, together with a covariantly
constant causal covector field $k_e$ on $\gamma$, \ie~$\nabla k_e =
0$, such that:
\begin{abclist}
\item if $e$ is an edge from $\nu$ to $\nu'$, then $\gamma ( e )$
  connects $x ( \nu )$ and $x ( \nu' )$;
\item if $e^{- 1}$ denotes the edge having the opposite direction
  compared to $e$, then the corresponding curve $\gamma ( e^{- 1} )$
  is the inverse of $\gamma ( e )$;
\item for every edge $e$ from $\nu$ to $\nu'$, $k_e$ is directed
  towards the future whenever $\nu < \nu'$;
\item $k_{e^{- 1}} = - k_e$.
\end{abclist}
With this notation the \emph{microlocal spectrum condition} for field
theories over a globally hyperbolic manifold (substituting the usual
Minkowski space spectrum condition) reads:
\begin{Def}[{$\mu$SC}]
  \label{def:muSC}
  Let $\Gamma_m \subseteq T^* \Mscr^m \setminus \set{0}$ denote the
  set of all $m^2$-tuples $( x_1 ,k_1 ; \dots ; x_m , k_m )$ with the
  property that there exist $G \in \Gscr_m$ and a corresponding
  immersion $(x , \gamma , k )$ of $G$ into $\Mscr$ such that
  \begin{deflist}
  \item\label{list:G1} $x_i = x ( i )$ for $i = 1$, \dots, $m$;
  \item\label{list:G2} $k_i = \sum_{\setf{e}{s ( e ) = i}} k_e ( x_i
    )$.
  \end{deflist}
  Then a state $\omega$ with $m$-point distributions $\omega_m$ is
  said to satisfy the {\bfseries Microlocal Spectrum Condition}
  ($\mu$SC) if and only if $\WF ( \omega_m ) \subseteq \Gamma_m$ for
  any $m$.
\end{Def}
\begin{Rem}
  Admitting piecewise \emph{causal} or \emph{lightlike} curves as
  images of edges instead of \emph{smooth} ones in the definition of
  immersions above yields stronger versions of the Microlocal Spectrum
  Condition. For every set of base points $( x_1 , \dots , x_m ) \in
  \singsupp ( \omega_m )$ the first nonzero direction $k_l$ in the
  wave front set is future directed.
\end{Rem}
\begin{Lem}
  \label{lem:Gamma_m_additivity}
  The sets $\Gamma_m$ are stable under addition for all $m \in \Nbb$,
  \ie,
  \begin{equation*}
    \Gamma_m \oplus \Gamma_m \subseteq \Gamma_m \text{.}
  \end{equation*}
\end{Lem}

The existence of nontrivial states which satisfy the Microlocal
Spectrum Condition is the statement of the following proposition.
\begin{Pro}
  \label{pro:qfHadam_muSC}
  Let $\omega$ denote a quasifree Hadamard state for the Klein-Gordon
  field on a globally hyperbolic manifold $( \Mscr , g )$, then
  $\omega$ satisfies the $\mu$SC.
\end{Pro}
\begin{proof}
  All odd $m$-point distributions vanish by assumption, hence
  $\omega_m$ satisfies the $\mu$SC trivially for odd $m$. The wave
  front set of the $2$-point distribution $\omega_2$ is explicitly
  given by \eqref{eq:wfhadamscalar} and obviously satisfies the
  $\mu$SC. A general even $m$-point distributions has the
  representation \eqref{eq:qfprop}, which states that $\omega_m$ is a
  sum of tensor products of $2$-point distributions. Hence, there
  exists a disconnected graph $G_m \in \Gscr_m$ together with an
  immersion $( x , \gamma , k )$ satisfying \ref{list:G1} and
  \ref{list:G2} of Definition~\ref{def:muSC}, \ie, $G_m$ consists of
  subgraphs $G_2 \in \Gscr_2$ such that the immersion $( x , \gamma ,
  k )$ restricted to these subgraphs is compatible with the wave front
  set of the corresponding $2$-point distribution.
\end{proof}
\begin{Rem}
  \begin{remalphalist}
  \item A complete analogue of this proposition should be valid in the
    case of the Dirac equation, since Hadamard states for the latter
    are obtainable by applying the adjoint of the Dirac operator to a
    suitable (auxiliary) Hadamard state of the ``squared'' Dirac
    equation. For fixed spinor indices the wave front set of the
    latter is contained in the right-hand side of
    \eqref{eq:wfhadamscalar}, and derivatives do not enlarge the wave
    front set. See also \cite{koehler:1995,dantoni/hollands:2004}.
  \item Junker and Schrohe \cite{junker/schrohe:2002} have recently
    constructed states with adiabatic finiteness condition related to
    Sobolev order. They proved that, besides their existence, the
    states of finite ``Sobolev order'' are not Hadamard, while those
    with infinite order indeed are. They made extensive use of Sobolev
    wave front sets.
  \end{remalphalist}
\end{Rem}

States satisfying the $\mu$SC obey the important properties laid down
in the following theorems.
\begin{The}
  \label{the:prod_n-pointDist}
  Let $\omega^1$ and $\omega^2$ be two states satisfying the $\mu$SC.
  Then the pointwise products of their corresponding $n$-point
  distributions exist and define a new state satisfying the $\mu$SC.
\end{The}
\begin{proof}
  By Theorem~\ref{the:prodDist}, it is sufficient to show that the
  sums of $\WF ( \omega^i_m )$, $i = 1$, $2$, do not intersect the
  zero section in order to prove that the products of the
  corresponding $n$-point distributions exist. Now, by assumption,
  $\WF ( \omega^1_m )$ and $\WF ( \omega^2_m )$ are both contained in
  the set $\Gamma_m$ which is stable under addition, according to
  Lemma~\ref{lem:Gamma_m_additivity}. Hence, $\WF ( \omega^1_m )
  \oplus \WF ( \omega^2_m ) \subseteq \Gamma_m \subseteq T^* \Mscr^m
  \setminus \set{0}$, as required for the products to exist. Moreover,
  this implies that they satisfy the $\mu$SC.
  
  In order to prove that these new $m$-point distributions yield a
  state, \ie~ satisfy Wightman positivity, we consider the
  \emph{tensor} product of $\omega^1$ and $\omega^2$. This is a state
  on the Borchers-Uhlmann algebra of two commuting scalar fields.
  Positivity of this state means that for all test functions $f_j \in
  \Czeroinf ( \Mscr^j )$, $g \in \Czeroinf ( \Mscr^2 )$,
  \begin{multline}
    \label{eq:postensorprod}
    0 \leq \sum_{m , n} \int ( \omega_1 )_{n + m} ( x_n , \dots , x_1
    , x'_1 , \dots ,x'_m ) ( \omega_2 )_{n + m} ( y_n , \dots ,
    y_1 , y'_1 , \dots , y'_m ) \\
    \times \overline{f_n ( x_1 , \dots , x_n )} f_m ( x'_1 , \dots ,
    x'_m ) \prod_{i = 1}^n \overline{g ( x_i , y_i )} \prod_{i = 1}^m
    g ( x'_i , y'_i ) \text{.}
  \end{multline}
  Choose a family of real test functions $\set{g_\epsilon}_{0 <
    \epsilon \leqslant 1} \subseteq \Czeroinf ( \Mscr^2 , \Omega^2_1
  )$ such that the limit for $\epsilon \rightarrow 0$ is just the
  Dirac $\delta$-distribution. Inserting them into
  \eqref{eq:postensorprod}, the limit for $\epsilon \rightarrow 0$
  that exists by the above consideration is
  \begin{equation*}
    \label{eq:limitprodomega2}
       \sum_{m , n} \int \big( ( \omega_1 )_{n + m} ( \omega_2)_{n +
           m} \big) ( x_n , \dots , x_1 , x'_1 , \dots ,x'_m)
       \overline{f_n ( x_1 , \dots , x_n )} f_m ( x'_1 , \dots , x'_m
       ) \text{.}
  \end{equation*}
  From sequential continuity according to Theorem~\ref{the:prodDist}
  we conclude that this expression is nonnegative, which is the
  desired positivity.
\end{proof}

Let $\omega$ be a state on the Borchers-Uhlmann algebra $\BM$ with
associated GNS quadruple $( \Hscr_\omega , \Dscr_\omega , \phi_\omega
, \Omega_\omega )$. We recall from the previous chapter that 
the \emph{folium} of $\omega$ consists of finite
convex linear combinations $\tilde{\omega}$ of states induced by
vectors in $\Dscr_\omega$:
\begin{equation}
  \label{eq:omegafolium}
  \tilde{\omega} ( A ) = \trace \big( \rho \pi_\omega ( A ) \big)
  \text{,} \quad A \in \BM \text{,}
\end{equation}
where $\trace$ denotes the trace in $\Hscr_\omega$, $\pi_\omega$ is
the representation of $\BM$ associated with $\omega$ and
\begin{equation*}
  \rho = \sum_{i = 1}^N \ket{\Psi_i}\bra{\Psi_i} \text{,} \quad \Psi_i
  \in \Dscr_\omega \text{,}
\end{equation*}
is some density matrix. Contrary to the usual spectrum condition, the
$\mu$SC does not characterize a distinguished state, but is a property
of the full folium instead. We recall that this characterization has
been helpful for the functorial description of the state space in the
previous chapter.
\begin{The}
  \label{the:mufol}
  Let $\omega$ be a state satisfying the $\mu$SC. Then the $\mu$SC is
  satisfied for all states in the folium of $\omega$.
\end{The}
\begin{proof}
  Consider a state $\tilde\omega$ in the folium of $\omega$. All the
  associated $m$-point distributions $\tilde{\omega}_m$ are finite
  linear combinations of $(l + m)$-point distributions of $\omega$
  smeared with suitable test functions from both sides, \ie,
  \begin{equation*}
    \tilde{\omega}_m ( x_1 , \dots , x_m ) = \sum_{l} \omega_{l + m} (
    f_{j_1} , \dots , f_{j_k} , x_1 , \dots , x_m , f_{j_{k + 1}} ,
    \dots , f_{j_l} ) \text{.}
  \end{equation*}
  Therefore, it is sufficient to show that for all $m \in \Nbb$
  \begin{equation}
    \label{eq:WFomega_n_smeared}
    \Gamma \doteq \WF \big( \omega_{l + m} ( f_{j_1} , \dots , f_{j_k}
    , x_1 , \dots , x_m , f_{j_{k + 1}} , \dots , f_{j_l} ) \big)
    \subseteq \Gamma_m \text{,}
  \end{equation}
  $\Gamma_m$ as defined in Definition~\ref{def:muSC}.
  
  Using \cite[Theorem~8.2.13]{hoermander:2003}, one obtains for the
  left-hand side of \eqref{eq:WFomega_n_smeared},
  \begin{multline}
    \label{eq:WFomega_n_smeared_pf}
    \Gamma \subseteq \Bsetf{( x_1 , k_1 ; \dots ; x_m , k_m ) \in T^*
      \Mscr^m \setminus \set{0}}{ \\
      ( y_1 , 0 ; \dots ; y_k , 0 ; x_1 , k_1 ; \dots ; x_m , k_m ;
      y_{k + 1} , 0 ; \dots ; y_l,0 ) \in \WF (\omega_{l + m})
      \subseteq \Gamma_{l + m}} \text{.}
  \end{multline}
  Moreover, since by assumption $\omega_{l + m}$ satisfies the
  microlocal spectrum condition, to every element $( y_1 , 0 ; \dots ;
  x_i , k_i ; \dots ; y_l , 0 ) \in \WF ( \omega_{l + m} )$ there
  correspond a graph $G_{l + m} \in \Gscr_{l + m}$ and an immersion $(
  x , \gamma , k )$ such that the covector fields $k_e$ are zero
  whenever $\gamma ( e )$ does \emph{not} connect two points in
  $\set{x_1 , \dots , x_m}$. For this statement note that the
  direction associated to $y_1$ vanishes by
  \eqref{eq:WFomega_n_smeared_pf}. Moreover, all causal covector
  fields associated to curves $\gamma$ \emph{starting} at $y_1$ are
  directed towards the future by the definition of an immersion,
  hence, using property \ref{list:G2} of Definition~\ref{def:muSC},
  $k_e = 0$ whenever $\gamma$ starts \emph{or ends} at $y_1$.
  
  Consider now the point $y_2$. By assumption, the direction
  associated to it is again zero. Using the properties of the
  immersion and the previous result for covector fields along curves
  \emph{ending} at $y_1$, one sees that the covector fields $k_e$ for
  all curves \emph{starting} at $y_2$ are either future directed or
  zero. As in the previous case this implies $k_e = 0$ for \emph{all}
  curves starting or ending at $y_2$. By induction, this result
  extends to all points up to $y_k$ and, analogously, to all points
  from $y_l$ down to $y_{k + 1}$. Therefore, all points $y_1 , \dots ,
  y_k$, $y_{k + 1} , \dots , y_l$ together with all lines starting or
  ending at them can be removed from the graph $G_{l + m}$. The result
  is another graph $G_m \in \Gscr_m$ together with an immersion $( x ,
  \gamma , k )$ such that \ref{list:G1} and \ref{list:G2} of
  Definition~\ref{def:muSC} are satisfied. This completes the proof.
\end{proof}

The following theorem shows that the microlocal spectrum condition is
compatible with the usual Minkowski space spectrum condition.
\begin{The}
  \label{the:SCimpliesmuSC}
  Let $\omega$ be a state for a quantum field theoretical model on
  Minkowski space, whose $m$-point distributions $\omega_m$ satisfy
  the G{\aa}rding-Wightman axioms. Then $\omega$ satisfies the
  $\mu$SC.
\end{The}
It is not yet known whether, \vv, the microlocal spectrum condition
$\mu$SC is also sufficient for the usual spectrum condition to hold in
Minkowski spacetime.

\section[The Free Scalar Klein-Gordon Field as a Natural
Transformation]{The Free Scalar Klein-Gordon Field as a Natural\\
  Transformation}
\label{sec:KG-natural}

In this section we give an example for fields as natural
transformations by use of the Borchers-Uhlmann algebra $\BM$
associated with the spacetime $\Mscr$
(cf.~Section~\ref{sec:microlocal}).

An endomorphism $\psi \in \Hom_\Loc ( \Mscr_1 , \Mscr_2 )$ between two
spacetimes can be lifted to an algebraic endomorphism $\alpha_\psi :
\BMone \rightarrow \BMtwo$ of the corresponding Borchers-Uhlmann
algebras by setting
\begin{equation*}
  \alpha_\psi \big( ( f_n ) \big) \doteq \big( \psi_*^{( n )} f_n
  \big) \text{,} 
\end{equation*}
where $\psi_*^{( n )}$ denotes the $n$-fold push-forward defined via
$\big( \psi_*^{( n )} f_n \big) ( y_1 , \dots , y_n ) \doteq f_n \big(
\psi^{- 1} ( y_1 ) , \dots , \psi^{- 1} ( y_n ) \big)$. We define a
covariant functor $\Ascr$ between $\Loc$ and $\TAlg$ setting $\Ascr (
\Mscr ) \doteq \BM$ and $\Ascr ( \psi ) \doteq \alpha_\psi$. A locally
covariant quantum field in the sense of Definition~\ref{def:nattrans}
may then be obtained via the following definition for $\Mscr \in \obj
( \Loc )$ and $f \in \Dscr ( \Mscr ) = C_0^{\infty} ( \Mscr )$,
\begin{equation*}
  \Phi_\Mscr ( f ) \doteq ( f_n ) \text{,}
\end{equation*}
where $( f_n ) \in \Ascr ( \Mscr ) = \BM$ is the sequence with $f_1 =
f$ and $f_n = 0$ for all $n \ne 1$. It is straightforward to check
that this definition satisfies all conditions for a natural
transformation with respect to the functors $\Dscr$ and $\Ascr$.

\section{Interacting Fields}
\label{sec:interact}

One of the most recent crucial results is the possibility to cast the
usual formalism of perturbation theory into the setting we are
reviewing.

It has always been a source of diffidence among physicists that
algebraic quantum field theory was only (apparently, in retrospective)
managing structural aspects and never addressed the task of
constructing explicitly interacting models that form the core of
modern quantum field theory as QED, the standard model, or QCD. One of
the authors, in collaboration with Fredenhagen
\cite{brunetti/fredenhagen:1997,brunetti/fredenhagen:2000}, launched a
program which aimed at showing that algebraic quantum field theory can
indeed be used to construct models at the perturbative level. It has
been proved that the usual classification of the renormalization
program can be obtained within the algebraic setting and that,
moreover, one is able to produce results that were sought for a long
time, like the ultraviolet classification of perturbative quantum
field theory on curved spacetimes. Naively speaking, this last aspect
was expected on the ground of the equivalence principle, since at very
small scales (ultraviolet) one knows that the singularities arising in
the renormalization procedure are \emph{essentially} the same as in
flat Minkowski spacetime. However, the details hide several
difficulties, and the strategy for establishing the result required
some ingenuity and devices that went far beyond the usual bag of tools
that theoretical physicists commonly use. Indeed, one of the main
problems is that the usual renormalization procedure makes use of the
Feynman propagator (\ie~the Feynman graphic language) which is an
ambiguous concept on curved spacetimes (and also in time-dependent
external fields in general), not being uniquely defined. One of the
commonly envisaged ways out was to use artificial boundary conditions,
but in the general case not even this can be used. Notice that the
same criticism applies to the path integral approach.

The use of microlocal analysis to overcome these difficulties was
crucial. Indeed, wave front set properties of Wightman functions, the
use of distinguished parametrices, Fourier integral operators and
other tools form the essential language that allowed for the
classification. Recently, Hollands and Wald
\cite{hollands/wald:2001,hollands/wald:2002} have refined to a large
extent the structure of the renormalization program. They introduced
as a basic input the principle of local covariance according to
Definition~\ref{def:covfunctor} (plus some further technicalities)
with the remarkable result that they can prove that the interacting
fields are indeed locally covariant, \ie~natural transformation as in
the free case considered in the previous section.

Since the bulk of these results can by themselves constitute a (long)
review paper, we have to restrict ourselves to informing the
interested reader about the relevant points in the literature.
\begin{abclist}
\item Brunetti, Fredenhagen (1997) \cite{brunetti/fredenhagen:1997}:
  First complete nontechnical description of the perturbative
  approach in the self-interacting scalar field case.
\item D\"utsch, Fredenhagen (1999) \cite{duetsch/fredenhagen:1999}:
  Description of the perturbative setting in the case of QED on
  Minkowski spacetime.
\item Brunetti, Fredenhagen (2000) \cite{brunetti/fredenhagen:2000}:
  Completion of the technical description of the perturbative approach
  for the self-interacting scalar field. Re\-normalization as
  extension of distributions.
\item Hollands, Wald (2001) \cite{hollands/wald:2001}: Perturbative
  approach in the locally covariant case for the self-interacting
  scalar field.
\item Hollands, Wald (2002) \cite{hollands/wald:2002}: Proof of the
  existence of the time-ordered functions in the locally covariant
  case.
\end{abclist}

\chapter{Charges}
\label{chap:charges}

\section{Sharply Localized Charges and Particle Weights}
\label{sec:DHR-weights}

One of the aims of local quantum physics is to understand the nature
and the features of charges of elementary particles in terms of
superselection sectors of the observable net. The term \emph{charges}
here not only refers to charges of electromagnetic type, but also to
the baryonic number, the leptonic number and the isospin. A common
feauture of these charges is that they can be added, and that to any
charge there corresponds an opposite one according to the
particle-antiparticle symmetry. Furthermore, every charge has a
statistics (Fermi-Bose alternative), that of the particle carrying
that charge. Superselection sectors are the unitary equivalence
classes of irreducible representations of the observable net, and
\emph{quantum numbers} are the labels distinguishing different
sectors.  The main task is to single out, from the infinite number of
representations of the observable net, those describing the charges of
particles.

\subsection{DHR analysis}
\label{subsec:dhr-analysis}

The first class of superselection sectors, the DHR sectors, has been
established in a series of papers by Doplicher, Haag and Roberts
\cite{doplicher/haag/roberts:1971,doplicher/haag/roberts:1974}, a
method that came to be known as \emph{DHR analysis}. Let
$\Ascr_{\Kscr^{d c} ( \Mbb^4 )}$ be the observable net indexed by the
set $\Kscr^{d c} ( \Mbb^4 )$ of double cones in $4$-dimensional
Minkowski space $\Mbb^4$. Double cones arise as intersections of open
forward and backward cones $x + \fwcone \cap y - \fwcone$, $x$, $y \in
\Mbb^4$. Doplicher, Haag and Roberts singled out those representations
$\pi$ of the observable net which can be characterized as ``sharp
excitations'' of the vacuum representation $\pi_0$ in the sense that,
when restricted to the algebra $\Afrak ( \Oscr' )$ pertaining to the
spacelike complement $\Oscr'$ of the double cone $\Oscr$, $\pi$ is
unitarily equivalent to $\pi_0$, \ie,
\begin{equation}
  \label{eq:sharp-exc}
  \pi \restriction \Afrak ( \Oscr' ) \simeq \pi_0 \restriction \Afrak
  ( \Oscr' ) \text{,} \quad \Oscr \in \Kscr^{d c} ( \Mbb^4 ) \text{.}
\end{equation}
This family of representations together with the corresponding
intertwining operators forms a $C^*$-category that turns out to be
equivalent to a $C^*$-category of endomorphisms $\rho$ of the
observable net defined in the vacuum representation. This is a crucial
fact, because it is in this category that the charge structure of
sectors arises. In fact, it is possible to introduce a tensor product
$\otimes$ representing the property of composition of charges. The
tensor product has a permutation symmetry which encodes the possible
statistics of sectors.  The statistics of an irreducible object $\rho$
is classified by means of a number $\lambda$, the \emph{statistics
  parameter}, which is an invariant of the equivalence class of
$\rho$. $\lambda$ is the product of two invariants,
\begin{equation*}
  \lambda = \chi \cdot d^{- 1} \text{,} \quad \text{where $\chi \in
  \set{- 1 , 1}$ and $d \in \Nbb \cup \set{\infty}$.}
\end{equation*}
In the case of finite $d < + \infty$, the possible statistics are
classified by the \emph{statistical phase} $\chi$ distinguishing
para-Bose $(1)$ and para-Fermi $(-1)$ statistics and by the
\emph{statistical dimension} $d$ giving the order of the
parastatistics. Ordinary Bose and Fermi statistics correspond to $d =
1$. Any object $\rho$ with finite statistics has a conjugate: there
exists an object $\overline{\rho}$ with the same statistics as $\rho$
such that $\rho \otimes \overline{\rho}$ contains the vacuum sector.
Furthermore, if one consider covariant sectors, it is possible to
construct asymptotic multiparticle states associated with charged
sectors and to establish a clear-cut connection between the statistics
of sectors and the spin of the corresponding particle state. Finally,
in the case of infinite statistics, $d = + \infty$, there is no charge
interpretation since these sectors do not possess conjugates.

The criterion \eqref{eq:sharp-exc} excludes charges of electromagnetic
type. Because of Gau\ss{}' law, the flux of the electric field of a
localized electric charge is nonvanishing in any double cone
containing the charge; however, it was believed by Doplicher, Haag and
Roberts that \eqref{eq:sharp-exc} should hold in theories with a mass
gap. Buchholz and Fredenhagen \cite{buchholz/fredenhagen:1982} have
shown that this is not true. Sectors which describe the charge of
particles in purely massive theories correspond to a localization in a
cone which extends to spacelike infinity. These sectors are known as
BF sectors. Except for the localization region, the charge structure
of DHR sectors also applies to these BF sectors.

It is a deep result by Doplicher and Roberts that the superselection
structure determines a compact global gauge group $G_{\Mbb^4}$ acting
on a field net $\Fscr_{\Mbb^4}$ \cite{doplicher/roberts:1990}. The
observables turn out to be the gauge-invariant elements of the field
net, while the sectors are in 1-1 correspondence with the irreducible
representations of the gauge group, a result that holds true both for
DHR and BF sectors.

\subsection{Particle Weights}
\label{subsec:weights}

As mentioned before, in contradistinction to sharply localized
charges, there exist other variants of charges that elude a treatment
within the Doplicher-Haag-Roberts approach. A prominent example are
electrically charged particles; the amount of their charge can be
determined by measuring the electric flux passing through an
arbitrarily large closed surface surrounding them. Therefore, these
states can be discerned from the vacuum by measurements at appropriate
large distances \cite{buchholz/fredenhagen:1982}. This problem,
associated with the long range of the electromagnetic interaction, is
also connected with the problem to exhibit a clear-cut theoretical
scheme for the assignment of a definite mass and spin to these charges
that are inevitably acompanied by clouds of soft photons.

Usually, the theoretical basis for assigning mass and spin to a
particle is the analysis of all unitary representations of the
Poincar\'e group by Eugene Wigner \cite{wigner:1939}. These
representations are characterized by two parameters, $m$ and $s$,
their real respectively integer and half-integer values being
interpreted as mass and spin of a particle whose state is described as
a vector in the corresponding representation space. Since Lorentz
symmetry is broken in theories with long-range interactions
\cite{buchholz:1986a}, this approach is no longer useful under these
circumstances, giving rise to the so-called \emph{infraparticle
  problem} (in discrimination to the \emph{Wigner particles}
aforementioned). A unified treatment of both Wigner particles and
infraparticles has been proposed by Buchholz \cite{buchholz:1986b} and
elaborated in \cite{buchholz/porrmann/stein:1991}.

The idea is to single out elements of the quasi-local algebra $\Afrak$
that can be interpreted as particle detectors. These elements are thus
required to be insensitive to the vacuum with concurrent appropriate
localization properties. Due to the Reeh-Schlieder Theorem
\cite{reeh/schlieder:1961}, a strictly local operator, \ie~an element
of a local algebra $\AO$, cannot annihilate the vacuum. Therefore
attention is restricted to \emph{almost local vacuum annihilation
  operators}, characterized by their annihilating states with energy
below a certain threshold together with the property that they can be
approximated by local operators in such a way that the norm of their
dislocalized part (outside a region of radius $r$) falls off more
rapidly than any power of $r$. When supplemented with an assumption on
their infinite differentiability with respect to Poincar\'e
transformations, these elements constitute a subspace $\Lfrak_0$ of
$\Afrak$, while multiplication from the left by quasi-local operators
yields a left ideal $\Lfrak$ in $\Afrak$. The algebra $\Cfrak$ of
detectors is then spanned by elements of the form ${L_1}^* L_2$,
$L_1$, $L_2 \in \Lfrak$, \ie~by quasilocal operators $A \in \Afrak$
that are sandwiched with almost local vacuum annihilation operators
$L_0$, $K_0 \in \Lfrak_0$: $C \doteq {L_0}^* A K_0$. Note that this
algebra does not contain a unit! Now, on the algebra $\Cfrak$ of
detectors thus defined one considers physical states of bounded energy
and the expectation values that they return when applied to detectors.
As in the approach of Araki and Haag \cite{araki/haag:1967}, which was
limited to massive theories, it can be established in the present more
general setting that the limits of these expectation values exist at
asymptotic times (in the future as well as in the past). In view of
the construction of the algebra of detectors, the existence of these
limits is interpreted as the signature of an asymptotic stable
constellation of particles. Since, due to their construction, the
corresponding asymptotic positive functionals would take the value
$\infty$ when evaluated on the unit $\algunit$ of $\Afrak$, they have
the characteristics of \emph{weights} (or \emph{extended positive
  functionals}) that have been introduced by Jacques Dixmier into the
theory of $C^*$-algebras \cite[Section~I.4.2]{dixmier:1981}.
Therefore, the sesquilinear forms on $\Lfrak \times \Lfrak$ that these
asymptotic functionals define are called \emph{particle weights}.

These sesquilinear forms exhibit properties expected from asymptotic
mixtures of stable particles; in fact they are interpreted as mixtures
of \emph{pure} particle weights pertaining to a definite
energy-momentum. One should have in mind at this point Dirac's notion
of improper energy-momentum eigenstates. These unnormalizable ket
vectors $\ket{p^0 , \pib}$ become decent Hilbert space vectors when
acted upon (localized) by application of a quasi-local operator $A \in
\Afrak$: $\ket{p^0 , \pib} \mapsto A \ket{p^0 , \pib} \in \Hscr$. In a
similar way, a general particle weight $\scp{\mnarg}{\mnarg} : \Lfrak
\times \Lfrak \rightarrow \Cbb$ is a singular object, while, upon
localization with an operator $L \in \Lfrak$, the mapping $\Afrak \ni
A \mapsto \scp{L}{AL} \in \Cbb$ is a bounded linear functional on the
$C^*$-algebra $\Afrak$. Generic particle weights allow for a GNS
construction yielding a highly reducible representation of the
quasi-local algebra. It has been demonstrated that this representation
can be decomposed in terms of a direct integral of irreducible
representations (in fact a standard result of the theory of
$C^*$-algebras), where the sesquilinear forms arising in this way are
in fact again particle weights, \ie, they exhibit all their
characteristic features. These are the pure particle weights already
mentioned at the beginning of this paragraph. They are assigned a
definite energy-momentum and permit a clear-cut definition of mass and
spin, this time for Wigner particles as well as infraparticles. A
thorough presentation of the theory of particle weights can be found
in \cite{porrmann:2004a,porrmann:2004b}. 

The particle weights considered as sesquilinear forms on $\Lfrak$
constitute a positive cone in an infinite-dimensional locally convex
space. Given a convex base of this cone (defined by evaluation on
certain elements of $\Cfrak$), it is natural to ask for the
decomposition of a generic particle weight in terms of extremal ones.
The result of this barycentric decomposition will be an integral with
support on the extreme boundary (Choquet disintegration)
\cite{alfsen:1971,phelps:1966}. It is an obvious question if and how
the pure particle weights arising from the spatial disintegration of
the preceding paragraph correspond to the extremal elements of the
cone. In this guise, the particle spectrum of a quantum field theory
is expected to be encoded in the geometrical structure of the positive
cone of particle weights. How this viewpoint can be related to the
analyses expounded in other parts of our review is an open question.

\section{Net Cohomology}
\label{sec:net-cohom}

Net cohomology arose as an equivalent approach to the theory of
superselection sectors in Minkowski space \cite{roberts:1976}. The
basic idea is that the physical content of DHR sectors is completely
encoded in the charge transporters. These turn out to be $1$-cocycles
of the partially ordered set (\emph{poset}) formed by the set of
indices of the observable net when ordered by inclusion. The relevance
of this cohomological approach becomes clear in the extension of the
theory of superselection sectors to curved spacetimes, where one has
to deal with spacetimes having nontrivial topologies. In this context
net cohomology appears as a very powerful tool, both for analysing the
charge structure of sectors and, in particular, to establish a clear
connection between topology and sectors.

By the term \emph{net cohomology} we mean the cohomology of a poset
$\Pscr$ with coefficients in a net of local algebras $\Ascr_\Pscr$
(the observable net) indexed by elements of $\Pscr$. So it is a
nonabelian cohomology, and the language of $n$-categories is an
essential ingredient if one is interested in degrees of the cohomology
higher than $1$. It is a matter of fact that, up to now, only
$1$-cocycles have a direct physical interpretation. Therefore, this
section is focused on $1$-cocycles, and the term net cohomology refers
to $1$-cohomology. After some preliminaries the main topics are
discussed in full generality in the following two subsections,
\viz~the first homotopy group of a poset, the connection between
homotopy and net cohomology, and the behaviour of net cohomology under
a change of the index set. The last subsection is devoted to study the
case of a poset being the basis for the topology of a topological
space. References for this section are
\cite{roberts:1990,guido/longo/roberts/verch:2001,roberts:2004,%
ruzzi:2005b}.

\subsection{Preliminaries: the Simplicial Set and the Set of Paths}
\label{subsec:simp-net-co}

\paragraph*{The simplicial set} A \emph{poset} $( \Pscr , \leqslant )$
is a partially ordered set, \ie, the binary relation $\leqslant$ on the
nonempty set $\Pscr$ satisfies for $\Oscr$, $\Oscr_1$, $\Oscr_2$,
$\Oscr_3 \in \Pscr$:
\begin{align*}
  & & \quad \Oscr \leqslant \Oscr & \qquad \text{(reflexivity)}
  \text{,} \\
  \Oscr_1 \leqslant \Oscr_2 \quad \text{and} \quad \Oscr_2 \leqslant
  \Oscr_1 & \qquad \Rightarrow & \quad
  \Oscr_1 = \Oscr_2 & \qquad \text{(antisymmetry)} \text{,} \\
  \Oscr_1 \leqslant \Oscr_2 \quad \text{and} \quad \Oscr_2 \leqslant
  \Oscr_3 & \qquad \Rightarrow & \quad \Oscr_1 \leqslant \Oscr_3 &
  \qquad \text{(transitivity)} \text{.}
\end{align*}
A poset is said to be \emph{directed} if for any pair $\Oscr_1$,
$\Oscr_2 \in \Pscr$ there exists $\Oscr_3 \in \Pscr$ such that
$\Oscr_1$, $\Oscr_2 \leqslant \Oscr_3$. For our purposes, important
examples of posets are provided by the standard simplices. A
\emph{standard} $n$-simplex is defined as
\begin{equation*}
  \Delta_n \doteq \bsetf{( \lambda_0 , \dots , \lambda_n ) \in \Rbb^{n
      + 1}}{\lambda_0 + \cdots + \lambda_n = 1 \text{,} \quad
    \lambda_i \in [ 0 , 1 ]} \text{.}
\end{equation*}
Note that $\Delta_0$ is a point, $\Delta_1$ a closed interval \etc
The \emph{inclusion maps} $d^n_i$ between standard simplices are maps
$d^n_i : \Delta_{n - 1} \rightarrow \Delta_n$ defined as
\begin{equation*}
  d^n_i ( \lambda_0 , \dots , \lambda_{n - 1} ) \doteq ( \lambda_0 ,
  \lambda_1 , \dots , \lambda_{i - 1} , 0 , \lambda_i , \dots
  \lambda_{n - 1} )
\end{equation*}
for $n \geqslant 1$ and $0 \leqslant i \leqslant n - 1$. Now,
regarding a standard $n$-simplex as a partially ordered set with
respect to the inclusion of its subsimplices, a \emph{singular
  $n$-simplex} of a poset $\Pscr$ is an order preserving map $f :
\Delta_n \rightarrow \Pscr$. We denote by $\Sigma_n ( \Pscr )$ the
collection of singular $n$-simplices of $\Pscr$ and by $\Sigma_* (
\Pscr )$ the collection of all singular simplices of $\Pscr$.
$\Sigma_* ( \Pscr )$ is the \emph{simplicial set} of $\Pscr$. The
inclusion maps $d^n_i$ between standard simplices are extended to maps
$\partial^n_i : \Sigma_n ( \Pscr ) \rightarrow \Sigma_{n - 1} ( \Pscr
)$ called \emph{boundaries} by setting $\partial^n_i f \doteq f \circ
d^n_i$. One can easily check, by definition of $d^n_i$, that the
following relations hold:
\begin{equation}
  \label{eq:chain}
  \partial^{n - 1}_i \circ \partial^n_j = \partial^{n - 1}_j \circ
  \partial^n_{i + 1} \text{,} \quad  i \geqslant j \text{.} 
\end{equation}
From now on we will omit the superscript from the symbol
$\partial^n_i$ and denote the composition $\partial_i\circ \partial_j$
by $\partial_{ij}$, $0$-simplices by the letter $a$, $1$-simplices by
$b$ and $2$-simplices by $c$. Note that a $0$-simplex $a$ is nothing
but an element of $\Pscr$. A $1$-simplex $b$ is formed from two
$0$-simplices $\partial_0 b$, $\partial_1 b$ and an element $\abs{b}$
of $\Pscr$, called the \emph{support} of $b$, such that $\partial_0
b$, $\partial_1 b \leqslant \abs{b}$; a $2$-simplex $c$ is formed from
three $1$-simplices $\partial_0 c$, $\partial_1 c$, $\partial_2c$,
whose $0$-boundaries are chained according to \eqref{eq:chain}, and
from a $0$-simplex $\abs{c}$, the \emph{support} of $c$, such that
$\partial_0 c$, $\partial_1 c$, $\partial_2 c \leqslant \abs{c}$.
\begin{figure}[!ht]
\begin{center}
  \includegraphics[scale=0.6]{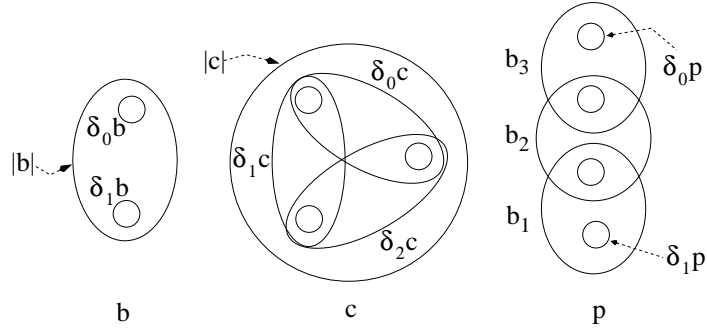}
     \caption{$b$ is a $1$-simplex, $c$ is a $2$-simplex, 
     $p = \set{b_3 , b_2 , b_1}$ is a path. 
     The symbol $\delta$ stands for $\partial$.} 
\end{center}
\end{figure}
\newline
The \emph{reverse} of a $1$-simplex $b$ is the $1$-simplex
$\overline{b}$ defined by
\begin{equation*}
  \partial_0 \overline{b} = \partial_1 b \text{,} \quad \partial_1
  \overline{b} = \partial_0 b \text{,} \quad \abs{\overline{b}} =
  \abs{b} \text{.}
\end{equation*}
Finally, a $1$-simplex $b$ is said to be \emph{degenerate to} a
$0$-simplex $a_0$ whenever
\begin{equation*}
  \partial_0 b = a_0 = \partial_1 b \text{,} \quad a_0 = \abs{b}
  \text{.} 
\end{equation*}
We will denote by $b ( a_0 )$ the $1$-simplex degenerate to $a_0$.

\paragraph*{Paths} Given a pair $a_0$, $a_1$ of $0$-simplices 
a \emph{path from $a_0$ to $a_1$} is a finite ordered sequence $p =
\set{b_n , \dots , b_1}$ of $1$-simplices with the relations
\begin{equation*}
  \partial_1 b_1 = a_0 \text{,} \quad \partial_0 b_i = \partial_1 b_{i
    + 1} \text{~for~} i \in \set{1 , \dots , n - 1} \text{,} \quad
  \partial_0 b_n =a_1 \text{}.
\end{equation*}
The \emph{starting point} of $p$, $\partial_1 p$, is the $0$-simplex
$a_0$, while the \emph{endpoint} of $p$, $\partial_0p$, is the
$0$-simplex $a_1$. We denote by $\Pscr ( a_0 , a_1 )$ the set of paths
from $a_0$ to $a_1$. The poset $\Pscr$ is said to be
\emph{pathwise-connected} if $\Pscr ( a_0 , a_1 ) \ne \emptyset$ for
any pair $a_0$, $a_1$ of $0$-simplices. The set of paths is equipped
with the following operations: Consider a path $p = \set{b_n , \dots ,
  b_1} \in \Pscr ( a_0 , a_1 )$. Its \emph{reverse}, $\overline{p}$,
is the path
\begin{equation*}
  \overline{p} \doteq \set{\overline{b}_1 , \dots , \overline{b}_n}
  \in \Pscr ( a_1 , a_0 ) \text{.}
\end{equation*}
The \emph{composition} of $p$ with a path $q = \set{b'_k , \dots ,
  b'_1}$ of $\Pscr ( a_1 , a_2 )$ is defined as
\begin{equation*}
  q * p \doteq \set{b'_k , \dots , b'_1 , b_n , \dots , b_1} \in \Pscr
  ( a_0 , a_2 ) \text{.}
\end{equation*}
Note that forming the reverse is an involutive mapping, while the
composition $*$ is associative.

\subsection{The First Homotopy Group of a Poset}
\label{subsec:hom-po}

The logical steps necessary to define the first homotopy group of
posets are the same as in the case of topological spaces. We first
recall the definition of a homotopy of paths. Then we prove that the
reverse of a path and the composition of paths are well-behaved
mappings under the homotopy equivalence relation. Finally, the first
homotopy group of a poset is defined.

An \emph{elementary deformation} of a path $p$ consists in replacing a
$1$-simplex $\partial_1 c$ of the path by a pair, $\partial_0 c$,
$\partial_2 c$, where $c$ is a singular $2$-simplex in $\Sigma_2 (
\Pscr )$, or, conversely, in replacing a consecutive pair $\partial_0
c$, $\partial_2 c$ of $1$-simplices of $p$ by a single $1$-simplex
$\partial_1 c$.
\begin{figure}[!ht]
  \begin{center}
    \includegraphics[scale=0.7]{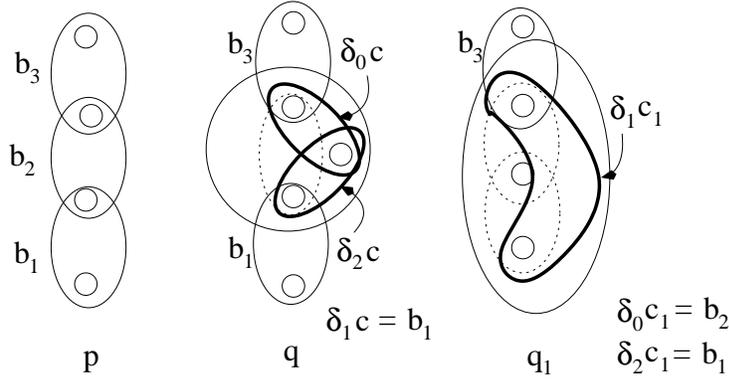}
    \caption{$q$ and $q_1$ are two different elementary deformations 
      of the path $p$. The symbol $\delta$ stands for $\partial$.}
  \end{center}
\end{figure}
\newline Two paths with the same endpoints are \emph{homotopic} if
they can be obtained from one another by a finite number of elementary
deformations. Homotopy thus defines an equivalence relation $\sim$ on
the set of paths with the same end points.

The reverse and the composition of paths are compatible with the
homotopy equivalence relation. To be precise, for any $p$, $q \in
\Pscr ( a_0 , a_1 )$ and $p_1$, $q_1 \in \Pscr ( a_1 , a_2 )$, we have
\begin{subequations}
  \label{eq:hom-equiv1}
  \begin{align}
    p \sim q & \Leftrightarrow \overline{p} \sim \overline{q} \text{,}
    \\ 
    p \sim q \text{,} \quad p_1 \sim q_1 & \Rightarrow p_1 * p \sim
    q_1 * q \text{.} 
  \end{align}
\end{subequations}
Furthermore, for any $p \in \Pscr ( a_0 , a_1 )$, the following
relations hold:
\begin{subequations}
  \label{eq:hom-equiv2}
  \begin{align}
    p * b ( a_0 ) \sim p & \text{~and~} p \sim b ( a_1 ) * p \text{,}
    \\ 
    \overline{p} * p \sim b ( a_0 ) & \text{~and~} b ( a_1 ) \sim p *
    \overline{p} \text{,}
  \end{align}
\end{subequations}
where $b ( a_0 )$ is the $1$-simplex degenerate to $a_0$.

We now are in a position to define the first homotopy group of a
poset. Fix a base $0$-simplex $a_0$ and consider the set of closed
paths $\Pscr ( a_0 )$. Note that the composition and the reverse are
internal operations of $\Pscr ( a_0 )$ and that $b ( a_0 ) \in \Pscr (
a_0 )$. We define
\begin{equation}
  \label{eq:fi-hom-po}
  \pi_1 ( \Pscr , a_0 ) \doteq  \Pscr ( a_0 ) / \sim \text{,}
\end{equation}
where $\sim$ is the homotopy equivalence relation. Let $[ p ]$ denote
the homotopy class of an element $p$ of $\Pscr ( a_0 )$. We equip
$\pi_1 ( \Pscr , a_0 )$ with the product
\begin{equation*}
  [ p ] * [ q ] \doteq [ p * q ] \text{,} \quad [ p ] , [ q ] \in
  \pi_1 ( \Pscr  , a_0 ) \text{.}
\end{equation*}
The operation $*$ is associative, and it easily follows from previous
results that $\pi_1 ( \Pscr , a_0 )$ endowed with $*$ is a group. The
identity $1$ of the group is $[ b ( a_0 ) ]$ and the inverse $[ p ]^{-
  1}$ of $[ p ]$ is $[ \overline{p} ]$. Now assume that $\Pscr$ is
pathwise-connected. Given a $0$-simplex $a_1$, let $q$ be a path from
$a_0$ to $a_1$. Then the map
\begin{equation*}
  \pi_1 ( \Pscr , a_0 ) \ni [ p ] \mapsto [ q * p * \overline{q} ]
  \in \pi_1 ( \Pscr , a_1 )
\end{equation*}
is a group isomorphism. On the basis of these facts we give the
following
\begin{Def}
  \label{def:fi-ho-gr}
  We call $\pi_1 ( \Pscr , a_0 )$ the \textbf{first homotopy group} of
  $\Pscr$ with base $a_0 \in \Sigma_0 ( \Pscr )$. If $\Pscr$ is
  pathwise-connected, we denote this group by $\pi_1 ( \Pscr )$ and
  call it the \textbf{fundamental group} of $\Pscr$. If $\pi_1 ( \Pscr
  ) = 1$ we will say that $\Pscr$ is \textbf{simply connected}.
\end{Def}
We have the following result:
\begin{Pro}
  \label{pro:connec}
  If $\Pscr$ is directed, then $\Pscr$ is pathwise- and simply
  connected.
\end{Pro}

\subsection{Net Cohomology of a Poset}
\label{subsec:net-cohom-poset}

We introduce the net cohomology of a poset equipped with a causal
disjointness relation. Throughout it is assumed that the poset is
pathwise-connected.
\paragraph*{Causal disjointness and net of local algebras} 
Given a poset $\Pscr$, a \emph{causal disjointness relation} on
$\Pscr$ is a symmetric binary relation $\perp$ on $\Pscr$ satisfying
the following properties:
\begin{subequations}
  \label{eq:perp}
  \begin{align}
    & \Oscr_1 \in \Pscr \Rightarrow \text{there exists~} \Oscr_2 \in
    \Pscr \text{ such that }\Oscr_1 \perp \Oscr_2 \text{;} \\
    & \Oscr_1 \leqslant \Oscr_2 \text{~and~} \Oscr_2 \perp \Oscr_3
    \Rightarrow \Oscr_1 \perp \Oscr_3 \text{.}
  \end{align}
\end{subequations}
Given a subset $P \subseteq \Pscr$, the \emph{causal complement} of
$P$ is the subset $P^\perp$ of $\Pscr$ defined as
\begin{equation*}
  P^\perp \doteq \setf{\Oscr \in \Pscr}{\Oscr \perp \Oscr_1 , \quad
    \Oscr_1 \in P} \text{.}
\end{equation*}
Note that if $P_1 \subseteq P$, then $P^\perp \subseteq P^\perp_1$.
Now, assume that $\Pscr$ is a pathwise-connected poset equipped with a
causal disjointness relation $\perp$. A \emph{net of local algebras}
indexed by $\Pscr$ is a correspondence
\begin{equation*}
  \Afrak_\Pscr : \Pscr \ni \Oscr \mapsto \AO \subseteq \BHzero
  \text{,}
\end{equation*}
associating with any $\Oscr$ a von Neumann algebra $\AO$ defined on a
fixed Hilbert space $\Hscr_0$ and satisfying
\begin{align*}
  \text{isotony:} \quad \Oscr_1 \leqslant \Oscr_2 & \Rightarrow
  \AOone \subseteq \AOtwo \text{,} \\
  \text{causality:} \quad \Oscr_1 \perp \Oscr_2 & \Rightarrow \AOone
  \subseteq \AOtwo' \text{,}
\end{align*}
where the prime denotes the commutant of the algebra. The algebra
$\AOperp$ associated with the causal complement $\Oscr^\perp$ of
$\Oscr \in \Pscr$, is the $C^*$-algebra generated by all the algebras
$\AOone$ with $\Oscr_1 \in \Pscr$ and $\Oscr_1 \perp \Oscr$.  The net
$\Afrak_\Pscr$ is said to be \emph{irreducible} whenever, given $T \in
\BHzero$ with $T \in \AO'$ for any $\Oscr \in \Pscr$, then $T = c
\cdot \algunit$, $c \in \Cbb$.
\paragraph*{The category of $1$-cocycles} We refer the reader to
\cite{ghez/lima/roberts:1985} for the definition of a $C^*$-category.
Let $\Pscr$ be a poset with a causal disjointness relation $\perp$,
and let $\Afrak_\Pscr$ be an irreducible net of local algebras. A
\emph{$1$-cocycle} $z$ of $\Pscr$ with values in $\Afrak_\Pscr$ is a
field $z : \Sigma_1 ( \Pscr ) \ni b \mapsto z ( b ) \in \BHzero$ of
unitary operators satisfying the $1$-cocycle identity,
\begin{equation*}
  z ( \partial_0 c ) \cdot z ( \partial_2 c ) = z ( \partial_1 c)
  \text{,} \quad c \in \Sigma_2 ( \Pscr ) \text{,}
\end{equation*}
and the locality condition, $z ( b ) \in \Afrak ( \abs{b} )$, for any
$1$-simplex $b$. An \emph{intertwiner} $t \in ( z , z_1 )$ between a
pair of $1$-cocycles $z$ and $z_1$ is a field $t : \Sigma_0 ( \Pscr )
\ni a \mapsto t_a \in \BHzero$ satisfying the relation
\begin{equation*}
  t_{\partial_0 b} \cdot z ( b ) = z_1 ( b ) \cdot t_{\partial_1 b}
  \text{,} \quad b \in \Sigma_1 ( \Pscr ) \text{,}
\end{equation*}
and the locality condition, $t_a \in \Afrak ( a )$, for any
$0$-simplex $a$. The \emph{category of $1$-cocycles} $\Zup$ is the
category whose objects are $1$-cocycles and whose arrows are the
corresponding intertwiners. The composition between $s \in ( z , z_1
)$ and $t \in ( z_1 , z_2 )$ is the arrow $t \cdot s \in ( z , z_2 )$
defined as
\begin{equation*}
  ( t \cdot s )_a \doteq t_a \cdot s_a \text{,} \quad a \in \Sigma_0 (
  \Pscr ) \text{.}
\end{equation*}
Note that the arrow $1_z$ of $( z , z )$ defined as $( 1_z )_a \doteq
\algunit$ for any $a \in \Sigma_0 ( \Pscr )$ is the identity of $( z ,
z )$. It turns out that $\Zup$ is a $C^*$-category (for details
cf.~\cite{ruzzi:2005b}). Two $1$-cocycles $z$ and $z_1$ are
\emph{equivalent} (or \emph{cohomologous}) if there exists a unitary
arrow $t \in ( z , z_1 )$. A $1$-cocycle $z$ is \emph{trivial} if it
is equivalent to the \emph{identity} cocycle $\iota$ defined as $\iota
( b ) = \algunit$ for any $1$-simplex $b$.  Note that, since
$\Afrak_{\Pscr}$ is irreducible, $\iota$ is irreducible: $( \iota ,
\iota ) = \Cbb \algunit$.
\paragraph*{Equivalence in $\BHzero$ and path independence} A weaker
form of equivalence between $1$-cocycles is the following. $z$ and
$z_1$ are said to be \emph{equivalent in $\BHzero$} if there exists a
field $V : \Sigma_0 ( \Pscr ) \ni a \mapsto V_a \in \BHzero$ of
unitary operators such that
\begin{equation*}
  V_{\partial_0 b} \cdot z ( b ) = z_1 ( b ) \cdot V_{\partial_1 b}
  \text{,} \quad b \in \Sigma_1 ( \Pscr ) \text{.}
\end{equation*}
Note that the field $V$ is not an arrow of $( z , z_1 )$ because it is
not required that $V$ satisfies the locality condition. A $1$-cocycle
is \emph{trivial in $\BHzero$} if it is equivalent in $\BHzero$ to the
trivial $1$-cocycle $\iota$. We denote by $\Zutp$ the set of
$1$-cocycles that are trivial in $\BHzero$, and with the same symbol
we denote the full $C^*$-subcategory of $\Zup$ whose objects are the
$1$-cocycles trivial in $\BHzero$. Triviality in $\BHzero$ is related
to the notion of path independence. The evaluation of a $1$-cocycle
$z$ on a path $p = \set{b_n , \dots , b_1}$ is defined as
\begin{equation*}
  z ( p ) \doteq z (b_n ) \cdots z ( b_2 ) \cdot z ( b_1 ) \text{.}
\end{equation*}
$z$ is said to be \emph{path-independent} whenever 
\begin{equation}
  \label{eq:path-ind}
  z ( p ) = z ( q )  \text{ for any } p , q \in \Pscr ( a_0 , a_1 )
  \text{.} 
 \end{equation}
 As $\Pscr$ is pathwise-connected, a $1$-cocycle is trivial in
 $\BHzero$ if and only if it is path-independent
 \cite{guido/longo/roberts/verch:2001}.
\paragraph*{Connection between homotopy and net cohomology} Let us
consider a poset $\Pscr$ equipped with a causal disjointness relation
$\perp$, and let $\Afrak_\Pscr$ be an irreducible net of local
algebras. In this section we establish the relation between $\pi_1 (
\Pscr )$ and the set $\Zup$.

To begin with, note that $1$-cocycles are \emph{invariant 
for homotopic paths}. This means that, given 
$z \in \Zup$ and two paths $p$, $q$  with the same endpoints,
\begin{equation}
  \label{eq:hom-end}
  p \sim q \Rightarrow z ( p ) = z ( q ) \text{.} 
\end{equation}
This is easily seen because $1$-cocycles are invariant for elementary
deformations and homotopic paths are finite sequences of elementary
deformations of each other. Furthermore, by invariance of $1$-cocycles
for homotopic paths and by relations \eqref{eq:hom-equiv1} and
\eqref{eq:hom-equiv2}, it turns out that
\begin{subequations}
  \begin{align}
    \label{eq:hom-list}
    & z ( b ( a ) ) = \algunit \text{~for any $0$-simplex~} a \text{;}
    \\ 
    & z ( \overline{p} ) = z ( p )^* \text{~for any path~} p \text{.}
  \end{align}
\end{subequations}
We are now in the position to show the connection between the
fundamental group of $\Pscr$ and $\Zup$.
\begin{The}
  \label{the:coc-urep}
  Given $z \in \Zup$ and a $0$-simplex $a_0$, define 
  \begin{equation}
    \label{eq:pi-zed}
    \pi_z ( [ p ] ) \doteq z ( p ) \text{,} \quad [ p ] \in \pi_1 (
    \Pscr , a_0 ) \text{.}
  \end{equation}
  Then, $\pi_z$ is a unitary representation of $\pi_1 ( \Pscr , a_0 )$
  in $\Hscr_0$. The correspondence $z \mapsto \pi_z$ maps $1$-cocycles
  equivalent in $\BHzero$ into equivalent unitary representations of
  $\pi_1 ( \Pscr , a_0 )$ in $\Hscr_0$. Up to equivalence, this map is
  injective. If $\pi_1 ( \Pscr ) = 1$, then $\Zup = \Zutp$.
\end{The}
\begin{proof}
  Observe that the definition is well-posed as $z$ is invariant for
  homotopic paths. By \eqref{eq:hom-list} and \eqref{eq:hom-end}), we
  have $\pi_z ( 1 ) = 1$, $\pi_z ( [ p ]^{- 1} ) = \pi_z ( [ p ] )^*$,
  and $\pi_z ( [ p ] * [ q ] ) = \pi_z ( [ p ] ) \cdot \pi_z ( [ q ]
  )$. Hence $\pi_z$ is a unitary representation of $\pi_1 ( \Pscr )$
  in $\Hscr_0$. Furthermore, if $z_1 \in \Zup$ and $u \in ( z , z_1 )$
  is unitary, then $u_{a_0} \cdot \pi_z ( [ p ] ) = \pi_{z_1} ( [ p ]
  ) \cdot u_{a_0}$.  So what remains to be proved is that the
  correspondonce is injective up to equivalence. To this end consider
  a unitary representation $\pi$ of $\pi_1 ( \Pscr , a_0 )$ on
  $\Hscr_0$. For any $0$-simplex $a$ denote by $p_a$ a path with
  $\partial_1 p_a = a$ and $\partial_0 p_a = a_0$. Let
 \begin{equation*}
   z_\pi ( b ) \doteq \pi \big( [ p_{\partial_0 b} * b *
   \overline{p_{\partial_1b }} ] \big) \text{,} \quad b \in \Sigma_1 (
   \Pscr ) \text{.}
 \end{equation*}
 Given a $2$-simplex $c$, we have
 \begin{align*}
   z_\pi ( \partial_0 c ) \cdot z_\pi ( \partial_2 c) & = \pi \big( [
   p_{\partial_{00} c} * \partial_0 c * \overline{p_{\partial_{10} c}}
   * p_{\partial_{02} c} * \partial_2 c * \overline{p_{\partial_{12}
       c}} ] \big) \\ 
   & = \pi \big( [ p_{\partial_{00} c} * \partial_0 c *
   \overline{p_{\partial_{10} c}} * p_{\partial_{10} c} * \partial_2
   c * \overline{p_{\partial_{12} c}} ] \big) \\
   & = \pi \big( [ p_{\partial_{00} c} * \partial_0 c * \partial_2 c *
   \overline{p_{\partial_{11} c}} ] \big) = \pi \big( [
   p_{\partial_{01} c} * \partial_1 c * \overline{p_{\partial_{11} c}}
   ] \big) \\ 
   & = z_\pi ( \partial_1 c ) \text{.}
 \end{align*}
 Hence $z_\pi$ satisfies the $1$-cocycle identity but in general
 $z_\pi \not\in \Zuk$ because $z_\pi ( b )$ does not have to belong to
 $\Afrak ( \abs{b} )$. However, note that if we consider $\pi_{z_1}$
 for some $z_1 \in \Zuk$, then
 \begin{equation*}
   z_{\pi_{z_1}} ( b ) = \pi_{z_1} \big( [ p_{\partial_0 b} * b *
   \overline{p_{\partial_1 b}} ] ) = z_1 ( p_{\partial_0 b}) \cdot
   z_1 ( b ) \cdot z_1( p_{\partial_0 b} \big)^* \text{.}
 \end{equation*}
 Therefore, $z_{\pi_{z_1}}$ is equivalent in $\BHzero$ to $z_1$. This
 entails that, if $\pi_z$ is equivalent to $\pi_{z_1}$, then $z$ is
 equivalent in $\BHzero$ to $z_1$. Finally, assume that $\pi_1 ( \Pscr
 ) = 1$, then $z ( p ) = \algunit$ for any closed path $p$.  Thus $z$
 is path-independent on $\Pscr$, hence $z$ is trivial in $\BHzero$.
\end{proof}
\paragraph*{Change of Index Set} We now study the behaviour  of net
cohomology under a change of the index set. By a \emph{subposet} of a
poset $\Pscr$ we mean a subset $\widehat{\Pscr}$ of $\Pscr$ equipped
with the same order relation as $\Pscr$.
\begin{Def}
 \label{def:subpo}
 Consider a subposet $\widehat{\Pscr}$ of $\Pscr$. We will say that
 $\widehat{\Pscr}$ is a \textbf{refinement} of $\Pscr$, if for any
 $\Oscr \in \Pscr$ there exists $\widehat{\Oscr} \in \widehat{\Pscr}$
 such that $\widehat{\Oscr} \leqslant \Oscr$. A refinement
 $\widehat{\Pscr}$ of $\Pscr$ is said to be \textbf{locally
   relatively connected} if, given $\Oscr \in \Pscr$, for any pair
 $\widehat{\Oscr}_1$, $\widehat{\Oscr}_2 \in \widehat{\Pscr}$ with
 $\widehat{\Oscr}_1$, $\widehat{\Oscr}_2 \leqslant \Oscr$ there is a
 path $\hat{p}$ in $\widehat{\Pscr}$ from $\widehat{\Oscr}_1$ to
 $\widehat{\Oscr}_2$ such that $\abs{\hat{p}} \leqslant \Oscr$.
\end{Def}

Let $\Pscr$ be a pathwise-connected poset and let $\perp$ be a causal
disjointness relation for $\Pscr$. Let $\Afrak_\Pscr$ be an
irreducible net of local algebras indexed by $\Pscr$ and defined on a
Hilbert space $\Hscr_0$. If $\widehat{\Pscr}$ is a locally relatively
connected refinement of $\Pscr$, then it is easily seen that
$\widehat{\Pscr}$ is pathwise-connected and that $\perp$ is a causal
disjointness relation on $\widehat{\Pscr}$. Furthermore, the
restriction of $\Afrak_\Pscr$ to $\widehat{\Pscr}$ is a net of local
algebras $\Afrak_{\widehat{\Pscr}}$ indexed by $\widehat{\Pscr}$. Let
$\Zscr^1_t \big( \Afrak_{\widehat{\Pscr}} \big)$ be the category of
$1$-cocycles of $\widehat{\Pscr}$, trivial in $\BHzero$, with values
in the net $\Afrak_{\widehat{\Pscr}}$.  Notice that
$\Afrak_{\widehat{\Pscr}}$ need not be irreducible. However, as
$\widehat{\Pscr}$ is a refinement of $\Pscr$, the trivial $1$-cocycle
$\hat{\iota}$ of $\Zscr^1_t \big( \Afrak_{\widehat{\Pscr}} \big)$ is
irreducible.
\begin{The}
 \label{the:cat-equiv}
 Let $\widehat{\Pscr}$ be a locally relatively connected refinement of
 $\Pscr$. Then the categories $\Zutp$ and $\Zscr^1_t (
 \Afrak_{\widehat{\Pscr}} )$ are equivalent.
\end{The}
\begin{proof}
  For any $z \in \Zutp$ and for any $t \in ( z , z_1 )$ define
  \begin{equation*}
    \RE ( z ) \doteq z \restriction \Sigma_1 ( \widehat{\Pscr} )
    \text{,} \quad \RE ( t ) \doteq t \restriction \Sigma_0 (
    \widehat{\Pscr} ) \text{.}
  \end{equation*}
  $\RE$ is a covariant and faithful functor from $\Zutp$ into
  $\Zscr^1_t ( \Afrak_{\widehat{\Pscr}} )$. We now define a functor
  from $\Zscr^1_t ( \Afrak_{\widehat{\Pscr}} )$ to $\Zutp$. To this
  end, choose a function $\f : \Pscr \rightarrow \widehat{\Pscr}$
  satisfying the following properties: given $\Oscr \in \Pscr$, then
  $\f ( \Oscr ) = \Oscr$ for $\Oscr \in \widehat{\Pscr}$, otherwise
  $\f ( \Oscr ) \leqslant \Oscr$. For any $b \in \Sigma_1 ( \Pscr )$
  we denote by $\hat{p} \big( \f ( \partial_0 b ) , \f ( \partial_1b )
  \big)$ a path of $\widehat{\Pscr}$ from $\f ( \partial_0 b )$ to $\f
  ( \partial_1 b )$ whose support is contained in $\abs{b}$; this is
  possible because $\widehat{\Pscr}$ is a locally relatively connected
  refinement of $\Pscr$. Given $\hat{z},\hat{z}_1 \in \Zscr^1_t (
  \Afrak_{\widehat{\Pscr}} )$ and $\hat{t} \in ( \hat{z} , \hat{z}_1
  )$ define
  \begin{align*}
    \F ( \hat{z} ) ( b ) & \doteq \hat{z} \big( \hat{p} \big( \f (
    \partial_0 b ) , \f ( \partial_1 b ) \big) \big) \text{,} & b
    \in \Sigma_1 ( \Pscr ) \text{.} \\
    \F ( \hat{t} )_a & \doteq \hat{t}_{\f ( a )} \text{,} & a \in
    \Sigma_0 ( \Pscr ) \text{.}
  \end{align*}
  $\F ( \hat{z} ) ( b ) \in \Afrak ( \abs{b} )$ for any $1$-simplex
  $b$. As $\hat{z}$ is path-independent, given $c \in \Sigma_2 ( \Pscr
  )$, we have
  \begin{align*}
    \F ( \hat{z} ) ( \partial_0 c ) \cdot \F ( \hat{z} ) ( \partial_2
    c ) & = \hat{z} \big( \hat{p} \big( \f ( \partial_{00} c ) , \f (
    \partial_{10} c ) \big) \big) \cdot \hat{z} \big( \hat{p} \big( \f
    ( \partial_{02} c ) , \f ( \partial_{12} c ) \big) \big) \\
    & = \hat{z} \big( \hat{p} \big( \f ( \partial_{01} c ) , \f (
    \partial_{02} c) \big) \big) \cdot \hat{z} ( \hat{p} ( \f (
    \partial_{02} c ) , \f ( \partial_{11} c ) \big) \big) \\
    & = \hat{z} \big( \hat{p} \big( \f ( \partial_{01} c ) , \f (
    \partial_{11} c ) \big) \big) = \F ( \hat{z} ) ( \partial_1 c )
    \text{.} 
  \end{align*}
  Hence $\F ( \hat{z} )$ satisfies the $1$-cocycle identity, and it is
  trivial in $\BHzero$ because so is $\hat{z}$. Therefore, $\F (
  \hat{z} ) \in \Zutp$. In an analogous fashion one can show that $\F
  ( \hat{t} ) \in \big( \F ( \hat{z} ), \F ( \hat{z}_1 ) \big)$.
  Hence, $\F$ is a covariant functor from $\Zscr^1_t \big(
  \Afrak_{\widehat{\Pscr}} \big)$ to $\Zutp$. Finally, it can easily
  be checked that the pair $\RE$, $\F$ constitutes an equivalence
  between $\Zutp$ and $\Zscr^1_t \big( \Afrak_{\widehat{\Pscr}} \big)$
  (for details cf.~ \cite{brunetti/ruzzi}).
\end{proof}

\subsection{The Poset as a Basis for a Topological Space}
\label{subsec:poset-top}

\begin{figure}[!ht]
  \begin{center}
    \includegraphics[scale=0.6]{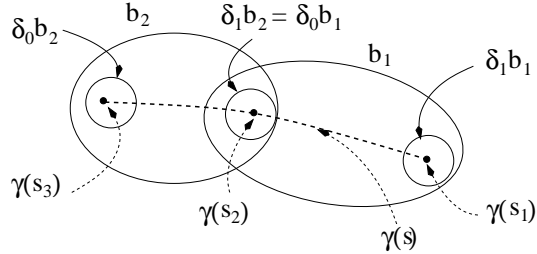}
     \caption{The path $\set{b_2 , b_1}$ is an approximation of the
       curve $\gamma$ (dashed). The symbol $\delta$ stands for
       $\partial$.} 
    \label{fig:poset-approx}
   \end{center}
\end{figure}

Consider a topological Hausdorff space $\Xscr$. The topics of the
previous sections are now investigated in the case that $\Pscr$ is a
basis for the topology of $\Xscr$ ordered under \emph{inclusion}
$\subseteq$. This allows us both to show the connection between the
notions for posets and the corresponding topological ones and to
understand how topology affects net cohomology.
\paragraph*{Homotopy} In what follows, by a curve $\gamma$ in $\Xscr$
we mean a continuous function from the interval $[ 0 , 1 ]$ into
$\Xscr$. We recall that the reverse of a curve $\gamma$ is the curve
$\overline{\gamma}$ defined as $\overline{\gamma} ( t ) \doteq \gamma
( 1 - t )$ for $t \in [ 0 , 1 ]$. If $\beta$ is a curve such that
$\beta ( 1 ) = \gamma ( 0 )$, the composition $\gamma * \beta$ is the
curve
\begin{equation*}
  ( \gamma * \beta ) ( t ) \doteq
  \begin{cases}
    \beta ( 2 t ) \text{,} & 0 \leqslant t \leqslant 1 / 2 \text{,}\\
    \gamma ( 2 t - 1 ) \text{,} &  1 / 2 \leqslant t \leqslant 1
    \text{.} 
  \end{cases}
\end{equation*}
Finally, the constant curve $e_x$ is the curve $e_x ( t ) = x$ for any
$t \in [ 0 , 1 ]$.
\begin{Def}
  \label{def:poset-approx}
  Given a curve $\gamma$. A path $p = \set{b_n , \dots , b_1}$ is said
  to be a \textbf{poset approximation} of $\gamma$ (or simply an
  \textbf{approximation}) if there is a partition $0 = s_0 < s_1 <
  \dots < s_n =1$ of the interval $[ 0 , 1 ]$ such that
  (Fig.~\ref{fig:poset-approx}) 
  \begin{equation*}
    \gamma ( [ s_{i - 1} , s_i ] ) \subseteq \abs{b_i} \text{,} \quad
    \gamma ( s_{i - 1} ) \in \partial_1 b_i \text{,} \quad \gamma (
    s_i ) \in \partial_0 b_i \text{,} \qquad i = 1 , \dots , n
    \text{.} 
  \end{equation*}
  By $\App ( \gamma )$ we denote the set of approximations of
  $\gamma$.
\end{Def}
Since $\Pscr$ is a basis for the topology of $\Xscr$, $\App ( \gamma )
\ne \emptyset$ for any curve $\gamma$. The converse, \ie, that for a
given path $p$ there is a curve $\gamma$ such that $p$ is an
approximation of $\gamma$, holds if the elements of $\Pscr$ are
arcwise-connected sets of the topological space $\Xscr$. The
approximations of curves have the following properties:
\begin{subequations}
  \begin{align}
    \label{eq:approx-prop}
    p \in \App ( \gamma ) & \Leftrightarrow  \overline{p} \in \App (
    \overline{\gamma} ) \text{,} \\
    p \in  \App ( \sigma ) \text{,~} q \in \App ( \beta ) &
    \Rightarrow p * q \in \App ( \sigma * \beta ) \text{,}
  \end{align}
\end{subequations}
where $\beta ( 1 ) = \sigma ( 0 )$, $\partial_0 q = \partial_1 p$. As
already said, we can find an approximation for any curve $\gamma$.

We now compare the connectedness relations of $\Xscr$ with those of
$\Pscr$. If the elements of $\Pscr$ are arcwise-connected sets of
$\Xscr$, then an open set $X \subseteq \Xscr$ is arcwise-connected in
$\Xscr$ if and only if the poset $\Pscr_X$, defined as
\begin{equation}
  \label{eq:pos-def}
  \Pscr_X \doteq \setf{\Oscr \in \Pscr}{\Oscr \subseteq X} \text{,} 
\end{equation}
is pathwise-connected. Note that the set $\Pscr_X$ is a \emph{sieve}
of $\Pscr$, \ie~a subfamily $S$ of $\Pscr$ such that, if $\Oscr \in
S$ and $\Oscr_1 \subseteq \Oscr$, then $\Oscr_1 \in S$. Now, assume
that $P$ is a sieve of $\Pscr$. Then $P$ is pathwise-connected in
$\Pscr$ if and only if the open set $\Xscr_P$, defined as
\begin{equation}
  \label{eq:open-def}
  \Xscr_P \doteq \bigcup \setf{\Oscr \subseteq \Xscr}{\Oscr \in P}
  \text{,} 
\end{equation}
is arcwise-connected in $\Xscr$.

For simply connectedness, assume that the elements of $\Pscr$ are
arcwise- and simply connected subsets of $\Xscr$. Given two curves
$\gamma$ and $\beta$ with the same endpoints, let $p \in \App ( \gamma
)$ and $q \in \App ( \beta )$ have the same endpoints. Then
\begin{equation}
  \label{eq:approx-endp-hom}
  p \sim q \Leftrightarrow \gamma \sim \beta \text{.}
\end{equation}
By using this approximation result we get
\begin{The}
  \label{the:top-basis}
  Let $\Xscr$ be a Hausdorff, arcwise-connected topological space.
  Assume that there is a basis $\Pscr$ for the topology of $\Xscr$
  whose elements are arcwise- and simply connected subsets of $\Xscr$.
  Then $\pi_1 ( \Xscr ) \simeq \pi_1 ( \Pscr )$.
\end{The}
\begin{proof}
  Fix a base $0$-simplex $a_0$ and a base point $x_0 \in a_0$. Define
  \begin{equation*}
    \pi_1 ( \Xscr , x_0 ) \ni [ \gamma ] \mapsto [ p ] \in \pi_1 (
    \Pscr , a_0 ) \text{,}
  \end{equation*}
  where $p$ is an approximation of $\gamma$. By \eqref{eq:approx-prop}
  and  \eqref{eq:approx-endp-hom}, this map is a group isomorphism.
\end{proof}
\begin{Cor}
  \label{cor:top-nonsimp}
  Let $\Xscr$ and $\Pscr$ be as in the previous theorem. If $\Xscr$ is
  not simply connected, then $\Pscr$ is not directed under inclusion.
\end{Cor}
\paragraph*{Net cohomology} The aim here is to apply the results 
of the previous section to the case where the topological space is a
globally hyperbolic spacetime $\Mscr$. In particular we will consider
the set of globally hyperbolic regions $\Kscr^h ( \Mscr )$ and the set
of diamonds $\Kscr^d ( \Mscr )$ introduced in Section~\ref{sec:pre}.
As seen there, both of them are bases for the topology of $\Mscr$
consisting of arcwise-connected elements. However, while diamonds are
simply connected, this is not true for globally hyperbolic regions. So
we conclude:
\begin{Pro}
  \label{pro:triv-coc}
  For any $\Mscr \in \Loc$ the following assertions hold.
  \begin{proplist}
  \item\label{list:triv-coc1} The fundamental group of $\Mscr$ and
    that of the poset $\Kscr^d ( \Mscr )$ are isomorphic.
  \item\label{list:triv-coc2} The fundamental group of the poset
    $\Kscr^h ( \Mscr )$ is trivial.
  \end{proplist}  
\end{Pro}
\begin{proof}
  \ref{list:triv-coc1} follows from Theorem \ref{the:top-basis}.
  Concerning \ref{list:triv-coc2}, cf.~\cite{brunetti/ruzzi:2005}.
\end{proof}
So the poset $\Kscr^h ( \Mscr )$ does not encode the global
topological features of $\Mscr$.

We now turn to the cohomological consequences. First, note that
$\Kscr^d ( \Mscr )$ is a locally relatively connected refinement of
$\Kscr^h ( \Mscr )$, since the elements of $\Kscr^h ( \Mscr )$ are
arcwise-connected and $\Kscr^d ( \Mscr )$ is a basis for the topology
of $\Mscr$. Consider an irreducible net of local algebras
$\Afrak_{\Kscr^h ( \Mscr )}$ defined on a Hilbert space $\Hscr_0$, and
denote by $\Afrak_{\Kscr^d ( \Mscr )}$ the restriction of
$\Afrak_{\Kscr^h ( \Mscr )}$ to the set of diamonds $\Kscr^d ( \Mscr
)$. It is easy to check that $\Afrak_{\Kscr^d ( \Mscr )}$ is
irreducible.
\begin{Pro}
  \label{pro:irr-cat-equiv}
  The following assertions hold.
  \begin{proplist}
  \item\label{list:irr-cat-equiv1} If $\Mscr$ is simply connected,
    then $\Zscr^1_t \big( \Afrak_{\Kscr^d ( \Mscr )} \big) = \Zscr^1
    \big( \Afrak_{\Kscr^d ( \Mscr )} \big)$.
  \item\label{list:irr-cat-equiv2} $ \Zscr^1_t \big( \Afrak_{\Kscr^h (
      \Mscr )} \big) = \Zscr^1 \big( \Afrak_{\Kscr^h ( \Mscr )} \big)$
    for any spacetime $\Mscr$.
  \item\label{list:irr-cat-equiv3} $\Zscr^1_t \big( \Afrak_{\Kscr^h (
      \Mscr )} \big)$ is equivalent to $\Zscr^1_t \big(
    \Afrak_{\Kscr^d ( \Mscr )} \big)$.
  \end{proplist}
\end{Pro}
\begin{proof}
  \ref{list:irr-cat-equiv1} and \ref{list:irr-cat-equiv2} follow from
  Theorem~\ref{the:coc-urep} and from Proposition~\ref{pro:triv-coc}.
  \ref{list:irr-cat-equiv3} Since $\Kscr^d ( \Mscr )$ is a locally
  relatively connected refinement of $\Kscr^h ( \Mscr )$, the proof
  then follows by using Theorem~\ref{the:cat-equiv}.
\end{proof}
Some observations are in order at this point.
\begin{abclist}
\item The main difference between the net cohomology of $\Kscr^d (
  \Mscr )$ and that of $\Kscr^h ( \Mscr )$ is that in the former case
  there may exist $1$-cocycles which are not trivial in $\BHzero$.
  Examples of this kind of $1$-cocycles arise in $2$-dimensional
  chiral conformal quantum field theories
  \cite{fredenhagen/rehren/schroer:1992}.
\item As already said, DHR sectors are described in terms of
  $1$-cocycles that are trivial in $\BHzero$. By
  Proposition~\ref{pro:irr-cat-equiv}\ref{list:irr-cat-equiv3}, the
  properties of sectors do not depend on the choice of the index set.
\end{abclist}

\section{DHR and Local Covariance}
\label{sec:DHR-cov}

Charged superselection sectors in Minkowski space $\Mbb^4$ are the
unitary equivalence classes of irreducible representations of a net of
local observables which are ``local excitations'' of the vacuum
representation. We can distinguish two types of charged sectors
according to the class of regions in spacetime which are used as index
sets of the net. On the one hand, charged sectors of
Doplicher-Haag-Roberts type, where one considers double cones of
$\Mbb$, and, on the other hand, charges of Buchholz-Fredenhagen type
associated with a particular class of noncompact regions like
spacelike cones. In both cases, sectors define a $C^*$-category in
which the charge structure manifests itself by the existence of a
tensor product, a permutation symmetry and a conjugation(see
Subsection~\ref{subsec:dhr-analysis}).
\paragraph*{The Reference State Space} Our aim is to study the
superselection sectors of Doplicher-Haag-Roberts type in the framework
of a locally covariant quantum field theory defined via the functor
$\Ascr$. The first step is to introduce the notion of a reference
state space taking on the role of the vacuum representation. To this
end, note that the vacuum representation serves as a reference
representation that singles out charged sectors, and that it is enough
to consider a vacuum representation that has the Borchers property and
satisfies Haag duality\footnote{It is possible to attenuate the
  Borchers property \cite{ruzzi:2003}.  Conversely, up to now, for the
  theory of superselection sectors Haag duality or a weaker form of it
  \cite{roberts:1990} seems to be an essential requirement that the
  vacuum representation has to meet.}  \cite{doplicher/roberts:1990}.
\begin{Def}  
  \label{def:ref-state-space}
  We call a \textbf{reference state space} for $\Ascr$ a locally
  quasi-equivalent state space $\Sscr_0$ on globally hyperbolic
  regions such that for any $\Mscr \in \Loc$ there is a state
  $\omega \in \Sscr_0 ( \Mscr)$ such that
  \begin{deflist}
  \item the net $\omega^* \Ascr_{\Kscr^h ( \Mscr )}$ satisfies the
    Borchers property;
  \item the net $\omega^* \Ascr_{\Kscr^d ( \Mscr )}$, indexed by
    diamonds, is irreducible and satisfies punctured Haag duality.
  \end{deflist}
\end{Def}
Let us briefly explain the meaning of this definition. Given $\omega
\in \Sscr_0 ( \Mscr )$, the net $\omega^* \Ascr_{\Kscr^h ( \Mscr )}$
satisfies the \textbf{Borchers property} if, given a set $\Oscr \in
\Kscr^h ( \Mscr )$ for any diamond $\Oscr_1 \in \Kscr^h ( \Mscr )$
with $\overline{\Oscr_1} \subset \Oscr$ and a nonzero orthogonal
projection $E$ of $\Afrak_\omega ( \Oscr_1 )$, there exists an
isometry $V \in \Afrak_\omega ( \Oscr )$ such that $V V^* = E$. The
net $\omega^* \Ascr_{\Kscr^d ( \Mscr )}$ satisfies \textbf{punctured
  Haag duality} if for any $x \in \Mscr$ and for any diamond $\Oscr$
with $\overline{\Oscr} \perp x$ we have
\begin{equation}
  \label{eq:punc-Haag}
  \Afrak_\omega ( \Oscr ) = \bigcap \bsetf{\Afrak_\omega ( \Oscr_1
    )'}{\overline{\Oscr_1} \perp x , \quad \Oscr_1 \perp \Oscr}
  \text{.} 
\end{equation}
It is clear that punctured Haag duality entails Haag duality. 
If $\omega^* \Ascr_{\Kscr^d ( \Mscr )}$ is irreducible and satisfies 
punctured Haag duality then it is \textbf{locally definite}, \ie, for
any $x \in \Mscr$ we have  
\begin{equation}
  \label{eq:loc-def}
  \Cbb \cdot \algunit = \bigcap \bsetf{\Afrak_\omega ( \Oscr )}{\Oscr
    \in \Kscr^d ( \Mscr ) , \quad x \in \Oscr} \text{.} 
\end{equation}
As $\Kscr^h ( \Mscr )$ is a basis for the topology of $\Mscr$, the net
$\omega^* \Ascr_{\Kscr^h ( \Mscr )}$ is locally definite as well. An
example of a locally covariant quantum field theory with a state space
satisfying the properties of Definition~\ref{def:ref-state-space} is
provided by the Klein-Gordon scalar field and by the space of
quasi-free states satisfying the microlocal spectrum condition
\cite{brunetti/fredenhagen/verch:2003,ruzzi:2005a}. We stress that we
require the existence for any $\Mscr \in \Loc$ of at least one state
$\omega \in \Sscr_0 ( \Mscr )$ satisfying punctured Haag duality. This
property seems to be the correct generalization of Haag duality to
deal with the nontrivial topology of arbitrary globally hyperbolic
spacetimes \cite{ruzzi:2005b}. The reason for punctured Haag duality
to be assumed for the net indexed by $\Kscr^d ( \Mscr )$ is that
$\Kscr^h ( \Mscr )$ contains elements which are not simply connected
as well as elements whose causal complement is not arcwise-connected.
Therefore, punctured Haag duality (and also Haag duality) might not
hold for a net indexed by $\Kscr^h ( \Mscr )$
(cf.~\cite{ruzzi:2005a}).

Let us consider the main properties of such a state space.

\emph{First}, as a consequence of local quasi-equivalence, for any
pair $\omega$, $\sigma \in \Sscr_0 ( \Mscr )$ the nets $\omega^*
\Ascr_{\Kscr^h ( \Mscr )}$ and $\sigma^* \Ascr_{\Kscr^h ( \Mscr )}$
are isomorphic, \ie, there is a collection
\begin{equation}
  \label{eq:net-flip-1}
  \rho_{\omega , \sigma} = \bsetf{\rho_\Oscr : \Afrak_\omega ( \Oscr )
    \rightarrow \Afrak_\sigma ( \Oscr )}{\Oscr \in \Kscr^h ( \Mscr )}
\end{equation}
consisting of $^*$-isomorphisms of von Neumann algebras which are compatible 
with the net structure, 
\begin{equation}
  \label{eq:net-flip-2}
  \rho_\Oscr \restriction \Oscr_1 = \rho_{\Oscr_1} \text{,} \quad
  \Oscr_1 \subseteq \Oscr \text{.}
\end{equation}
It is clear that, by restricting $\rho_{\omega , \sigma}$ to the set
of diamonds $\Kscr^d ( \Mscr )$, one gets a net isomorphism
$\rho_{\omega , \sigma} : \omega^* \Ascr_{\Kscr^d ( \Mscr )}
\rightarrow \sigma^* \Ascr_{\Kscr^d ( \Mscr )}$.

\emph{Second}, note that in $\Sscr_0 ( \Mscr )$ there is a state
$\omega$ satisfying punctured Haag duality, the Borchers property and,
as seen above, local definiteness. Since the Borchers property and
local definiteness are local properties, the fact that for any $\sigma
\in \Sscr_0 ( \Mscr )$ $\rho_{\omega , \sigma}$ is a net isomorphism
implies that also the net $\sigma^* \Ascr_{\Kscr^h ( \Mscr )}$
satisfies the Borchers property and local definiteness.

\emph{Third}, consider the mapping $\psi \in \hom_\Loc ( \Mscr_1 ,
\Mscr )$ and the associated injective $C^*$-morphism $\alpha_\psi :
\Ascr ( \Mscr_1 ) \rightarrow \Ascr ( \Mscr )$. Any state $\omega \in
\Sscr ( \Mscr )$ induces two different representations of $\Ascr (
\Mscr_1 )$. On the one hand, if $\pi_\omega$ is the GNS representation
associated with $\omega$, then $\pi_\omega \alpha_\psi$ is a
representation of $\Ascr ( \Mscr_1 )$.  On the other hand, by local
covariance $\omega \alpha_\psi \in \Sscr_0 ( \Mscr_1 )$ so that the
GNS representation $\pi_{\omega \alpha_\psi}$ associated with $\omega
\alpha_\psi$ is again a representation of $\Ascr ( \Mscr_1 )$. As
noticed before, any element of $\Sscr_0 ( \Mscr )$ for arbitrary
$\Mscr \in \Loc$ satisfies the Borchers property and local
definiteness. Using the Borchers property, it turns out that the
representations $\pi_\omega \alpha_\psi$ and $\pi_{\omega
  \alpha_\psi}$ are locally quasi-equivalent (for details
cf.~\cite{brunetti/ruzzi:2005}). In particular, if we define
\begin{equation}
  \label{eq:net-emb-1}
  \tau^\omega_\psi \big( \pi_\omega \alpha_\psi ( A ) \big) \doteq
  \pi_{\omega \alpha_\psi} ( A ) \text{,} \quad A \in \Ascr ( \Mscr_1
  ) \text{,}
\end{equation}
it turns out that 
\begin{equation}
  \label{eq:net-emb-2}
  \tau^\omega_\psi : \omega^* \Ascr_{\psi \left( \Kscr^h ( \Mscr_1 )
    \right)} \rightarrow ( \omega \alpha_\psi )^* \Ascr_{\Kscr^h (
    \Mscr_1 )}
\end{equation}
is a net isomorphism, where $\omega^* \Ascr_{\psi \left( \Kscr^h (
    \Mscr_1 ) \right)}$ is the restriction of $\omega^* \Ascr_{\Kscr^h
  ( \Mscr )}$ to the poset $\psi \big( \Kscr^h ( \Mscr_1 ) \big)$.

\emph{Fourth}, the net isomorphisms just introduced satisfy the
following commutation relation. Given $\psi \in \hom_\Loc ( \Mscr_1 ,
\Mscr )$, $\phi \in \hom_\Loc ( \Mscr_2 , \Mscr_1 )$, then
\begin{subequations}
  \label{eq:comm-rel}
  \begin{align}
    \rho_{\omega \alpha_\psi , \sigma \alpha_\psi} \tau^\omega_\psi &
    = \tau^\sigma_\psi \rho_{\omega , \sigma} \text{,} & \sigma ,
    \omega \in \Sscr_0 ( \Mscr ) \text{,} \\
    \tau^\omega_{\psi \phi} & = \tau^{\omega \alpha_\psi}_\phi
    \tau^\omega_\psi \text{,} & \omega \in \Sscr_0 ( \Mscr ) \text{.}
\end{align} 
\end{subequations}
\paragraph*{The Selection Criterion}
\begin{Def}
  \label{def:charge-supersel}
  The \textbf{charged superselection sectors} of $\Ascr$ with respect
  to the reference state space $\Sscr_0$ are the unitary equivalence
  classes of the irreducible elements of the categories $\Zscr^1_t
  \big( \omega , \Kscr^d ( \Mscr ) \big)$ as $\omega$ varies in
  $\Sscr_0 ( \Mscr )$ and as $\Mscr$ varies in $\Loc$.
\end{Def}
Our aim is, first to understand the charge structure of sectors of the
categories $\Zscr^1_t \big( \omega , \Kscr^d ( \Mscr ) \big)$ on a
fixed spacetime background $\Mscr$ as $\omega$ varies in $\Sscr_0 (
\Mscr )$, and second to inspect the locally covariant structure of
sectors. This means that we will study the connection of sectors
associated with different spacetime backgrounds which are
isometrically embedded.

At this point some observations concerning the definition of
superselection sectors in a locally covariant quantum field theory are
in order.
\begin{abclist}
\item The above definition of superselection sectors in terms of net
  cohomology is equivalent to the usual one given in terms of
  representations of the net of local observables which are ``sharp
  excitations'' of a reference representation.  In particular, it is
  shown in \cite{brunetti/ruzzi:2005} that for any spacetime $\Mscr$
  and any $\sigma \in \Sscr_0 ( \Mscr )$, to any $1$-cocycle $z \in
  \Zscr^1_t \big( \sigma , \Kscr^d ( \Mscr ) \big)$ there corresponds
  a unique representation $\pi^z$ (up to equivalence) of the net of
  local observables which is a ``sharp excitation'' of a
  representation $\pi_\omega$ associated with a state $\omega \in
  \Sscr_0 ( \Mscr )$ satisfying punctured Haag duality.
\item There are several reasons why we choose to study $1$-cocycles of
  the poset $\Kscr^d ( \Mscr )$ instead of $1$-cocycles of $\Kscr^h (
  \Mscr )$. On the one hand, $\Kscr^d ( \Mscr )$ fits topological and
  causal properties of $\Mscr$ better than $\Kscr^h ( \Mscr )$: The
  fundamental group of $\Kscr^d ( \Mscr )$ is the same as that of
  $\Mscr$ and any diamond has an arcwise-connected causal complement.
  These two properties belong to the key ingredients in
  \cite{ruzzi:2005b} where, in the Haag-Kastler framework, the charge
  structure of sharply localized sectors in a fixed background
  spacetime has been established. On the contrary, $\Kscr^h ( \Mscr )$
  is simply connected irrespective of the topology of $\Mscr$
  (Proposition~\ref{pro:triv-coc}), and it has elements with a
  nonarcwise-connected causal complement. On the other hand,
  Proposition~\ref{pro:irr-cat-equiv} states that there is no loss of
  generality in studying path-independent $1$-cocycles of $\Kscr^d (
  \Mscr )$ instead of those of $\Kscr^h ( \Mscr )$, since the
  corresponding categories are equivalent. We have to mention,
  however, that the cited result provides an equivalence of
  $C^*$-categories, but it does not concern the tensorial structure of
  the categories. This topic is analysed in \cite{brunetti/ruzzi},
  where a symmetric tensor equivalence between the categories
  associated with $\Kscr^d ( \Mscr )$ and $\Kscr^h ( \Mscr )$ is
  given.
\item Since $\Sscr_0$ satisfies the Borchers property, by a routine
  calculation (see \cite{roberts:2004}), it turns out that the
  category $\Zscr^1_t \big( \omega , \Kscr^d ( \Mscr ) \big)$ is
  closed under direct sums and subobjects for any $\omega \in \Sscr_0
  ( \Mscr )$ and any $\Mscr \in \Loc$. As observed above, $\Sscr_0$ is
  locally definite. Then, by \cite[Lemma~4.6]{brunetti/ruzzi:2005},
  the identity cocycle of $\Zscr^1_t \big( \omega , \Kscr^d ( \Mscr )
  \big)$ is irreducible for any $\omega \in \Sscr_0 ( \Mscr )$ and any
  $\Mscr \in \Loc$.
\end{abclist}

\subsection{Fixed Spacetime Background}
\label{subsec:fixed-st-back}

In the present subsection we investigate the charge structure of
superselection sectors in a fixed spacetime background $\Mscr \in
\Loc$. We start with the remark that for a state $\omega \in \Sscr_0 (
\Mscr )$ satisfying punctured Haag duality the corresponding category
has a tensor product and a permutation symmetry and that its objects
with finite statistics have conjugates. Afterwards, we will show that
all the constructions can coherently be extended to the categories
$\Zscr^1_t \big( \sigma , \Kscr^d ( \Mscr ) \big)$ for any $\sigma \in
\Sscr_0 ( \Mscr )$. We conclude by studying the behaviour of these
categories under restriction to subregions of $\Mscr$.
\paragraph*{Independence of the choice of reference states} 
As a starting point, we apply the results of the analysis of sharply
localized sectors on a fixed spacetime background $\Mscr$, carried out
in the Haag-Kastler framework, to our present setting. In particular we
have
\begin{The}[\cite{ruzzi:2005b}]
  \label{the:tensor-permutation}
  Let $\omega \in \Sscr_0 ( \Mscr )$ satisfy punctured Haag duality.
  Then
  \begin{proplist}
  \item $\Zscr^1_t \big( \omega ,\Kscr^d ( \Mscr ) \big)$ has a tensor
    product and a permutation symmetry;
  \item the category has left inverses and a notion of statistics of
    objects;
  \item the objects with finite statistics have conjugates.
  \end{proplist}
\end{The}
The existence of at least one state $\omega \in \Sscr_0 ( \Mscr )$
satisfying punctured Haag duality for any $\Mscr \in \Loc$ is a
cornerstone for our analysis. In fact, the tensor product and the
permutation symmetry of $\Zscr^1_t \big( \omega , \Kscr^d ( \Mscr )
\big)$ can be extended to the category $\Zscr^1_t \big( \omega ,
\Kscr^d ( \Mscr ) \big)$ for any state $\sigma \in \Sscr_0 ( \Mscr )$
by means of the net isomorphism $\rho_{\omega , \sigma}$ introduced in
the previous subsection. All these categories turn out to be symmetric
tensor $^*$-isomorphic \cite{brunetti/ruzzi:2005}. In this review we
do not consider in detail the tensor structure, but will prove that
all the categories are $^*$-isomorphic.

Given any pair $\sigma$, $\omega \in \Sscr_0 ( \Mscr )$, consider 
\begin{equation*}
  \rho_{\omega , \sigma} : \omega^* \Ascr_{\Kscr^h ( \Mscr )}
  \rightarrow \sigma^* \Ascr_{\Kscr^h ( \Mscr )} \text{,}
\end{equation*}
the net isomorphism \eqref{eq:net-flip-1}. We stress that, in spite of
our considering the categories associated with the set $\Kscr^d (
\Mscr )$, the fact that $\rho_{\omega , \sigma}$ is a net isomorphism
of the nets indexed by $\Kscr^h ( \Mscr )$ is of crucial importance to
establish the claimed isomorphism. Given any pair $z$, $z_1 \in
\Zscr^1_t \big( \omega , \Kscr^d ( \Mscr ) \big)$ and $t \in ( z , z_1
)$, define
\begin{subequations}
  \label{eq:def-indep}
  \begin{align}
    \Fscr_{\omega , \sigma} ( z ) ( b ) & \doteq \rho_{\abs{b}} ( z (
    b )
    ) \text{,} & b \in \Sigma_1 \big( \Kscr^d ( \Mscr ) \big) \text{,}
    \\ 
    \Fscr_{\omega , \sigma} ( t )_a & \doteq \rho_a ( t_a ) \text{,} &
    a \in \Sigma_0 \big( \Kscr^d ( \Mscr ) \big) \text{.}
  \end{align}
\end{subequations}
The purpose is to show that $\Fscr_{\omega , \sigma} : \Zscr^1_t \big(
\omega , \Kscr^d ( \Mscr ) \big) \to \Zscr^1_t \big( \sigma , \Kscr^d
( \Mscr ) \big)$ is a $^*$-isomorphism of $C^*$-categories.
\begin{The}
  \label{the:tensor-cat}
  For any $\omega \in \Sscr_0 ( \Mscr )$ the category $\Zscr^1_t \big(
  \omega , \Kscr^d ( \Mscr ) \big)$ is a (symmetric tensor)
  $C^*$-category with left inverses. Any object with finite statistics
  has conjugates. For any $\sigma \in \Sscr_0 ( \Mscr )$
  \begin{equation*}
    \Fscr_{\omega , \sigma} : \Zscr^1_t \big( \omega , \Kscr^d ( \Mscr
    ) \big) \rightarrow \Zscr^1_t \big( \sigma , \Kscr^d ( \Mscr )
    \big)  
  \end{equation*}
  is a covariant (symmetric tensor) $^*$-isomorphism.
\end{The}
\begin{proof}
  As already mentioned, we will only prove that $\Fscr_{\omega ,
    \sigma}$ is a $^*$-isomorphism for any pair $\omega$, $\sigma \in
  \Sscr_0 ( \Mscr )$. Given $z\in \Zscr^1_t \big( \omega , \Kscr^d (
  \Mscr ) \big)$, it is clear that $\Fscr_{\omega , \sigma} ( z ) ( b
  ) \in \Afrak_\sigma ( \abs{b} )$ for any $1$-simplex $b$. Given a
  $2$-simplex $c$ we have
  \begin{align*}
    \Fscr_{\omega , \sigma} ( z ) ( \partial_0 c ) & \cdot
    \Fscr_{\omega , \sigma} ( z ) ( \partial_2 c ) =
    \rho_{\abs{\partial_0 c}} ( z ( \partial_0 c ) ) \cdot
    \rho_{\abs{\partial_2 c}} ( z ( \partial_2 c
    ) ) \\
    & = \rho_{\abs{c}} ( z ( \partial_0 c ) ) \cdot \rho_{\abs{c}} ( z
    ( \partial_2 c ) ) = \rho_{\abs{c}} ( z ( \partial_0 c ) \cdot z (
    \partial_2 c ) ) = \rho_{\abs{\partial_1 c}} ( z ( \partial_1 c )
    )
    \\
    & = \Fscr_{\omega , \sigma} ( z ) ( \partial_1 c ) \text{.}
  \end{align*}
  Hence $\Fscr_{\omega , \sigma} ( z )$ is a $1$-cocycle of $\Kscr^d (
  \Mscr )$. If $\Mscr$ is simply connected, then, by
  Proposition~\ref{pro:irr-cat-equiv}\ref{list:irr-cat-equiv1},
  $\Fscr_{\omega , \sigma} ( z )$ is a path-independent $1$-cocycle.
  For the general case consider a closed path $p$ of $\Kscr^d ( \Mscr
  )$ with endpoint $a_0$. We can find a closed path $q$ of $\Kscr^d (
  \Mscr )$, with endpoint $a_0$ and an element $\Oscr \in \Kscr^h (
  \Mscr )$ such that $p \sim q$ and $\abs{q} \subseteq \Oscr$ (see
  \cite[Lemma~4.4]{brunetti/ruzzi:2005}). Assume that $q$ is of the
  form $\set{b_n , \dots , b_1}$. By homotopic invariance of
  $1$-cocycles, we have
  \begin{align*}
    \Fscr_{\omega , \sigma} ( z ) ( p ) & = \Fscr_{\omega , \sigma} (
    z ) ( q ) = \rho_{\abs{b_n}} ( z ( b_n ) ) \cdots \rho_{\abs{b_1}}
    ( z ( b_1 ) ) \\
    & = \rho_\Oscr ( z ( b_n ) ) \cdots \rho_\Oscr ( z ( b_1 ) ) =
    \rho_\Oscr ( z ( b_n ) \cdots z ( b_1 ) ) \\
    & = \rho_\Oscr ( z ( q ) ) = \rho_\Oscr ( \algunit) = \algunit
    \text{,} 
  \end{align*}
  where path independence of $z$ has been used. Therefore,
  $\Fscr_{\omega , \sigma} ( z )$ is a path-inde\-pen\-dent
  $1$-cocycle. Now, making use of \eqref{eq:def-indep}, one can easily
  check that $\Fscr_{\omega , \sigma} ( t ) \in \big( \Fscr_{\omega ,
    \sigma} ( z ) , \Fscr_{\omega , \sigma} ( z_1 ) \big)$ for any $t
  \in ( z , z_1 )$. Moreover, since $\rho_{\omega , \sigma}$ is a net
  isomorphism, $\Fscr_{\omega , \sigma}$ is a $^*$-isomorphism.
  Indeed, given the functor $\Fscr_{\sigma , \omega}$ which is
  associated with the net isomorphism $\rho_{\sigma , \omega}$ ( the
  inverse of $\rho_{\omega , \sigma}$ ), one can easily see that
  $\Fscr_{\sigma , \omega} \circ \Fscr_{\omega , \sigma} =
  \id_{\Zscr^1_t \left( \omega , \Kscr^d ( \Mscr ) \right)}$ and that
  $\Fscr_{\omega , \sigma} \circ \Fscr_{\sigma , \omega} =
  \id_{\Zscr^1_t \left( \sigma , \Kscr^d ( \Mscr ) \right)}$.
\end{proof}
We will refer to the functor $\Fscr_{\omega , \sigma}$ as the
\textbf{flip functor}.
\paragraph*{Restriction to subregions}
Let $\Nscr \subset \Mscr$ be an open arcwise-connected subset of
$\Mscr$, such that for any pair $x_1$, $x_2 \in \Nscr$ the set $J^+ (
x_1 ) \cap J^- ( x_2 )$ is either empty or contained in $\Nscr$. This
property says that $\Nscr$ is a globally hyperbolic spacetime. As
$\Nscr$ is isometrically embedded in $\Mscr$ and diamonds are stable
under isometric embeddings (Lemma~\ref{lem:emb-diamond}) we have
\begin{equation*}
  \Kscr^d ( \Mscr) \restriction \Nscr \doteq \bsetf{\Oscr \in \Kscr^d
    ( \Mscr )}{\overline{\Oscr} \subset \Nscr} = \Kscr^d ( \Nscr )
  \text{.} 
\end{equation*}
Let $\Ascr_{\Kscr^d ( \Nscr )}$ be the net of local algebras with
indices in $\Kscr^d ( \Nscr )$, obtained by restricting
$\Ascr_{\Kscr^d ( \Mscr )}$ to $\Kscr^d ( \Nscr )$. Let $\omega \in
\Sscr_0 ( \Mscr )$, then $\omega^* \Ascr_{\Kscr^d ( \Nscr )}$ inherits
from $\omega^* \Ascr_{\Kscr^d ( \Mscr )}$ the Borchers property and
the local definiteness. However, it need not be irreducible. Let
$\Zscr^1_t \big( \omega , \Kscr^d ( \Nscr ) \big)$ be the category of
path-independent $1$-cocycles of $\Kscr^d ( \Nscr )$ with values in
$\omega^* \Ascr_{\Kscr^d ( \Nscr )}$. This is a $C^*$-category closed
under direct sums and subobjects, and, by local definiteness, the
identity cocycle is irreducible. The aim now is to show that the
restriction functor $\RE : \Zscr^1_t \big( \omega , \Kscr^d ( \Mscr
) \big) \rightarrow \Zscr^1_t \big( \omega , \Kscr^d ( \Nscr ) \big)$,
defined in Theorem~\ref{the:cat-equiv}, is a full and faithful
covariant $^*$-functor.

Let $\omega \in \Sscr_0 ( \Mscr )$ and recall that the restriction
functor is defined for any $z$, $z_1 \in \Zscr^1_t \big( \omega ,
\Kscr^d ( \Mscr ) \big)$ and $t \in ( z , z_1 )$ as
\begin{align*}
  \RE ( z ) ( b ) & \doteq z ( b ) \text{,} & b \in \Sigma_1 \big(
  \Kscr^d ( \Nscr ) \big) \text{,} \\
  \RE ( t ) ( a ) & \doteq t_a \text{,} & a \in \Sigma_0 \big(
  \Kscr^d ( \Nscr ) \big) \text{.}
\end{align*}
Note that, if we take $\sigma \in \Sscr_0 ( \Mscr )$, then it can
easily be shown that the following diagram is commutative
\begin{equation*}
  \xymatrix{
  \Zscr^1_t \big( \omega , \Kscr^d ( \Mscr ) \big) \ar[d]_\RE
  \ar[r]^{\Fscr_{\omega , \sigma}} & \Zscr^1_t \big( \sigma , \Kscr^d
  ( \Mscr ) \big) \ar[d]^\RE \\
  \Zscr^1_t \big( \omega , \Kscr^d ( \Nscr ) \big)
  \ar[r]^{\Fscr_{\omega , \sigma}} & \Zscr^1_t \big( \sigma , \Kscr^d
  ( \Nscr) \big)} 
\end{equation*}
Therefore, if $\RE$ is full and faithful for a particular choice of
$\omega$, then it is also full and faithful for any other element
$\sigma \in \Sscr_0 ( \Mscr )$.
\begin{The}
  \label{the:functor-full-faith}
  $\RE$ is a full and faithful $^*$-functor.
\end{The}
\begin{proof}
  In the first part of the proof we follow
  \cite[Theorem~30.2]{roberts:2004}. As observed above, it is enough
  to prove the assertion in the case that $\omega$ satisfies punctured
  Haag duality. Given $z$, $z_1 \in \Zscr^1_t \big( \omega , \Kscr^d (
  \Mscr ) \big)$, let $t \in \big( \RE ( z ) , \RE ( z_1 ) \big)$.
  This means that
  \begin{equation*}
    t_{\partial_0 b} \cdot z ( b ) = z_1 ( b ) \cdot t_{\partial_1 b}
    \text{,} \quad b \in \Sigma_1 \big( \Kscr^d ( \Nscr ) \big)
    \text{.} 
  \end{equation*}
  We want prove that there exists $\hat{t} \in ( z , z_1 )$ such that
  $\hat{t}_a = t_a$ whenever $a \in \Sigma_0 \big( \Kscr^d ( \Nscr )
  \big)$. Fix $a_0 \in \Kscr^d ( \Nscr )$, define
  \begin{equation*}
    \hat{t}_a \doteq z_1 ( p_a ) \cdot t_{a_0} \cdot z (p_a )^*
    \text{,} \quad a \in \Sigma_0 \big( \Kscr^d ( \Mscr ) \big)
    \text{,} 
  \end{equation*}
  where $p_a$ is a path in $\Kscr^d ( \Mscr )$ from $a_0$ to $a$. This
  definition does not depend on the chosen $a_0 \in \Sigma_0 \big(
  \Kscr^d ( \Nscr ) \big)$ and on the chosen path $p_a$. Moreover,
  $\hat{t}_{\partial_0 b} \cdot z ( b ) = z_1 ( b ) \cdot
  t_{\partial_1 b}$ for any $b \in \Sigma_1 \big( \Kscr^d ( \Mscr )
  \big)$, and
  \begin{equation*}
    \hat{t}_a = t_a \text{,} \quad a \in \Sigma_0 \big( \Kscr^d (
    \Nscr ) \big) \text{.}
  \end{equation*}
  What remains to be shown is that $t$ satisfies the locality
  condition, \ie, $\hat{t}_a \in \Afrak_{\omega} ( a )$ for any $a \in
  \Sigma_0 \big( \Kscr^d ( \Mscr ) \big)$. From now on the proof is
  very similar to the proof of \cite[Proposition 4.19]{ruzzi:2005b}.
  Let $x_0 \in \Nscr$, we show that $\hat{t}_a \in \Afrak_{\omega} ( a
  )$ for any $a \in \Sigma_0 \big( \Kscr^d ( \Mscr ) \big)$ whose
  closure $\overline{a}$ is causally disjoint from $\set{x_0}$. Fix a
  $0$-simplex $a_1$ of $\Kscr^d ( \Mscr )$ such that $\overline{a_1}
  \perp \set{x_0}$ and $a_1 \perp a$. First note that we can always
  find $a_0 \in \Sigma_0 \big( \Kscr^d ( \Nscr ) \big)$ such that $a_0
  \perp a_1$ and $\overline{a_0} \perp \set{x_0}$. Furthermore, since
  the causal complement of $a_1$ is arcwise-connected, there is a path
  $p_a$ which lies in the causal complement of $a_1$. Therefore,
  \begin{equation*}
    \hat{t}_a \cdot A = z_1 ( p_a ) \cdot t_{a_0} \cdot z ( p_a )^*
    \cdot A = z_1 ( p_a ) \cdot t_{a_0} \cdot A \cdot z ( p_a )^* = A
    \cdot \hat{t}_a
  \end{equation*}
  for any $A \in \Afrak_{\omega} ( a_1 )$. Hence, $\hat{t}_a \in
  \Afrak_{\omega} ( a_1 )'$ for any $a_1 \perp a$ and $\overline{a_1}
  \perp x$. By punctured Haag duality, $\hat{t}_a \in \Afrak_{\omega}
  ( a )$. Thus, we have shown that $\hat{t}_a \in \Afrak_{\omega} ( a
  )$ for any $0$-simplex $a$ with $\overline{a} \perp \set{x_0}$. By
  \cite[Proposition 4.19]{ruzzi:2005b}, the proof is complete.
\end{proof}
\begin{Rem}
  Two comments on Theorem~\ref{the:functor-full-faith} are in order.
  \begin{remalphalist}
  \item This is a key result. It will entail that the embedding of a
    sector into a different spacetime preserves the statistical
    properties (see Subsection~\ref{subsec:loc-cov-sectors}).
  \item Theorem~\ref{the:functor-full-faith} is nothing but the
    cohomological version of the equivalence between local and global
    intertwiners, a property that the superselection sectors which are
    preserved in the scaling limit fulfill
    \cite{dantoni/morsella/verch:2004} (see also \cite{roberts:1976}).
    We emphasize that in the present review this equivalence arises as
    a natural consequence of punctured Haag duality.
  \end{remalphalist}
\end{Rem}

\subsection{Locally Covariant Structure of Sectors}
\label{subsec:loc-cov-sectors}

In this subsection it is shown how the locally covariant structure of
superselection sectors arises. We introduce the embedding functor
which gives first important information on the covariant structure of
sectors. This structure is encoded in the superselection functor to be
analysed subsequently.
\paragraph*{The embedding functor}
Consider $\Mscr_1$, $\Mscr \in \Loc$ with $\psi \in \hom_\Loc (
\Mscr_1 , \Mscr )$, and let $\alpha_\psi : \Ascr ( \Mscr_1 )
\rightarrow \Ascr ( \Mscr )$ be the $C^*$-morphism associated with
$\psi$. Given $\omega \in \Sscr_0 ( \Mscr )$, let $\tau^\omega_\psi :
\omega^* \Ascr_{\psi \left( \Kscr^d ( \Mscr_1 ) \right)} \rightarrow (
\omega \alpha_\psi )^* \Ascr_{\Kscr^d ( \Mscr_1 )}$ be the
corresponding net isomorphism \eqref{eq:net-emb-1}. Given $z$, $z_1
\in \Zscr^1_t \big( \omega , \Kscr^d ( \Mscr ) \big)$ and $t \in ( z ,
z_1)$, we define the map of categories, $\Escr^\omega_\psi : \Zscr^1_t
\big( \omega , \Kscr^d ( \Mscr ) \big) \rightarrow \Zscr^1_t \big(
\omega \alpha_\psi , \Kscr^d ( \Mscr_1 ) \big)$, via
\begin{subequations}
  \begin{align}
    \Escr^\omega_\psi ( z ) ( b ) & \doteq \tau_\psi^\omega \big( z (
    \psi ( b ) ) \big) \text{,} & b \in \Sigma_1 \big( \Kscr^d (
    \Mscr_1 ) \big) \text{,} \\
    \Escr^\omega_\psi ( t )_a & \doteq \tau_\psi^\omega \big( t_{\psi
      ( a )} \big) \text{,} & a \in \Sigma_0 \big( \Kscr^d ( \Mscr_1 )
    \big) \text{.} 
  \end{align}
\end{subequations}
where $\psi ( b )$ is the $1$-simplex of $\Kscr^d ( \Mscr )$ defined
as $\babs{\psi ( b )} = \psi ( \abs{b} )$, $\partial_0 \psi ( b ) =
\psi ( \partial_0 b )$, $\partial_1 \psi ( b ) = \psi ( \partial_1 b
)$. Then, by using Theorem \ref{the:functor-full-faith}, it turns out
that this mapping, $\Escr^\omega_\psi$, is a covariant symmetric
tensor $^*$-functor which is full and faithful. We call
$\Escr^\omega_\psi$ the \textbf{embedding functor}. By
\eqref{eq:comm-rel}, the embedding and the flip functor enjoy the
following relation:
\begin{equation}
  \label{eq:embedding-flip}
  \Escr^\omega_\psi \circ \Fscr_{\sigma , \omega} = \Fscr_{\sigma
    \alpha_\psi , \omega \alpha_\psi} \circ \Escr^\sigma_\psi
  \text{.} 
\end{equation}
\paragraph*{The superselection functor}
We can now establish the covariant structure of superselection
sectors. Let $\Sym$ be the category, whose objects are symmetric
tensor $C^*$-categories and whose arrows are the full and faithful,
symmetric tensor $^*$-functors. According to the approach to locally
covariant quantum field theory expounded in this review, the
superselection sectors are expected to exhibit the structure of a
functor from the category $\Loc$ into the category $\Sym$. We know
that the superselection sectors of any spacetime $\Mscr \in \Loc$
identify a family of categories within the same isomorphism class, any
such category $\Zscr^1_t \big( \omega , \Kscr^d ( \Mscr ) \big)$ is
labelled by an element of the reference state space, $\omega \in \Ss_0
( \Mscr )$.  Since there is no natural way to associate an element of
this isomorphism class to the spacetime $\Mscr$, $\Mscr$ varying in
$\Loc$, we are forced to make a choice.

Given a locally covariant quantum field theory $\Ascr$ and a reference
state space $\Ss_0$, let a \emph{choice of states} be given by
\begin{equation}
  \label{eq:state-choice}
  \underline{\omega} \doteq \bsetf{\omega_\Mscr}{\Mscr \in \Loc
    \text{,} \quad \omega_\Mscr \in \Ss_0 ( \Mscr )} \text{.}
\end{equation}
We define a map of categories via
\begin{equation}
  \label{eq:def-supersel-funct}
  \begin{cases}
    \Ss_{\underline{\omega}} ( \Mscr ) \doteq \Zscr^1_t \big(
    \omega_\Mscr , \Kscr^d ( \Mscr ) \big) \text{,} & \Mscr \in \Loc
    \text{,} \\ 
    \Ss_{\underline{\omega}} ( \psi ) \doteq \Fscr_{\omega_\Mscr
      \alpha_\psi , \omega_{\Mscr_1}} \circ \Escr^{\omega_M}_\psi
    \text{,} & \psi \in \hom_\Loc ( \Mscr_1 , \Mscr ) \text{,}
  \end{cases}
\end{equation} 
and call the mapping $\Ss_{\underline{\omega}} : \Loc \rightarrow
\Sym$ the \textbf{superselection functor} associated with the choice
$\underline{\omega}$. There holds the following theorem.
\begin{The}
  \label{the:supersel-contra}           
  Given a choice of states $\underline{\omega}$ the mapping
  \begin{equation*}
    \Ss_{\underline{\omega}} : \Loc \longrightarrow \Sym
  \end{equation*}
  is a contravariant functor. If $\underline{\sigma}$ is another
  choice of states, then the functors $\Ss_{\underline{\omega}}$ and
  $\Ss_{\underline{\sigma}}$ are isomorphic.
\end{The}
\begin{proof}        
  Let $\psi \in \hom_\Loc ( \Mscr_1 , \Mscr )$. Since
  $\Ss_{\underline{\omega}} ( \psi )$ is defined as the composition of
  the flip and of the embedding functor, the above discussion of the
  embedding functor and Theorem~\ref{the:tensor-cat} imply that
  $\Ss_{\underline{\omega}} ( \psi ) : \Ss_{\underline{\omega}} (
  \Mscr ) \rightarrow \Ss_{\underline{\omega}} ( \Mscr_1 )$ is a full
  and faithful, symmetric tensor $^*$-functor. Given $\phi \in
  \hom_\Loc ( \Mscr_2 , \Mscr_1 )$, by \eqref{eq:embedding-flip}, we
  have
  \begin{align*}      
    \Ss_{\underline{\omega}} ( \phi ) \circ \Ss_{\underline{\omega}} (
    \psi ) & = \Fscr_{\omega_{\Mscr_1} \alpha_\phi , \omega_{\Mscr_2}}
    \circ \Escr^{\omega_{\Mscr_1}}_\phi \circ \Fscr_{\omega_\Mscr
      \alpha_\psi , \omega_{\Mscr_1}} \circ \Escr^{\omega_\Mscr}_\psi
    \\ 
    & = \Fscr_{\omega_{\Mscr_1} \alpha_\phi , \omega_{\Mscr_2}} \circ
    \Fscr_{\omega_\Mscr\alpha_{\psi \phi} , \omega_{\Mscr_1}
      \alpha_\phi} \circ \Escr^{\omega_{\Mscr} \alpha_\psi}_\phi \circ
    \Escr^{\omega_\Mscr}_\psi \\
    & = \Fscr_{\omega_{\Mscr} \alpha_{\psi \phi} , \omega_{\Mscr_2}}
    \circ \Escr^{\omega_{\Mscr}}_{\psi \phi} \\
    & = \Ss_{\underline{\omega}} ( \psi \phi ) \text{.}
  \end{align*}       
  Finally, the definitions of the flip and of the embedding functors
  imply
  \begin{equation*}
    \Ss_{\underline{\omega}} ( \id_{\Mscr} ) = \Fscr_{\omega_{\Mscr} ,
      \omega_{\Mscr}} \circ \Escr^{\omega_{\Mscr}}_{\id_{\Mscr}} =
    \id_{\Zscr^1_t \left( \omega_\Mscr , \Kscr^d ( \Mscr ) \right)} =
    \id_{\Ss_{\underline{\omega}} ( \Mscr )} \text{,}
  \end{equation*}
  so that $\Ss_{\underline{\omega}}$ turns out to be a contravariant
  functor.  For the rest of the proof see \cite{brunetti/ruzzi:2005}
\end{proof}
This result establishes the covariance of charged superselection
sectors. If $\psi \in \hom_\Loc ( \Mscr_1 , \Mscr )$, then to any
sector of $\Mscr$ corresponds a unique sector of $\Mscr_1$ with the
same charge quantum numbers. To be precise, let $z \in
\Ss_{\underline{\omega}} ( \Mscr )$ be an irreducible object with
statistical parameter $\lambda ( [ z ] ) = \chi ( [ z ] ) \cdot d ( [
z ] )^{- 1}$, where $[ z ]$ denotes the equivalence class of $z$. Let
$\overline{z}$ be the conjugate of $z$. Then $\Ss_{\underline{\omega}}
( \psi ) ( z )$ is an irreducible object of $\Ss_{\underline{\omega}}
( \Mscr_1 )$ such that
\begin{equation*}
  \big[ \Ss_{\underline{\omega}} ( \psi ) ( z ) \big] =
  \Ss_{\underline{\omega}} ( \psi ) ( [ z ] ) \text{.}
\end{equation*}
Furthermore, $z$ and $\Ss_{\underline{\omega}} ( \psi ) ( z )$ have
the \emph{same} statistics, \ie,
\begin{equation*}
  \chi ( [ z ] ) = \chi \big( \big[ \Ss_{\underline{\omega}} ( \psi
  ) ( z ) \big] \big) \text{,} \quad d ( [ z ] ) = d \big( \big[
  \Ss_{\underline{\omega}} ( \psi ) ( z ) \big] \big) \text{.}
\end{equation*}
Moreover,
\begin{equation*}
  \Ss_{\underline{\omega}} ( \psi ) ( [ \overline{z} ] ) =
  \overline{\big[ \Ss_{\underline{\omega}} ( \psi ) ( z ) \big]}
  \text{,}  
\end{equation*}
\ie, $\Ss_{\underline{\omega}} ( \psi ) ( [ \overline{z} ] )$ is the
conjugate sector of $\Ss_{\underline{\omega}} ( \psi ) ( z )$.

\providecommand{\SortNoop}[1]{}

\end{document}